\documentclass[letterpaper, 10 pt, conference]{ieeeconf}  

\IEEEoverridecommandlockouts                              
\usepackage{amsmath, amssymb, url}
\usepackage{tikz}
\usepackage{xcolor,soul}
\usepackage{verbatim}
\usepackage{multirow}
\usepackage{colortbl}
\usepackage{pgfplots}
\DeclareUnicodeCharacter{2212}{−}
\usepgfplotslibrary{groupplots,dateplot}
\usepgfplotslibrary{fillbetween}
\usetikzlibrary{patterns,shapes.arrows}
\pgfplotsset{compat=newest}
\usetikzlibrary{shapes.arrows}
\usetikzlibrary{arrows.meta} 
\usetikzlibrary{positioning, fit, shapes.geometric}
\usetikzlibrary{backgrounds}
\usepackage{algorithm}
\usepackage{algpseudocode}
\usepackage{mathtools}
\usepackage{float} 

\definecolor{newblue}{RGB}{189,215,239} 
\definecolor{newgray}{RGB}{214,214,214} 
\definecolor{neworange}{RGB}{241,205,177} 
\definecolor{newyellow}{RGB}{255,193,19} 
\definecolor{newgreen}{RGB}{202,223,184} 

\usepackage{subcaption}
\usetikzlibrary{ext.paths.ortho}
\usepackage{listings}
\usepackage{tikz-3dplot}

\let\labelindent\relax
\usepackage{enumitem}

\newcommand{\param}{\boldsymbol{\theta}}
\newcommand{\paramr}{\boldsymbol{\theta}}
\newcommand{\paramc}{\boldsymbol{\theta}}

\newcommand{\eg}{\textit{e.g., }}
\newcommand{\Reward}{R^{[i]}(x(\cdot), u(\cdot); {\paramr^{[i]}})}
\newcommand{\Constraint}{\boldsymbol{\Phi}^{[i]}(x(\cdot), u(\cdot); {\paramr^{[i]}})}
\newcommand{\TReward}{R^{[i]}(x(t), u(t); {\paramr^{[i]}})}

\newcommand{\RewardS}{R^{[i]}(\cdot; \paramr^{[i]})}
\newcommand{\ConstraintS}{\boldsymbol{\Phi}^{[i]}(\cdot; \paramr^{[i]})}
\newcommand{\Subtask}{R^{[i]},\boldsymbol{\Phi}^{[i]}}
\newcommand{\myL}{\mathcal{L}(\{R^{[i]}, \boldsymbol{\Phi}^{[i]}, \param^{[i]}\}_{i \in \mathcal{I}})}

\algrenewcommand{\Return}{\State\algorithmicreturn~}
\algrenewcommand\algorithmicrequire{\textbf{Input:}}
\algrenewcommand\algorithmicensure{\textbf{Output:}}

\makeatletter
\newcommand{\ALOOP}[1]{\ALC@it\algorithmicloop\ #1%
  \begin{ALC@loop}}
\newcommand{\ENDALOOP}{\end{ALC@loop}\ALC@it\algorithmicendloop}
\renewcommand{\algorithmicrequire}{\textbf{Input:}}
\renewcommand{\algorithmicensure}{\textbf{Output:}}

\makeatother

\title{\LARGE \bf
Context-aware LLM-based Safe Control Against Latent Risks
}

\author{Xiyu Deng$^{1}$, 
        Quan Khanh Luu$^{2}$, 
        Van Anh Ho$^{2}$, 
        and Yorie Nakahira$^{1}$ 
\thanks{$^{1}$Authors are with the Department of Electrical and Computer Engineering, Carnegie Mellon University, Pittsburgh, Pennsylvania, United States.
        {\tt\small \{xiyud, ynakahir\}@andrew.cmu.edu}}%
\thanks{$^{2}$Authors are with the School of Materials Science, Japan Advanced Institute of Science and Technology, Nomi, Ishikawa, Japan.
        {\tt\small \{quan-luu, van-ho\}@jaist.ac.jp}}%
}

\begin{document}

\maketitle
\thispagestyle{empty}
\pagestyle{empty}



\begin{abstract}
Autonomous control systems face significant challenges in performing complex tasks in the presence of latent risks. To address this, we propose an integrated framework that combines Large Language Models (LLMs), numerical optimization, and optimization-based control to facilitate efficient subtask learning while ensuring safety against latent risks. The framework decomposes complex tasks into a sequence of context-aware subtasks that account for latent risks. These subtasks and their parameters are then refined through a multi-time-scale process: high-layer multi-turn in-context learning, mid-layer LLM Chain-of-Thought reasoning and numerical optimization, and low-layer model predictive control. The framework iteratively improves decisions by leveraging qualitative feedback and optimized trajectory data from lower-layer optimization processes and a physics simulator. We validate the proposed framework through simulated case studies involving robot arm and autonomous vehicle scenarios. The experiments demonstrate that the proposed framework can mediate actions based on the context and latent risks and learn complex behaviors efficiently.
\end{abstract}


\section{Introduction}
\label{sec:intro}
Autonomous control systems must manage complex tasks while anticipating potential risks. However, existing control techniques often rely on manually engineered design specifications and struggle to handle diverse contexts and latent risks not explicitly given at the design phase. Large language models (LLMs) and human inputs have demonstrated the potential to decompose complex tasks~\cite{Yu2023}, make informed guesses~\cite{zhao2023expel}, and reason about failures~\cite{huang2022inner}. Despite this potential, the decisions produced by LLMs are not always reliable or sufficient to ensure successful task completion.

The decisions must also ensure safety in the presence of latent risks--those associated with unobserved variables or unspecified in the design objectives. One example is when a risk-critical object is not directly observable, such as children hidden behind a school bus. Another example is when certain risk scenarios are left out of the design specifications. For example, a risk estimator may not differentiate individuals by age, but small children may exhibit unexpected/unsafe behaviors, such as suddenly leaping off a sidewalk.
While existing control techniques can handle well-specified risks associated with the observable hazardous objects or states, they often fall short when critical risk factors are unobservable or overlooked in the system’s design, possibly excluding these latent risks from the decision making process entirely. 

\begin{figure}[t]
  \begin{center}
    \definecolor{newblue}{RGB}{189,215,239} 
\definecolor{newgray}{RGB}{214,214,214} 
\definecolor{neworange}{RGB}{241,205,177} 
\definecolor{newyellow}{RGB}{251,231,163} 
\definecolor{newgreen}{RGB}{202,223,184} 

\pgfdeclarelayer{bg}    
\pgfsetlayers{bg,main}  

\begin{tikzpicture}[
    node distance = 4cm,
    node/.style={circle, fill=#1, inner sep=2pt, minimum size=0.7cm}, 
    edge/.style={-stealth, #1},
    arr/.style = {semithick, -Stealth}
]

\node[rectangle, draw = black, fill = white, line width=0.5mm, minimum height=0.7cm,minimum width=2.5cm,inner sep=-1pt](task){\scriptsize\begin{tabular}{cc}
Task instruction\\Evaluation metric
\end{tabular}};

\node[rectangle, draw, fill = white, line width=0.5mm, minimum height=0.7cm,minimum width=2.5cm,inner sep=-1pt][below = 0.3cm of task](context){\scriptsize\begin{tabular}{cc}
Verbalize context\\
Anticipate latent risk
\end{tabular}};

\node[rectangle, draw, fill = white, line width=0.5mm, minimum height=1cm,minimum width=2.5cm,inner sep=-1pt][below = 0.8cm of context](subtask){\scriptsize\begin{tabular}{cc}
\textit{Candidate} \\
Subtask sequence w/\\
subtask specifications
\end{tabular}};

\node[rectangle, draw, fill = white, line width=0.5mm, minimum height=1cm,minimum width=1.6cm,inner sep=1pt][right = 1.8cm of subtask](failure){\scriptsize\begin{tabular}{cc}
Failure or \\
suboptimal \\trials
\end{tabular}};
\node[rectangle, line width=0.5mm, minimum height=1.5cm,minimum width=2cm,inner sep=1pt][above right = -0.1cm and 3cm of subtask](blockfailure1){};
\node[rectangle, line width=0.5mm, minimum height=1cm,minimum width=2cm,inner sep=1pt][below right = -0.1cm and 3cm of subtask](blockfailure2){};


\draw [edge, line width=0.5mm, draw, newblue!300] (failure.south) -- ++(0,-0.65) node(lowerright){} -| (subtask.south);
\node[rectangle, fill = newblue!30, minimum height=0.5cm,minimum width=3cm,inner sep=-3pt][below left = 0.4cm and -0.4cm of failure](optim){\scriptsize\begin{tabular}{cc}
Parameter optimization 
\end{tabular}};

\node[rectangle, draw, fill = white, line width=0.5mm, minimum height=1.5cm,minimum width=2cm,inner sep=1pt][right = 0.3cm of failure](newsubtask){\scriptsize\begin{tabular}{cc}
\textit{Optimal}\\
Subtask \\
sequence w/\\
specifications
\end{tabular}};

\draw[edge,line width=0.5mm, neworange!300] (task) -> (context);
\draw[edge,line width=0.5mm, neworange!300] ([xshift = 15pt]context.south) -> ([xshift = 15pt]subtask.north);

\draw[edge,line width=0.5mm, neworange!300] ([xshift = -15pt]context.south) -> ([xshift = -15pt]subtask.north);
\draw[edge,line width=0.5mm, newgreen!300] (subtask) -> (failure);
\node[rectangle, fill = newgreen!30, line width=0.5mm, minimum height=1cm,minimum width=1cm,inner sep=-5pt][right = 0.1cm of subtask](obc){\scriptsize\begin{tabular}{cc}
Optimization\\ -based \\ controller
\end{tabular}};
\draw[edge,line width=0.5mm, newblue!300] (failure) -> (newsubtask);
\node[rectangle, fill = neworange!30, minimum height=0.5cm,minimum width=2.5cm,inner sep=-3pt][below = 0.1cm of context](decomp){\scriptsize\begin{tabular}{cc}\textcolor{neworange!300}{
Task}\\
\textcolor{neworange!300}{decomposition}
\end{tabular}};

\node[rectangle,rounded corners, minimum height=0.8cm,minimum width=2cm,inner sep=1pt][below right= 0.1cm and -0.2cm of context](corr){\scriptsize\begin{tabular}{cc}
\textcolor{newgreen!300}{Task} \\
\textcolor{newgreen!300}{correction}
\end{tabular}};

\draw[edge,line width=0.5mm, newgreen!300] (failure) |- (context);

\node[rectangle, fill = newgreen!30, minimum height=0.5cm,minimum width=2cm,inner sep=-1pt][above = 0.5cm of failure](err){\scriptsize\begin{tabular}{cc}
Subtask correction \\
(LLM error analysis)
\end{tabular}};

\begin{pgfonlayer}{bg}


\node[rectangle,draw = newblue, fill=newblue!30, fit=(context)(subtask)(err)(obc)(optim),  dashed, line width=0.1mm, inner sep=1.5mm] (point3) {};
\node[rectangle,draw = newgreen,fill=newgreen!30, fit=(context)(subtask)(err)(obc),  dashed, line width=0.1mm, inner sep=0.7mm] (point2) {};
\node[rectangle,draw = neworange, fill=neworange!30, fit=(context)(subtask),  dashed, line width=0.1mm, inner sep=0mm] (point1) {};

\node[rectangle, draw = black, fit=(task)(context)(blockfailure1)(decomp),  dashed, line width=0.4mm, inner sep=1.5mm] (block1) {};
\node[rectangle, minimum height=0.8cm,minimum width=2.4cm,inner sep=1pt][above left =-0.2cm and -2.5cm of block1](reasoning){\scriptsize{Qualitative reasoning}};
\node[rectangle, rounded corners, line width=0.2mm, minimum height = 0.4cm, minimum width = 2.5cm, inner sep=-3pt] [below = 2cm of decomp](blanktd) {};
\node[rectangle, draw = black, fit=(blockfailure2) (blanktd)(optim),  dashed, line width=0.4mm, inner sep=1.5mm] (block2) {};
\node[rectangle, minimum height=0.8cm,minimum width=2.4cm,inner sep=1pt][below right =-0.2cm and -4.2cm of block2](physical){\scriptsize{Quantitative optimization and validation}};

\end{pgfonlayer}




\end{tikzpicture}
  \end{center}
  \vspace{-0.4cm}
  \caption{Overview of the proposed framework for generating context-aware decisions that anticipate latent risks. The orange, green, and blue shaded areas correspond to the three central design components introduced in Section~\ref{sec:intro}.}
  \label{fig:process_diagram}
  \vspace{-0.7cm}
\end{figure}

In this paper, we introduce a novel framework for \emph{efficient subtask learning} and \emph{ensuring safety against latent risks} (Fig. \ref{fig:process_diagram}). Central to our framework are the following design choices. 
First, the proposed method breaks down complex tasks into a sequence of simpler subtasks while augmenting them with additional details. This decomposition makes the control of individual subtasks easier, as each subtask does not need to consider future ones and can therefore operate with a shorter outlook time horizon. Moreover, when generating these subtasks, the LLM may introduce additional specifications beyond those explicitly identified by the system designers, enhancing the flexibility and completeness of the task representation.
Second, our framework implements a multi-level optimization structure. A middle layer optimizes subtask sequences, while the lower layer optimizes control actions. The former optimization process use LLM chain-of-thought reasoning and numerical optimization. The latter procedure uses Model Predictive Control (MPC) to find the corresponding actions to achieve individual subtasks. 
A higher-layer process performs multi-turn in-context learning, guided by data generated from middle- and lower-layer optimizations. Third, this learning process integrates both qualitative feedback and optimized trajectory data from failed and successful trials. The use of multi-turn learning (prompting) allows incremental improvement based on the data from middle and low-layer optimization processes and the physics simulator. 
To balance contextual and risk-aware reasoning with real-time control, the three-layered architecture operates in distinct time scales—from slow-timescale in-context learning, to intermediate-timescale subtask refinement, and to fast-timescale control. The above-mentioned design dimension is summarized in Fig.~\ref{fig:cube}. The influence of these design dimensions is evaluated using ablation experiments of simulated environments (Section~\ref{sec:simulation}). 

\subsection*{Related work}

\textit{LLMs in control tasks.}
Recent research has shown that large language models (LLMs) can be used to decompose complex tasks into manageable subtasks and even generate reward functions for control problems. For example, LLMs can convert language commands into a quantitative design objective for optimization-based control~\cite{huang2023voxposer,Yu2023} and decompose complicated tasks into simpler subtasks~\cite{yang2023plug}. LLMs have been used to describe failure scenarios and potential causes~\cite{liu2023reflect}. 
LLMs are used to provide rough initial guesses and structured reasoning~\cite{zhao2024expelllmagentsexperiential}.
LLMs are also used to identify high-uncertainty areas~\cite{ren2023robots}.
In this paper, we integrate subtask decomposition, structured reasoning, and failure analysis into the framework of multi-timescale, multi-level optimization for real-time control in the presence of latent risks.


\textit{Handling unknowns in control.}
When the goal is to generate safe actions, safety specifications are usually incorporated into the objective function/constraints of optimization-based controllers or policy learning algorithms~\cite{huang2021toward,marvi2021safe}. 
While those approaches are adept at handling apparent risks, they struggle to account for contexts and latent risks when risk-critical variables are not observable (\eg the presence or absence of occluded objects is unknown). 
Thus, a critical challenge in autonomous systems is anticipating and managing latent risks. 
%
One approach for managing latent risk is to eliminate unknowns. For example, in the context of autonomous ground vehicles, improving V2V and V2X communication helps to explore occlusions~\cite{ahangar2021survey, muller2022motion}. 
Another approach is to explicitly quantify unknowns and risk~\cite{wang2023physics,hoshino2024physics,Wright2024propagation,wang2023generalizable}. When unknown risks are converted into apparent or well-specified risks, many techniques can be used to generate uncertainty- or risk-aware decisions and to design safety certificate~\cite{gangadhar2023occlusion,cao2022trustworthy}. Those safety certificates can be modularly attached to a black box controller (which uses neural networks, transformers, and language models) to modify unsafe actions~\cite{miyaoka2024cbfllmsafecontrolllm}.  In the face of a large risk, one can either halt the system or transfer control to human supervisors~\cite{biondi2019human}. 
However, these methods often require predefined structures of unknowns (\eg defining occlusions based on sensor geometry) and cannot utilize contextual descriptions in natural language. The proposed method leverages the large dataset used to train LLMs, covering risks that may be overlooked in design specifications. Unlike the techniques that rely on fixed safety certificates, the proposed method refines and validates subtasks and policies (determined by reward and constraints) through an automated design loop.

\textit{Prompts, feedback, and learning to improve LLM.}
Previous studies have combined text and image prompts for LLMs to generate task plans, such as cooking instructions with pictures~\cite{lu2023multimodal}. Verbal reflections on task outputs have been shown to improve coding performance~\cite{shinn2023reflexion}. Feedback mechanisms and solution trees are suggested to enhance LLMs' task completion and tool-usage capabilities~\cite{wu2024toolplanner}. Requiring LLMs to describe the plan execution process and analyze failures has also been found to improve output quality~\cite{wang2023describe}. Similarly, environmental feedback can be used as prompts to guide LLMs in generating actions, which are then used in supervised fine-tuning~\cite{feng2023llama}. Our study is inspired by this body of work but has the following differences and innovations. First, we focus on dynamic systems with continuous state and action spaces, unlike prior work that emphasizes language-based or discrete actions~\cite{lu2023multimodal, shinn2023reflexion, wu2024toolplanner, wang2023describe, feng2023llama}. Second, we tackle latent risks by using chain-of-thought prompting to anticipate and mitigate hazards from latent variables or unspecified safety criteria. Third, we introduce a multi-timescale optimization strategy that balances real-time control with long-term planning. Fourth, our approach integrates in-context learning with chain-of-thought prompting, unlike previous methods that treat action generation and fine-tuning separately~\cite{wang2023describe, wang2023voyager, lin2023swiftsage, feng2023llama}. Additionally, our feedback mechanism incorporates numerical optimization, reducing the LLM's need to find optimal task parameters or adapt to system dynamics through fine-tuning.


\section{Problem Statement} \label{sec:problem_statement}

The proposed framework takes task instruction in natural language, video measurements, and evaluation metrics to find the decisions that fulfill the task safely. We consider a discrete system represented as:
\begin{equation} \label{eq:system}
\begin{split}
    x(t+1) = f[x(t)] + g[x(t),u(t)],
\end{split}
\end{equation}
where $x\in\mathbb{R}^n$ is the system state, $u\in\mathbb{R}^m$ is the control action, $f: \mathbb{R}^n \mapsto \mathbb{R}^n$ and $g: \mathbb{R}^{n\times m} \mapsto \mathbb{R}^{n}$ determine system dynamics, and $t \in \mathbb{Z}_+$ denotes time.

The design specification of the controller is given by a language command $l \in \mathcal{T}$, which qualitatively specifies the objective; video measurements $I \in \mathcal{V}$ capturing the surrounding environment; and an evaluation metric $\mathcal{R}(x_{0:T}, u_{0:T})$ that quantitatively assesses performance and safety.
Here, $\mathcal{T}$ is a set of commands or descriptions in natural language, $\mathcal{V}$ represents the set of image sequences, and $\mathcal{R}$ is a function of state $x_{0:T} = \{x(t), t\in \{0,1,\cdots,T\}\}$ and control action $u_{0:T} = \{u(t), t\in \{0,1,\cdots,T\}\}$ up to the terminal time $T$.
Our goal is to design an effective layered control architecture to achieve the task specifications. For example, the specification may require a robot arm to transport a box along a prescribed trajectory, or an autonomous ground vehicle to navigate to a target location while mitigating latent risks.

\section{Proposed Method}
\begin{figure*}[!t]
  \centering
  \vspace{0.2cm}
  \input{figures/workflow}
  \caption{Workflow of the proposed framework (Algorithm~\ref{alg:proposed_framework}). The orange part denotes the subtask generation and contextual reasoning in Algorithm~\ref{alg:context-resoning}. The green part is for optimization-based control in Algorithm~\ref{alg:mpc}. The blue part illustrates self-adaptive reasoning with zeroth-order optimization in Algorithm~\ref{alg:proposed_framework}, lines 6-17. The module $\mathtt{LLM}_{\mathtt{coder}}$ utilizes in-context learning, while other LLM components operate in a zero-shot/few-shot setting.}
  \label{fig:workflow}
\end{figure*}


\textbf{Overview.} 
The overview of the proposed framework for offline design is illustrated in Fig.~\ref{fig:workflow} and Algorithm~\ref{alg:proposed_framework}, and for online control (execution) is presented in Algorithm~\ref{alg:online_framework}.
%
The framework first utilizes LLMs to decompose the task into a sequence of subtasks, each represented by a reward function and a set of constraint functions (Section~\ref{sec:subtask}, Algorithm~\ref{alg:proposed_framework}, line 3). This process uses \textit{LLM Chain-of-Thought} to identify latent risks that may be missed in the design specification of the controller. 
For each subtask, the framework employs an \textit{optimization-based control} to iteratively find actions that optimize its reward under the given constraints (Section~\ref{sec:mpc}, Algorithm~\ref{alg:proposed_framework} line 5).
If the subtask decomposition or the associated reward and constraint functions are inadequate, the task may result in failure. To mitigate such failures, the framework integrates \textit{multi-turn in-context reasoning and zeroth-order optimization} to identify valid subtask sequences and refine the corresponding parameters (Section~\ref{sec:llm_sgd}, Algorithm~\ref{alg:proposed_framework}, lines 6-17). 

The framework integrates three types of feedback loops operating at different temporal scales to achieve both responsiveness and adaptability (Fig.~\ref{fig:speed}). 
At the fast timescale, the low-level controller executes real-time actions using MPC, enabling immediate responsiveness to dynamic environmental changes.
At the intermediate timescale, a reasoning module monitors recent performance and context shifts, refining subtask specifications through language-based reasoning.
At the slowest timescale, the system aggregates subtask logs and reasoning traces for in-context learning, improving the LLM module to enable more reliable future updates.
\vspace{-0.5cm}
\begin{figure}[h]
  \centering
    \definecolor{newblue}{RGB}{189,215,239} 
\definecolor{newgray}{RGB}{214,214,214} 
\definecolor{neworange}{RGB}{241,205,177} 
\definecolor{newyellow}{RGB}{255,193,19} 
\definecolor{newgreen}{RGB}{202,223,184} 
\tikzstyle{line} = [draw, thick, edge]
\tikzstyle{loosely dashdotted}=      [dash pattern=on 3pt off 4pt on \the\pgflinewidth off 4pt]
\tikzstyle{densely dashdotted}=      [dash pattern=on 3pt off 1pt on \the\pgflinewidth off 1pt]

\begin{tikzpicture}[
    node distance = 4cm,
    node/.style={circle, fill=#1, inner sep=2pt, minimum size=0.7cm}, 
    edge/.style={-stealth, #1},
    arr/.style = {semithick, -Stealth},
    arr/.style = {semithick, -Stealth},
    cross/.style={path picture={ 
  \draw[black]
(path picture bounding box.south east) -- (path picture bounding box.north west) (path picture bounding box.south west) -- (path picture bounding box.north east);
}}
]
\node[draw=newgreen!300, circle, minimum size=1.5cm, line width=0.5mm] (mpc) {\scriptsize MPC};

\draw[line width=0.5mm, edge, draw=newblue!300]
  (mpc) to[out=150, in=210, looseness=8]
  node[midway, right] {\scriptsize \begin{tabular}{@{}c@{}}Refinement\\subtasks\end{tabular}} (mpc);

\node[circle, minimum size=1.5cm] at ([xshift=-1.5cm]mpc) (mpcghost) {}; 

\draw[line width=0.5mm, edge, draw=newyellow!100]
  (mpcghost) to[out=150, in=210, looseness=8] node[midway, right] {\scriptsize \begin{tabular}{@{}c@{}}In-context \\learning \\of LLM \end{tabular}} (mpcghost);

\shade[left color=newyellow!100, middle color=newblue!300, right color=newgreen!300]
    (-4, -1) rectangle (1, -0.95);
\node[below] at (-4, -1.05) {\scriptsize Slow};
\node[below] at (1, -1.05) {\scriptsize Fast};
  

\end{tikzpicture}
    \caption{Three types of feedback loops at different temporal scale.}
    \label{fig:speed}
    \vspace{-0.5cm}
\end{figure}
\begin{algorithm}[!ht]
  \caption{Proposed framework (offline)}\label{alg:proposed_framework}
  \begin{algorithmic}[1]
    \Require $l$: task instructions, $I$: environment video clips, $x(0)$: initial state, $\mathcal{R}(x_{0:T},u_{0:T})$: evaluation metric

    \Ensure $\{\Subtask, \param^{[i]}\}_{i \in \mathcal{I}}$: adjusted subtask sequence and optimized parameters
    
    \State \textbf{Initialize} $\mathcal{P}_{\mathtt{correction}},\mathcal{P}_{\mathtt{\mathtt{coder}}},\epsilon, \mathcal{B}_{\mathtt{f}}\gets \emptyset$
    \Statex \Comment{prompts, convergence threshold and memory buffer}
    \State \textbf{Initialize}  $\mathtt{LLM_{\mathtt{correction}}}(\mathcal{P}_{\mathtt{correction}}), \mathtt{LLM_{\mathtt{coder}}}(\mathcal{P}_{\mathtt{coder}})$
    \Statex \Comment{initialize LLMs with prompts}
    
    \State $\{R^{[i]}, \boldsymbol{\Phi}^{[i]}, \param^{[i]}\}_{i \in \mathcal{I}} \gets$ Algorithm \ref{alg:context-resoning}$(l, I)$ 

    \While {$\mathtt{True}$}
    \State $x_{0:T}, u_{0:T}, \boldsymbol{D}$
    \Statex $\qquad\qquad\qquad \gets\text{Algorithm~\ref{alg:mpc}}(\{R^{[i]}, \boldsymbol{\Phi}^{[i]}, \param^{[i]}\}_{i \in \mathcal{I}},x(0))$
    \Statex \Comment{run MPC for all the subtasks}
    \If{$\boldsymbol{D}$ is $\mathtt{False}$}
    \State $\infty \gets \mathcal{R}(x_{0:T},u_{0:T})$ \Comment{task failed}
    \State $\mathcal{F}\gets \mathtt{FeedbackTemplate}(x_{0:T}, u_{0:T})$
    \State $c\gets \mathtt{LLM_{correction}}(\mathcal{F})$
    \State $\mathcal{B}_{\mathtt{f}}\gets \text{CONCAT}(\{\mathcal{F}, c, \{\Subtask, \param^{[i]}\}_{i \in \mathcal{I}}\},\mathcal{B}_{\mathtt{f}})$
    \Statex \Comment{$\mathcal{F}$ is optional}
    \State $\{R^{[i]},\boldsymbol{\Phi}^{[i]}, \param^{[i]}\}_{i\in\mathcal{I}}\gets \mathtt{LLM_{coder}}(l, \mathcal{B}_{\mathtt{f}})$ 
    \Statex \Comment{the total number of subtask $N = |\mathcal{I}|$ may change}
    \ElsIf{gradient $\nabla L(\{R^{[i]}, \boldsymbol{\Phi}^{[i]}, \param^{[i]}\}_{i \in \mathcal{I}})> \epsilon$} 
    \Statex \Comment{the whole task succeeds, need optimization}

    \State $\{\param^{*[i]}\}_{i\in\mathcal{I}} \gets \mathtt{Optimize}_\epsilon(\{R^{[i]}, \boldsymbol{\Phi}^{[i]}, \param^{[i]}\}_{i \in \mathcal{I}})$
    \State $\{\param\}_{i\in\mathcal{I}} \gets \{\param^{*[i]}\}_{i\in\mathcal{I}}$
    
    \Else
    \State \textbf{break} \Comment{termination on optimized task success}
    \EndIf
    \If{$\mathtt{INTERRUPTION}$}  
    \textbf{break} 
    \Statex \Comment{termination from human or system-triggered event}
    \EndIf

\EndWhile

    

    
  \end{algorithmic}
  
\end{algorithm}

\subsection{Subtask generation and Chain-of-Thought} \label{sec:subtask}
This section describes the procedure for interpreting task instructions and processing video measurements into a sequence of subtasks (Algorithm~\ref{alg:context-resoning}). 
The framework first utilize $\mathtt{LLM_{vision}}: \mathcal{V} \mapsto \mathcal{T}$, initialized with the prompt $\mathcal{P}_{\mathtt{vision}}$, to translate the camera measurement $I \in \mathcal{V}$ into a language-based description $l_{I}\in\mathcal{T}$ (line 2). The description $l_{I}$ is then fed into the latent-risk query module, which employs $\mathtt{LLM_{latentobject}}: \mathcal{T} \times \mathcal{T} \mapsto \mathcal{O}$ with $\mathcal{P}_{\mathtt{latentobject}}$ to deduce the latent-risk object $o \in \mathcal{O}$, based on the task requirement $l$ and video description $l_{I}$ (line 3). Upon identification of the latent risk object $o$, the $\mathtt{LLM_{riskhandling}}: \mathcal{O}\mapsto\mathcal{T}$, initialized with $\mathcal{P}_{\mathtt{riskhandling}}$, generates a natural language command $s\in\mathcal{T}$ that prescribes a safety-preserving solution for handling the identified risk (line 4). 
The $\mathtt{LLM_{coder}}$, initialized with prompt $\mathcal{P}_{\mathtt{coder}}$, then generates a sequence of subtasks that explicitly accounts for these latent risks. The subtasks are indexed by $i\in\mathcal{I}=\{1, 2, \cdots, N\}$, where $N$ denotes the number of subtasks. 
Specifically, the $\mathtt{LLM_{coder}}$ takes the task requirement $l$ and solution $s$ as input to specify reward functions $\Reward:\mathbb{R}^{n}\times\mathbb{R}^{m}\mapsto\mathbb{R}$ and constraint functions $\Constraint := \{\Phi^{[i]}_k(x(\cdot), u(\cdot); \paramr^{[i]}_{k}):\mathbb{R}^{n}\times\mathbb{R}^{m}\mapsto\mathbb{R},\,\,\forall k\in\{1, 2, \cdots, M^{[i]}\}\}$, where $M^{[i]}$ is the number of safety constraint functions for subtask $i$ (line 5). These constraint functions can capture diverse safety considerations, such as maintaining safe distances, limiting contact forces, or avoiding high speed, including those that address latent risks. 
The sequence of reward and constraint functions $\{\Subtask\}_{i \in \mathcal{I}}$ is parameterized by $\{\paramc^{[i]}\}_{i \in \mathcal{I}}$. Here, a latent risk-critical object or variable $o$ can be captured by multiple constraint functions to mitigate its risks. For instance, if an autonomous vehicle (AV) is obstructed by a school bus, the latent-risk object can be assigned as $o\leftarrow\ $\textit{children}, leading to constraints that enforce slowing down and maintaining a safe distance from children. Example prompts for above LLM modules are provided in Appendix~\ref{sec:av_simulation}.

\subsection{Optimization-based control}\label{sec:mpc}
This section describes the procedure for executing the sequence of subtasks using optimization-based control and self-adaptive reasoning (Algorithm~\ref{alg:proposed_framework}, lines 4-11 and Algorithm~\ref{alg:mpc}). Specifically, a model predictive controller (MPC) is used to find an optimal control action at each time step for a given subtask $i$, which is defined by a reward function $\Reward$ and a set of constraint functions $\Constraint$, as follows:
\begin{equation} \label{eq:optimization-control}
\begin{split}
    u_{t:t_0^{[i]}+\omega^{[i]}}^{*} =\, &\underset{{u_{t:t_0^{[i]}+\omega^{[i]}}}}{\arg \min}\sum_{t=t}^{t_0^{[i]}+\omega^{[i]}} -\TReward,\\
    \text{s.t.}\quad &x(t+1) = f[x(t)] + g[x(t), u(t)], \\
    &\Phi^{[i]}_k(x(t), u(t); \paramr^{[i]}) \geq  0,
\end{split}
\end{equation}
for all $t\in\{t_0^{[i]},t_0^{[i]}+1,\cdots t_0^{[i]}+\omega^{[i]}\}$, $k\in\{1, 2, \cdots, M^{[i]}\}$, where $t_0^{[i]}$ denotes the beginning timestep, $\omega^{[i]}$ is the MPC time horizon. At each time step, only the first action $u(t) = u^*(t)$ from the optimal control sequence is executed. This process is defined as:
\begin{equation} \label{eq:mpc}
    u(t) = \text{MPC}[\RewardS,\,\ConstraintS, x(t)].
\end{equation}
For each subtask $i$, the MPC attempts to find a control sequence that achieves the completion condition within the time window.  
In our case, both the time window $\omega^{[i]}$ and the completion threshold $\varphi^{[i]}$ are encapsulated in the parameter set $\paramr^{[i]}$. If subtask $i$ fails to meet the completion threshold $\varphi^{[i]}$ within $\omega^{[i]}$, both the subtask completion flag $D^{[i]}$ and the overall task completion flag $\boldsymbol{D}$ are set to $\mathtt{False}$. Alternative methods for defining task completion, such as assertion checks on task-relevant state variables~\cite{singh2023progprompt} or visual grounding for monitoring execution progress~\cite{song2023llm}, can optionally be integrated depending on system design requirements.

When a subtask fails, the evaluation metric $\mathcal{R}(x_{0:T}, u_{0:T})$ is assigned an infinite value, and feedback $\mathcal{F}\in\mathcal{T}$ is generated from the executed trajectory $(x_{0:T}, u_{0:T})$ by populating a predefined $\mathtt{FeedbackTemplate}$ with the corresponding state and action information (Algorithm~\ref{alg:proposed_framework}, lines 7-8). Then, the $\mathtt{LLM_{correction}}$ module, initialized with prompt $\mathcal{P}_{\mathtt{correction}}$, processes the feedback $\mathcal{F}$ to a natural language correction command $c\in\mathcal{T}$ (Algorithm~\ref{alg:proposed_framework}, line 9). 
The correction command $c$ is then fed into $\mathtt{LLM_{coder}}$ with optional buffers $\mathcal{B}_{\mathtt{f}}$, which indicate buffers for failure cases. This buffer, along with the in-context learning component, will be described in detail in Section~\ref{sec:llm_sgd} (Algorithm~\ref{alg:proposed_framework}, lines 10-11).
After several loops of reasoning and correction, when all subtasks are successfully completed and the MPC has executed the entire sequence, the overall task completion flag $\boldsymbol{D}$ is set to $\mathtt{True}$. For instance, $\boldsymbol{D} = \mathtt{True}$ when an AV reaches its destination or when a robot arm delivers an object to the designated target location.
\begin{algorithm}[!h]
  \caption{Subtask generation and contextual reasoning (Section~\ref{sec:subtask})}\label{alg:context-resoning}
  \begin{algorithmic}[1]
    \Require $l$: task instruction, $I$: environment video clip, $c$: (optional) correction command
    \Ensure $\{\Subtask, \param^{[i]}\}_{i\in \mathcal{I}}$: subtask's reward, safety specification functions and corresponding parameters
    \State \textbf{Initialize}  $\mathcal{P}_{\mathtt{vision}}$, $\mathcal{P}_{\mathtt{latentobject}}$, $\mathcal{P}_{\mathtt{riskhandling}}$, $\mathcal{P}_{\mathtt{coder}}$, 
    \Statex \Comment{prompts}
    \Statex \textbf{Initialize} $\mathtt{LLM_{vision}}(\mathcal{P}_{\mathtt{vision}})$,  
    \Statex $\mathtt{LLM_{latentobject}}(\mathcal{P}_{\mathtt{latentobject}}),$
    \Statex $\mathtt{LLM_{riskhandling}}(\mathcal{P}_{\mathtt{riskhandling}}), \mathtt{LLM_{coder}}(\mathcal{P}_{\mathtt{coder}})$ 
    \Statex\Comment{initialize LLMs with prompts}
    
    \State $l_{I} \gets \mathtt{LLM_{vision}}(I)$ \Comment{describe the image in text}
    \State $o\gets \mathtt{LLM_{latentobject}}(l, l_{I})$ 
    \Statex \Comment{obtain (latent-) risk-critical objects}
    \State $s \gets \mathtt{LLM_{riskhandling}}(o)$ 
    \Statex \Comment{obtain a proposal to achieve safe action}
    \State $\{R^{[i]},\boldsymbol{\Phi}^{[i]}, \param^{[i]}\}_{i \in \mathcal{I}} \gets \mathtt{LLM_{coder}}(l, s, c)$ 
    \Statex \Comment{obtain subtask specifications, $c$ is optional}
  \end{algorithmic}
\end{algorithm}

\begin{algorithm}[!ht]
  \caption{Sequential MPC (Section \ref{sec:mpc})}\label{alg:mpc}
  \begin{algorithmic}[1]
    \Require$\{R^{[i]}, \boldsymbol{\Phi}^{[i]}, \param^{[i]}\}_{i \in \mathcal{I}}$: subtask functions and parameters, $x(0)$: initial state
    \Statex \Comment{execution time window $\omega^{[i]}$ is included in the $\param^{[i]}$}
    \Ensure $u_{0:T}, x_{0:T}$: state and control trajectory, $\boldsymbol{D}$: the whole task status
    \State $t \gets 0, \, \boldsymbol{D}\gets \mathtt{True}$ \Comment{reset clock and whole task status}
    \For{each substask $i \in \mathcal{I}$}
    \State $t_0^{[i]}\gets t, D^{[i]} \gets \mathtt{False}$ 
    \While{$t \leq t_0^{[i]}+\omega^{[i]}$}
    \State $u(t) = \text{MPC}[\RewardS,\,\ConstraintS,x(t)]$ 
    \State $x(t+1) = f[x(t)] + g[x(t), u(t)]$
    \State $t\gets t+1$
    \If{subtask completion condition is met}
    \State $D^{[i]}\gets \mathtt{True}$ \Comment{the subtask is completed}
    \State \textbf{break} \Comment{exit while-loop for the next subtask}
    \EndIf
    \EndWhile
    \If{$D^{[i]}$ is $\mathtt{False}$}
    \State $\boldsymbol{D}\gets \mathtt{False}$
    \State \textbf{break} \Comment{exit for-loop over $i$}
    \EndIf
    
    \EndFor
    \State $T \gets t$ \Comment{log the time for the whole trajectory}


  \end{algorithmic}
\end{algorithm}

\subsection{Multi-turn in-context reasoning and subtask optimization}\label{sec:llm_sgd}
The sequence of rewards $\{\Reward\}_{i \in \mathcal{I}}$ and constraints $\{\Constraint\}_{i \in \mathcal{I}}$, generated by the LLM may not complete the task very well. 
To address these issues, we propose a nested multi-turn reasoning procedure with optimization in a simulated environment, without requiring physical deployment (Algorithm~\ref{alg:proposed_framework}, lines 6-17).

Within this multi-turn procedure, in-context learning occurs within the $\mathtt{LLM_{coder}}$ module, which is responsible for optimizing the output and generating an improved subtask sequence, while other LLM modules operate in a zero-shot or few-shot manner, without accumulating contextual memory. 
%
We consider a buffer $\mathcal{B}_{\mathtt{f}}$ to pass in-context information during subtask replanning. This buffer accumulates failure-related information throughout the reasoning and correction process, where each entry takes the form $\{\mathcal{F}, c, \{\Subtask, \param^{[i]}\}_{i \in \mathcal{I}}\}$ (Algorithm~\ref{alg:proposed_framework}, lines 10–11). Each entry provides an in-context example that pairs a previously attempted subtask sequence with the corresponding failure trajectory and correction command. These examples serve as implicit supervision signals: the failure trajectory $\mathcal{F}$  and correction $c$ acts as feedback for the preceding subtask sequence, helping the LLM refine future predictions.

%
Despite the benefits of in-context learning, suboptimal outcomes can still arise due to poorly chosen parameters or inappropriate subtask sequence. 
While $\mathtt{LLM_{correction}}$ can adjust the logical subtask sequence and provide reasonable initial parameter seeds, it remains challenging for the LLM alone to find the optimal parameter values.
Therefore, a subsequent zeroth-order optimization process is employed to refine these parameters (Algorithm~\ref{alg:proposed_framework}, lines 12-14). 
Here we adopt a finite-difference-based deterministic gradient descent method to approximate the gradient $\nabla \myL$, where the loss is defined as $\myL = -\mathcal{R}(x_{0:T}, u_{0:T})$.
The parameters $\param$ are iteratively updated until the gradient approximation $\nabla \myL$ falls below a predefined threshold $\epsilon$.
Alternative zeroth-order optimization techniques—such as random search, evolutionary strategies, or bayesian optimization—can also be incorporated depending on the properties of the task and simulator. Here, we note that while the parameters $\{\param^{*[i]}\}_{i\in\mathcal{I}}$ are updated consecutively with zeroth-order optimization, the parametric forms of reward and constraint functions for the subtask sequence $\{\Subtask\}_{i \in \mathcal{I}}$, that generated by the LLM in the outer loop, remain fixed throughout each inner optimization.
Then, after inner optimization, the subtask sequence $\{\Subtask\}_{i \in \mathcal{I}}$ with updated parameters $\{\param^{*[i]}\}_{i \in \mathcal{I}}$ is passed back to the MPC (Algorithm~\ref{alg:proposed_framework}, line 5) for evaluation. If optimized parameter set does not work, the result is then fed into the multi-turn in-context reasoning loop (Algorithm~\ref{alg:proposed_framework}, lines 6–17), where the LLM may adjust the subtask structure if necessary.
Independently, the entire framework can be interrupted at any point by either a human-specified or system-triggered $\mathtt{INTERRUPTION}$ signal.

\begin{algorithm}[!ht]
  \caption{Proposed framework (online) (Section~\ref{sec:online})}
  \label{alg:online_framework}
  \begin{algorithmic}[1]
    \Require $l$: task instructions, $x(0)$: initial state, $\mathcal{R}(x_{0:T}, u_{0:T})$: evaluation metric, $\{R^{[i]}, \boldsymbol{\Phi}^{[i]}, \param^{[i]}\}_{i \in \mathcal{I}}$: subtask decomposition from offline framework
    \Ensure $\{R^{[i]}, \boldsymbol{\Phi}^{[i]}, \param^{[i]}\}_{i \in \mathcal{I}}$: adjusted subtask sequence 
    \State \textbf{Initialize} $\mathcal{P}_{\mathtt{correction}}$, $\mathtt{LLM_{\mathtt{correction}}}(\mathcal{P}_{\mathtt{correction}})$ 
    \State $t \gets 0$
    \State Obtain initial environment input $I$

      \For{each subtask $i \in \mathcal{I}$}
      \State $t_0^{[i]}\gets t, D^{[i]}\gets \mathtt{False}$ 
        \While{$t \leq t_0^{[i]}+\omega^{[i]}$}
          \State $u(t) \gets \text{MPC}[R^{[i]}(\cdot;\param^{[i]}), \boldsymbol{\Phi}^{[i]}(\cdot;\param^{[i]}), x(t)]$
          \State $x(t+1) \gets f[x(t)] + g[x(t), u(t)]$ 
          \State $t\gets t+1$
          \If{subtask completion condition is met}
            \State ${D}^{[i]} \gets \mathtt{True}$ \Comment{subtask is completed}
            \State \textbf{break} \Comment{exit while-loop for the next subtask}
          \EndIf
        \EndWhile
        \If{$D^{[i]} = \mathtt{False}$}
          \State Obtain updated environment inputs $I$
          \State $\mathcal{F} \gets \mathtt{FeedbackTemplate}(x_{0:t}, u_{0:t})$ 
          \State $c \gets \mathtt{LLM_{correction}}(\mathcal{F})$ 
          \State $\{R^{[j]}, \boldsymbol{\Phi}^{[j]}, \param^{[j]}\}_{j>i} \gets $Algorithm~\ref{alg:context-resoning}$(l, I, c)$
          \If{$\mathtt{verify}(\{R^{[j]}, \boldsymbol{\Phi}^{[j]}, \param^{[j]}\}_{j>i}) = \mathtt{False}$}
          \State $\{R^{[j]}, \boldsymbol{\Phi}^{[j]}, \param^{[j]}\}_{j>i} \gets$ Algorithm~\ref{alg:proposed_framework}$(l,I,x(t),\mathcal{R})$ 
          \State Pause until a valid set of $\{R^{[j]}, \boldsymbol{\Phi}^{[j]}, \param^{[j]}\}_{j>i}$ is obtained from Algorithm~\ref{alg:proposed_framework}
          
          \EndIf
        \EndIf
      \EndFor
  \end{algorithmic}
\end{algorithm}

          

\subsection{Online implementation}\label{sec:online}

The online framework retains the structure of subtask decomposition and multi-level optimization, while adapting the execution flow for time-sensitive environments (Algorithm~\ref{alg:online_framework}). Given the task instructions $l \in \mathcal{T}$, a subtask sequence $\{\boldsymbol{\Phi}^{[i]}, \param^{[i]}\}_{i \in \mathcal{I}}$ validated in the offline framework, an evaluation metric $\mathcal{R}(x_{0:T}, u_{0:T})$, the MPC executes the subtasks sequentially.
Unlike the offline setting, online execution does not allow subtask restarts. Instead, when a failure occurs, the framework reacquires the surrounding environment input $I \in \mathcal{V}$ and invokes Algorithm~\ref{alg:context-resoning} to replan the remaining subtasks based on the feedback $\mathcal{F}$ and correction command $c$ (line 18). The updated subtask sequence is then passed to the simulator for validation (line 20). If the simulated execution succeeds, the revised specifications are deployed online; otherwise, the framework pauses until a new subtask sequence is generated by the offline Algorithm~\ref{alg:proposed_framework} (line 21).

\section{Simulation and Result} \label{sec:simulation}
In this section, we validate the proposed framework through two case studies. The first case focuses on a robot arm executing a physically interactive task involving an object with partially unknown physical properties (Section~\ref{sec:fastadaptation}). We present the task setup and corresponding grasping results, while additional experiments are provided in Appendix~\ref{sec:robot_simulation} and~\ref{sec:grasp_simulation}. This task demands an effective learning process capable of jointly optimizing the subtask sequence and the associated parameters to fulfill task objectives under uncertainty.
The second case study focuses on an AV navigating toward a target location while maintaining safety and efficiency in the presence of latent risks (Section~\ref{sec:latentrisk}), with additional experiments provided in Appendix~\ref{sec:av_simulation}. Achieving these goals requires the ability to reason over partially observable contexts, where critical risk-related variables may be unobservable and safe behaviors depend heavily on situational nuance. This task thus necessitates a context-aware learning process that can infer latent risks and adapt control policies accordingly.
All LLM-based components were implemented using GPT-4o accessed via the OpenAI API, without any fine-tuning. The complete source code is available upon request, and all prompt templates used in the experiments are documented in the Appendices.

\subsection{Effective learning of robot arm policies}
\label{sec:fastadaptation}

\textbf{System model.} 
We consider a simulation in which a 2-degree-of-freedom robot arm is tasked with grasping a set of items. The full system state at time $t$ is denoted by
$x(t) = [x_a(t), v_a(t), \{x_k(t), \mathbf{F}_k(t), \mathtt{status}_k(t)\}_{k\in\mathcal{K}}]^\top$, 
where $x_a(t) \in \mathbb{R}^2$ and $v_a(t) \in \mathbb{R}^2$ represent the end-effector’s position and velocity, respectively. Each object $k \in \mathcal{K}$ is characterized by its position $x_k(t) \in \mathbb{R}^2$, the contact force $\mathbf{F}_k(t) \in \mathbb{R}^2$ exerted by the robot, and a discrete interaction status $\mathtt{status}_k(t) \in \{\mathtt{free}, \mathtt{grasped}, \mathtt{pushed}\}$, indicating its current mode of interaction with the arm.
This formulation captures both the internal configuration of the robot and the external object dynamics, enabling unified reasoning over physically grounded multi-object interactions.
The control input $u(t) =v_a(t)$ represents the desired joint velocities. The controller operates on the robot internal state $[x_a(t), u_a(t)]$, follow the formulation described in~\ref{sec:robot_dynamics}.
These continuous-time dynamics are discretized using a fixed sampling interval $\Delta t$ for simulation and control.

\textbf{Experimental settings.}
The simulation environment includes several objects, and the task involves manipulating one object in relation to another—for example, moving an apple into a box. In the environment setup visualized in Fig.~\ref{fig:grasp_simulations_main}, the red circle represents the apple, the green square represents the box, and the black circle denotes an obstacle. All configurations are given the same task instruction l: move the apple to the box, and please consider the energy efficiency of the whole rollout. The evaluation metric $\mathcal{R}$ is defined as:
\begin{equation}
\mathcal{R}(x_{0:T}, u_{0:T}) =
\begin{cases}
\begin{aligned}
    &\|\boldsymbol{x}_{\text{apple}} - \boldsymbol{x}_{\text{box}}\|_2 \\
    &+ \alpha \|x_{\text{apple}}(T) - x_{\text{box}}(T)\|_2 \\
    &+ \beta\|\boldsymbol{u}\|_2, \quad \text{if } \|x_{\text{apple}}(T) - x_{\text{box}}(T)\|_2 \leq \varphi
\end{aligned} \\
\infty, \qquad\quad\ \text{otherwise}
\end{cases}
\end{equation}
where $\alpha, \beta$ are weighting factors, $\varphi$ is a threshold that defines successful task completion, and $\{\text{apple}, \text{box}\} \subset \mathcal{K}$ and $\mathcal{K}$ denotes the set of all objects in the scene. Alse  $\boldsymbol{x} = x_{0:T}$ and $\boldsymbol{u} = u_{0:T}$ represent the full state and control trajectories, respectively. The energy consumption $\mathcal{E} = \|\boldsymbol{u}\|_2$, is recorded throughout each execution.
Here, we evaluate our algorithm across three distinct environment configurations to assess its adaptability and robustness under varying spatial arrangements and object placements. Based on these settings, we compare six different approaches, as illustrated in Fig.\ref{fig:legend}:
%
\begin{enumerate}[label=\textit{\alph*)}] 
    \item MPC: MPC is used to find actions that directly optimize the evaluation metrics $\mathcal{R}$. 
    \item Single-turn LLM: LLM is used to decompose the task into a sequence of subtasks. For each subtasks, MPC is used to find the optimal actions based on the cost and constraints associated with each subtask. 
    \item Single-turn LLM with subtask parameter optimization: LLM is used to decompose the task into a sequence of subtasks. The parameters of subtasks (cost and constraint functions) are jointly optimized for all subtasks, and MPC is used to find the optimal actions for each subtask.
    \item Multi-turn LLM: LLM is used to decompose the task into a candidate subtask sequence, and the outcome of the last candidate is fed back into LLM to refine the subtask sequence.
    \item Multi-turn in-context LLM: LLM is used to decompose the task into a candidate subtask sequence. The outcome of all past candidates is fed back into LLM to refine the subtask sequence. 
    \item Proposed technique (Algorithm~\ref{alg:proposed_framework}).

\end{enumerate}
\begin{figure}[htbp]
    \centering
    \includegraphics[width=0.8\linewidth]{figures/legend.pdf}
    \caption{Legend illustrating the six approaches compared in our experiments.}
    \label{fig:legend}
\end{figure}

\textbf{Results.} 
The results in Fig.~\ref{fig:grasp_simulations_main} show that approach \textit{a)}, which does not incorporate task decomposition, fails to complete the task. 
This is because the overall evaluation metric does not provide guidance on the necessary immediate steps.
%
In contrast, LLM-assisted approaches (approaches \textit{b)–f)}) overcome this limitation by decomposing the task into two subtasks in the first round of reasoning: (1) during subtask window $\omega^{[1]}$, approaching the apple, and (2) during subtask window $\omega^{[2]}$, moving the apple into the box. 
However, due to the emergence of an obstacle midway through the task, this initial decomposition fails to achieve the overall goal, as the original plan does not account for the unexpected obstruction. The multi-turn approaches (\textit{d)–f)}) address this issue by diagnosing the failure and initiating a second round of reasoning to replan the task with additional subtasks:
(1) during subtask window $\omega^{[1]}$, approaching the obstacle;
(2) during subtask window $\omega^{[2]}$, relocating the obstacle;
(3) during subtask window $\omega^{[3]}$, approaching the apple; and
(4) during subtask window $\omega^{[4]}$, moving the apple into the box.
Although the LLM can suggest a reasonable alternative location for the obstacle, it cannot identify an optimal placement that balance performance and energy consumption, and may occasionally suggest unreasonable locations that cause the task to fail again (hallow circles in the Fig.~\ref{fig:grasp_simulations_main}). The proposed method (\textit{f}) addresses this limitation by leveraging zeroth-order optimization to refine the LLM-suggested location, enabling energy-efficient execution while preserving task feasibility.

\subsection{Situational-aware risk mitigation}
\label{sec:latentrisk}
\textbf{System model.} We consider an AV operating in scenarios that involve latent risks. The full system state at time $t$ is defined as $x(t) = [x_{e}(t),\, v_{e}(t),\, \mathbf{F}_{r}(t), \, x_{r}(t)],$
where $x_e(t) \in \mathbb{R}^2$ and $v_e(t) \in \mathbb{R}^2$ represent the ego vehicle's position, velocity, respectively, $\mathbf{F}_r(t) \in \mathbb{R}^2$ denotes the net resistive force acting against motion (aerodynamic drag and rolling resistance), and $x_r(t) \in \mathbb{R}^2$ indicates the position of the closest risk-critical object in the environment. If no such object is detected at time $t$, we set $x_r(t) \rightarrow [\infty, \infty]$ to encode its absence. This results in a 8-dimensional state space that captures both the ego vehicle’s internal configuration and the external environmental context. The system dynamics are defined in continuous time. The ego vehicle follows second-order Newtonian dynamics: $\dot{x}_e(t) = v_e(t), \dot{v}_e(t) = \frac{1}{m_e} \left( u(t) - \mathbf{F}_r(t) \right)$, where $u(t) \in \mathbb{R}^2$ is the control input representing the force applied to the vehicle. These continuous-time dynamics are discretized using a fixed sampling interval $\Delta t$ for simulation and control.

\textbf{Experimental settings.} 
The AV aims to reach a target location efficiently and safely, starting from a fixed position and moving along a straight line. The environment may include occluding objects that hide latent hazards, such as children unexpectedly emerging and entering the AV’s path, creating safety risks not visible through standard sensors.
We define the evaluation metric as: 
\begin{equation}
\mathcal{R}(x_{0:T}, u_{0:T}) =\\
\begin{cases}
\sum_0^T\|X^*-x_{e}(t)\|_2 \\
    + \alpha \|X^*-x_{e}(T)\|_2 , \quad 
    \\ \qquad \qquad \text{if } \|X^* -x_{e}(T)\|_2 \leq \varphi\\
    \infty, \qquad\ \text{otherwise}
\end{cases}
\end{equation}
where $X^*$ is the target location and $varphi$ is the threshold that defines successful task completion. 
We demonstrate three examples here include: (i) encountering a school bus, (ii) interacting with teenagers near the roadway, and (iii) observing adults walking along the road (see Fig.~\ref{fig:sim1_example_img}). In these scenarios, we assign a probability of $p=0.5$ for a latent risk, such as an occluded child behind the school bus or a teenager, to become an apparent risk (\emph{i.e.}, suddenly jumping into the road), while the probability for an adult is set to $p=0$. From this, we evaluate three approaches: 
\begin{itemize}
    \item Typical strategy: imposing safety specifications on apparent risks without latent-risk considerations;
    \item Conservative strategy: taking worst-case uncertainties for unobserved variables without contextual consideration;
    \item Proposed framework: Algorithm \ref{alg:proposed_framework}.
\end{itemize}
Their performance in safety handling and operational efficiency is evaluated using two metrics: the minimum distance to latent risk-critical objects and the time to reach the target location. Final evaluation is based on the average of these metrics across scenarios (i)-(iii). Example video data from the driver's perspective is sourced from the PSI dataset~\cite{chen2021psi}.

\textbf{Results.} The results demonstrate that the proposed framework ensures a high level of safety, which is substantiated by a greater minimum distance to risky objects (Fig.~\ref{fig:sim1_distance}). This is attributed to the effective imposition of safety constraints on latent objects in scenarios (i) and (ii) by slowing down before children or teenagers jump out onto the street.
Also, the proposed framework maintains satisfactory performance without overly conservative behavior, as illustrated in Fig.~\ref{fig:sim1_time} by keeping a constant speed when facing adults who are considered less likely to become an apparent risk.
As seen in Fig.~\ref{fig:sim1_distance}, the typical strategy is prone to risky behavior in scenarios such as school bus occlusion or teenagers walking alongside the street, often resulting in accidents, as indicated by a minimum distance of zero and an unbounded travel time. This is due to its inability to anticipate latent risks, such as teenagers unexpectedly entering the AV’s path.
In contrast, the conservative strategy, which lacks contextual reasoning, tends to compromise efficiency by unnecessarily slowing down when encountering adults, who pose minimal or no risk of unpredictable behaviors (Fig.~\ref{fig:sim1_time}).
Thus, the proposed framework, capable of predicting latent-risk objects characterized by highly unpredictable behaviors, can ensure a high safety level without compromising efficiency.

\begin{figure*}[htbp]
    \centering

    \subfloat[Config. $1$]{\includegraphics[width=0.28\linewidth]{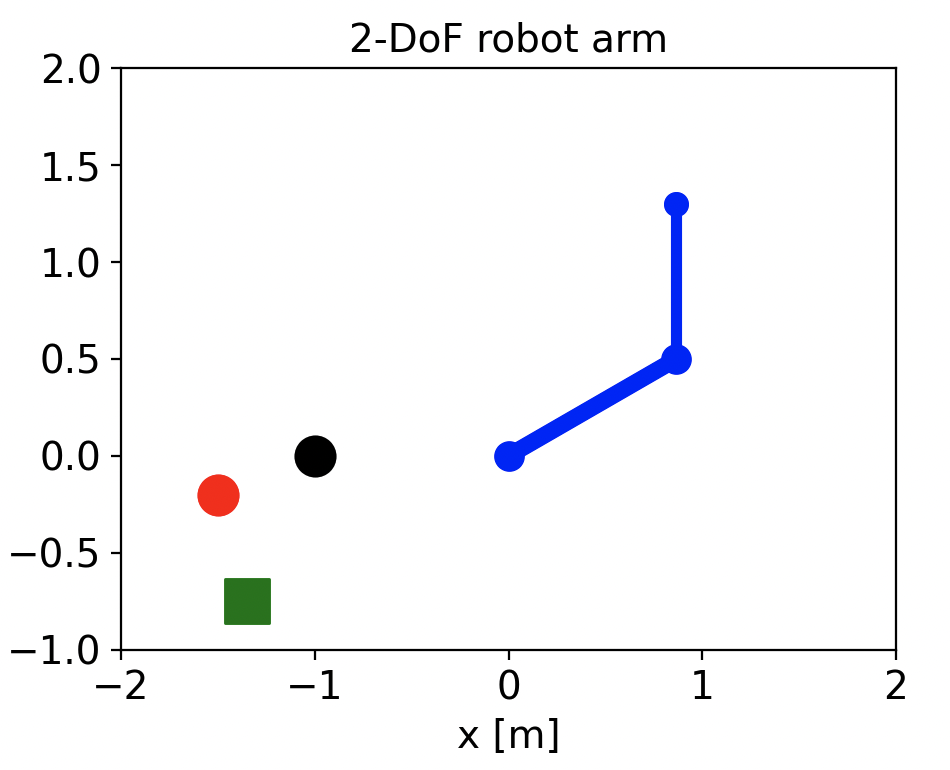}}\hfill
    \subfloat[Cost $\mathcal{L}$ vs. $\#$data point for config. 1]{    \begin{tikzpicture}
        \begin{axis}[
            xlabel={\# Data point},
            ylabel={Loss $\mathcal{L}$},
            ylabel shift=6pt,
            grid=both,
            grid style={dashed,gray!30},
            legend pos=north east,
            width=0.3\linewidth,
            height=0.25\linewidth,
            xmin=1, xmax=10,
            ytick={300,400,500,600},
            yticklabels={300,400,500,$\infty$},
            legend style={font=\scriptsize,draw=none, fill = none},
            legend style={font=\scriptsize, at={(axis cs:3.2,1.5)},anchor=south west},
            every axis label/.append style={font=\scriptsize},
            every tick label/.append style={font=\scriptsize}
        ]
        \addplot[dashdotted, violet, line width=1.2pt, domain=1:10] {600}; 
        \addplot[loosely dashed,newgreen!300, line width=1.2pt,, domain=1:10] {600}; 


        \addplot [magenta, mark=o, mark size=1.5, forget plot, mark options={fill=white, line width=0.8pt}] table [x=iter, y=D] {s2_1_loss.dat}; 
        \addplot [magenta, only marks, mark=*, mark size=1.5]
            table {
                iter    D
2	392.36935373313100
3	428.3908612277720
4	447.1090549460360
5	435.4709826352280
6	457.645864730493
7	443.56320434122200
8	462.74733481329500
9	464.4996666469980
10	488.51401619535800
            };
        \addplot [newblue!300, mark=o, mark size=1.5, forget plot, mark options={fill=white, line width=0.8pt}] table [x=iter, y=E] {s2_1_loss.dat}; 
        \addplot [newblue!300, only marks, mark=*, mark size=1.5]
            table {
                iter    E
2	392.36935373313100
3	428.3908612277720
4	449.72936604363800
5	498.60431307007000
6	515.3624105828050
7	516.6747197245190
8	518.8975819834830
9	518.40299271282
10	518.898097198999
            };
        \addplot [red, mark=o, mark size=1.5, forget plot] table [x=iter, y=P] {s2_1_loss.dat}; 
        \addplot [red, only marks, mark=*, mark size=1.5]
            table {
                iter    P
2	392.36935373313100
3	384.07359965808100
4	375.4352151606630
5	366.3228529529050
6	358.2440715903860
7	349.1587438166090
8	340.70708793185100
9	323.25805611947900
10	323.25805611947900
            };
        
        \end{axis}
    \end{tikzpicture}}\hfill
    \subfloat[Energy $\mathcal{E}$ vs. $\#$data point for config. 1]{    \begin{tikzpicture}
        \begin{axis}[
            xlabel={\# Data point},
            ylabel={Energy $\mathcal{E}$},
            ylabel shift=-6pt,
            grid=both,
            grid style={dashed,gray!30},
            legend pos=north east,
            width=0.3\linewidth,
            height=0.25\linewidth,
            xmin=1, xmax=10,
            legend style={font=\scriptsize,draw=none, fill = none},
            legend style={font=\scriptsize, at={(axis cs:3.2,1.5)},anchor=south west},
            every axis label/.append style={font=\scriptsize},
            every tick label/.append style={font=\scriptsize}
        ]
        \addplot[dashdotted, violet, line width=1.2pt, domain=1:10] {2.2338078045140800}; 
        \addplot[loosely dashed,newgreen!300, line width=1.2pt,, domain=1:10] {2.2338078045140800}; 


        \addplot [magenta, mark=o, mark size=1.5, forget plot, mark options={fill=white, line width=0.8pt}] table [x=iter, y=D] {s2_1_energy.dat}; 
        \addplot [magenta, only marks, mark=*, mark size=1.5]
            table {
                iter    D
2	3.7620846886366900
3	3.8861065416548800
4	3.9870630228641100
5	4.005269985435550
6	4.123158238852750
7	4.045422183814980
8	4.168324309633360
9	4.077278687041300
10	4.1868319816085900
            };
        \addplot [newblue!300, mark=o, mark size=1.5, forget plot, mark options={fill=white, line width=0.8pt}] table [x=iter, y=E] {s2_1_energy.dat}; 
        \addplot [newblue!300, only marks, mark=*, mark size=1.5]
            table {
                iter    E
2	3.7620846886366900
3	3.8861065416548800
4	4.053673273981880
5	4.214475072524030
6	4.465602153356850
7	4.4586693481354800
8	4.453809811679960
9	4.459012035412840
10	4.453829076532970
            };
        \addplot [red, mark=o, mark size=1.5, forget plot] table [x=iter, y=P] {s2_1_energy.dat}; 
        \addplot [red, only marks, mark=*, mark size=1.5]
            table {
                iter    P
2	3.7620846886366900
3	3.711702736025080
4	3.6527502455950900
5	3.5942741791700000
6	3.5364775871507400
7	3.4779959365512500
8	3.4138598501015100
9	3.2882062198511700
10	3.2882062198511700
            };
        
        \end{axis}
    \end{tikzpicture}}\par\medskip

    \subfloat[Config. $2$]{\includegraphics[width=0.28\linewidth]{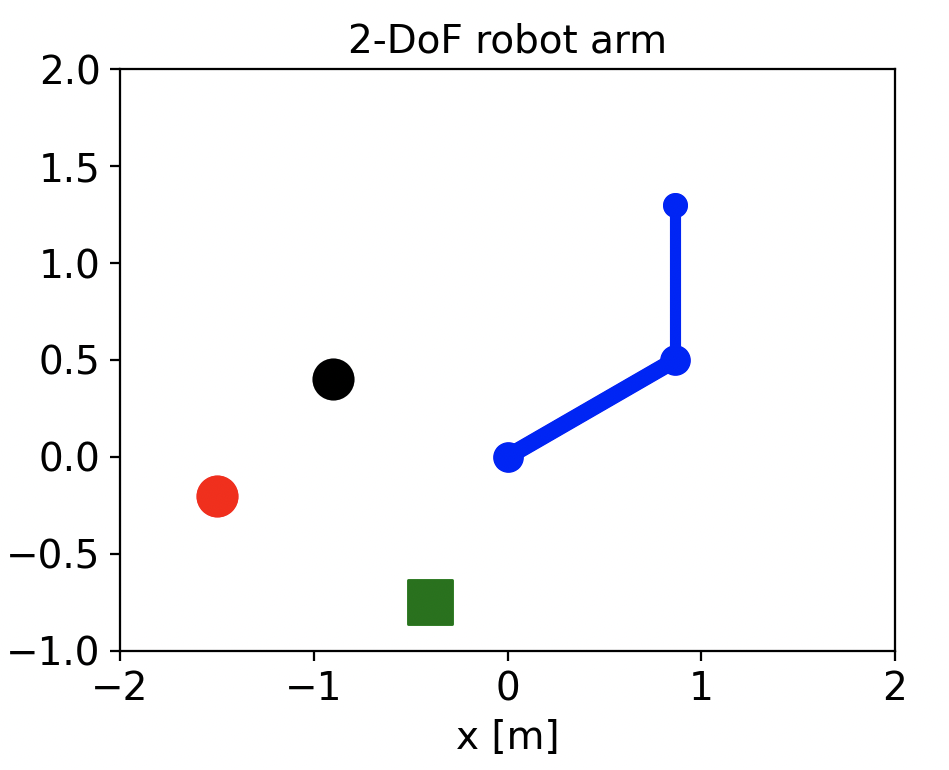}}\hfill
    \subfloat[Cost $\mathcal{L}$ vs. $\#$data point for config. 2]{    \begin{tikzpicture}
        \begin{axis}[
            xlabel={\# Data point},
            ylabel={Loss $\mathcal{L}$},
            ylabel shift=6pt,
            grid=both,
            grid style={dashed,gray!30},
            legend pos=north east,
            width=0.3\linewidth,
            height=0.25\linewidth,
            xmin=1, xmax=10,
            ytick={400,600,800,1000,1200},
            yticklabels={400,600,800,1000,$\infty$},
            legend style={font=\scriptsize,draw=none, fill = none},
            legend style={font=\scriptsize, at={(axis cs:3.2,1.5)},anchor=south west},
            every axis label/.append style={font=\scriptsize},
            every tick label/.append style={font=\scriptsize}
        ]
        \addplot[dashdotted, violet, line width=1.2pt, domain=1:10] {1200}; 
        \addplot[loosely dashed,newgreen!300, line width=1.2pt,, domain=1:10] {1200}; 


        \addplot [magenta, mark=o, mark size=1.5, forget plot, mark options={fill=white, line width=0.8pt}] table [x=iter, y=D] {s2_2_loss.dat}; 
        \addplot [magenta, only marks, mark=*, mark size=1.5]
            table {
                iter    D
2	606.9671192515750
3	619.0260378665920
4	641.1572317189870
5	682.9922779888870
6	739.9236914518870
7	865.4677295143570
8	908.159613903028
9	927.3354483834940
10	922.4343368616980
            };
        \addplot [newblue!300, mark=o, mark size=1.5, forget plot, mark options={fill=white, line width=0.8pt}] table [x=iter, y=E] {s2_2_loss.dat}; 
        \addplot [newblue!300, only marks, mark=*, mark size=1.5]
            table {
                iter    E
2	606.9671192515750
3	619.0260378665920
4	713.2564425693360
5	706.1526050127490
6	752.4439218719180
7	770.2266528391340
8	941.6651052673340
9	820.4321512248430
10	767.8719721978380
            };
        \addplot [red, mark=o, mark size=1.5, forget plot] table [x=iter, y=P] {s2_2_loss.dat}; 
        \addplot [red, only marks, mark=*, mark size=1.5]
            table {
                iter    P
2	789.109565402962
3	606.9671192515750
4	575.6397253139740
5	546.3670836940290
6	539.8854184543910
7	520.8482308023980
8	519.7233159456620
9	506.76313139341600
10	496.68356943591500
            };
        
        \end{axis}
    \end{tikzpicture}}\hfill
    \subfloat[Energy $\mathcal{E}$ vs. $\#$data point for config. 2]{\begin{tikzpicture}
        \begin{axis}[
            xlabel={\# Data point},
            ylabel={Energy $\mathcal{E}$},
            ylabel shift=-6pt,
            grid=both,
            grid style={dashed,gray!30},
            legend pos=north east,
            width=0.3\linewidth,
            height=0.25\linewidth,
            xmin=1, xmax=10,
            legend style={font=\scriptsize,draw=none, fill = none},
            legend style={font=\scriptsize, at={(axis cs:3.2,1.5)},anchor=south west},
            every axis label/.append style={font=\scriptsize},
            every tick label/.append style={font=\scriptsize}
        ]
        \addplot[dashdotted, violet, line width=1.2pt, domain=1:10] {1.8258486272990000}; 
        \addplot[loosely dashed,newgreen!300, line width=1.2pt,, domain=1:10] {1.8258486272990000}; 


        \addplot [magenta, mark=o, mark size=1.5, forget plot, mark options={fill=white, line width=0.8pt}] table [x=iter, y=D] {s2_2_energy.dat}; 
        \addplot [magenta, only marks, mark=*, mark size=1.5]
            table {
                iter    D
2	3.767798234427710
3	3.804114030390090
4	3.9257782408178300
5	3.9898814291811400
6	4.058765415845680
7	4.236900794111510
8	4.5474703992075600
9	4.6040350288787900
10	4.582812652784310
            };
        \addplot [newblue!300, mark=o, mark size=1.5, forget plot, mark options={fill=white, line width=0.8pt}] table [x=iter, y=E] {s2_2_energy.dat}; 
        \addplot [newblue!300, only marks, mark=*, mark size=1.5]
            table {
                iter    E
2	3.767798234427710
3	3.804114030390090
4	4.050092217241560
5	4.014715818416740
6	3.9807722333963100
7	4.155814968329060
8	4.633350606815250
9	4.307228046354780
10	4.182428746932540
            };
        \addplot [red, mark=o, mark size=1.5, forget plot] table [x=iter, y=P] {s2_2_energy.dat}; 
        \addplot [red, only marks, mark=*, mark size=1.5]
            table {
                iter    P
2	4.212785491725090
3	3.767798234427710
4	3.6768436114963200
5	3.5121062388586800
6	3.558550504481870
7	3.414790800994610
8	3.4797303304118300
9	3.4301082685046800
10	3.4322041002424200
            };
        
        \end{axis}
    \end{tikzpicture}}\par\medskip

    \subfloat[Config. $3$]{\includegraphics[width=0.28\linewidth]{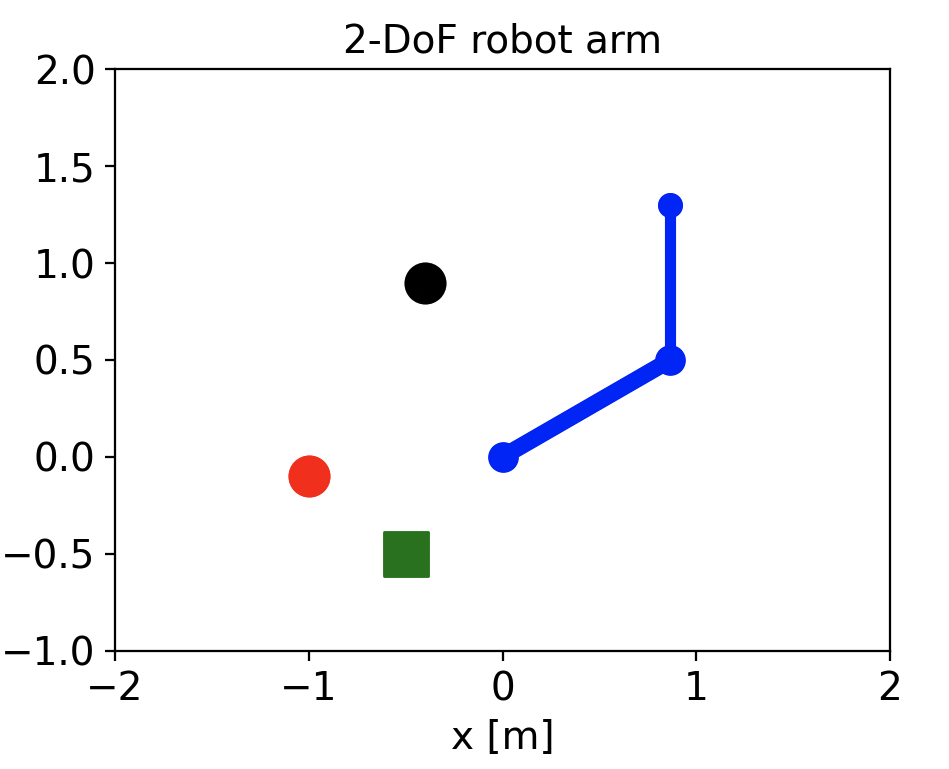}}\hfill
    \subfloat[Cost $\mathcal{L}$ vs. $\#$data point for config. 3]{    \begin{tikzpicture}
        \begin{axis}[
            xlabel={\# Data point},
            ylabel={Loss $\mathcal{L}$},
            ylabel shift=-6pt,
            grid=both,
            grid style={dashed,gray!30},
            legend pos=north east,
            width=0.3\linewidth,
            height=0.25\linewidth,
            xmin=1, xmax=10,
            ytick={300,350,400,450},
            yticklabels={300,350,400,$\infty$},
            legend style={font=\scriptsize,draw=none, fill = none},
            legend style={font=\scriptsize, at={(axis cs:3.2,1.5)},anchor=south west},
            every axis label/.append style={font=\scriptsize},
            every tick label/.append style={font=\scriptsize}
        ]
        \addplot[dashdotted, violet, line width=1.2pt, domain=1:10] {450}; 
        \addplot[loosely dashed,newgreen!300, line width=1.2pt,, domain=1:10] {450}; 


        \addplot [magenta, mark=o, mark size=1.5, forget plot, mark options={fill=white, line width=0.8pt}] table [x=iter, y=D] {s2_3_loss.dat}; 
        \addplot [magenta, only marks, mark=*, mark size=1.5]
            table {
                iter    D
2	392.4730368067170
3	322.81538202911500
4	329.21392700893500
6	313.7660584461410
7	391.2708352572530
9	313.7660584461410
            };
        \addplot [newblue!300, mark=o, mark size=1.5, forget plot, mark options={fill=white, line width=0.8pt}] table [x=iter, y=E] {s2_3_loss.dat}; 
        \addplot [newblue!300, only marks, mark=*, mark size=1.5]
            table {
                iter    E
2	392.4730368067170
3	380.64325483292600
4	380.87043786446300
5	338.2687307720940
7	392.4730368067170
9	392.4730368067170
10	389.3500419030890
            };
        \addplot [red, mark=o, mark size=1.5, forget plot] table [x=iter, y=P] {s2_3_loss.dat}; 
        \addplot [red, only marks, mark=*, mark size=1.5]
            table {
                iter    P
2	313.7660584461410
3	304.31614722345500
4	302.7374655242370
5	303.07839576800700
6	302.55501017231
7	302.37414394012900
8	302.21798888690300
9	302.0394486192560
10	301.1853231106850
            };
        
        \end{axis}
    \end{tikzpicture}}\hfill
    \subfloat[Energy $\mathcal{E}$ vs. $\#$data point for config. 3]{\begin{tikzpicture}
        \begin{axis}[
            xlabel={\# Data point},
            ylabel={Energy $\mathcal{E}$},
            ylabel shift=-6pt,
            grid=both,
            grid style={dashed,gray!30},
            legend pos=north east,
            width=0.3\linewidth,
            height=0.25\linewidth,
            xmin=1, xmax=10,
            legend style={font=\scriptsize,draw=none, fill = none},
            legend style={font=\scriptsize, at={(axis cs:3.2,1.5)},anchor=south west},
            every axis label/.append style={font=\scriptsize},
            every tick label/.append style={font=\scriptsize}
        ]
        \addplot[dashdotted, violet, line width=1.2pt, domain=1:10] {1.653509291499300		}; 
        \addplot[loosely dashed,newgreen!300, line width=1.2pt,, domain=1:10] {1.653509291499300		}; 


        \addplot [magenta, mark=o, mark size=1.5, forget plot, mark options={fill=white, line width=0.8pt}] table [x=iter, y=D] {s2_3_energy.dat}; 
        \addplot [magenta, only marks, mark=*, mark size=1.5]
            table {
                iter    D
2	3.5333776086013200
3	3.237114360664790
4	3.1573991878178800
6	3.2051775362212800
7	3.407226523867160
9	3.2051775362212800
            };
        \addplot [newblue!300, mark=o, mark size=1.5, forget plot, mark options={fill=white, line width=0.8pt}] table [x=iter, y=E] {s2_3_energy.dat}; 
        \addplot [newblue!300, only marks, mark=*, mark size=1.5]
            table {
                iter    E
2	3.5333776086013200
3	3.3861585278277300
4	3.3623582170878500
5	3.2286213304317100
7	3.5333776086013200
9	3.5333776086013200
10	3.3802126944713000
            };
        \addplot [red, mark=o, mark size=1.5, forget plot] table [x=iter, y=P] {s2_3_energy.dat}; 
        \addplot [red, only marks, mark=*, mark size=1.5]
            table {
                iter    P
2	3.2051775362212800
3	3.072895081235760
4	3.0574843248096400
5	3.0495751571216000
6	3.0432315294353800
7	3.04041139046437
8	3.032782434186140
9	3.0288571256281400
10	3.024487601125470
            };
        
        \end{axis}
    \end{tikzpicture}}\par\medskip

    \caption{Simulation configurations and results for the robot arm tasks described in Section~\ref{sec:fastadaptation}. Corresponding design space and legend are shown in Fig.~\ref{fig:legend}, with the curve colors above matching the respective approaches. In all loss and energy consumption plots, solid dots denote successful task executions, while hollow circles indicate failed attempts.}
    \label{fig:grasp_simulations_main}
\end{figure*}

\begin{figure*}[htbp]
    \begin{minipage}{0.65\textwidth}
        \subfloat[Latent objects and their likelihood level to become apparent \\ risks suggested by $\texttt{GPT-4}$. ]{\input{figures/sim1_example_img}\label{fig:sim1_example_img}}\hfill%
    \end{minipage}
    \begin{minipage}{0.3\textwidth}
        \subfloat[Average minimum distance (safety).]{  \begin{tikzpicture}
    \begin{axis}[
      scale only axis,
      ybar,
      bar width=0.6, 
      axis lines=middle,
      ylabel style={align=left},
      inner axis line style={=>},
      width=3.5cm,height=1.7cm,
      ymin=0,ymax=3,
      xmin=0.5,xmax=3.5,
      axis line style=thick,
      xtick={1,2,3},
      ytick={0,1,2,3},
      xticklabels={Typical\\strategy,Conservative\\strategy,Proposed},
      xticklabel style={align=center, font=\scriptsize},
      yticklabel style={font=\scriptsize}, 
      every axis x label/.style={at={(current axis.right of origin)},anchor=west, font=\scriptsize},
      every axis y label/.style={at={(current axis.north)},above=0.5mm, font=\scriptsize},
      axis on top,
      nodes near coords,
      nodes near coords style={font=\scriptsize}, 
      nodes near coords align={vertical},
      ]
      \addplot table [ 
          x expr=\coordindex+1, 
          y=Height 
      ] {sim1_distance.dat};
      
    \end{axis}
  \end{tikzpicture}\label{fig:sim1_distance}}\vfill%
        \subfloat[Average time-to-travel (performance).]{\pgfplotsset{
    discard if/.style 2 args={
        x filter/.code={
            \edef\tempa{\thisrow{#1}}
            \edef\tempb{#2}
            \ifx\tempa\tempb
                \def\pgfmathresult{inf}
            \fi
        }
    },
    discard if not/.style 2 args={
        x filter/.code={
            \edef\tempa{\thisrow{#1}}
            \edef\tempb{#2}
            \ifx\tempa\tempb
            \else
                \def\pgfmathresult{inf}
            \fi
        }
    }
}

  \begin{tikzpicture}
    \begin{axis}[
      scale only axis,
      ybar,
      bar width=0.6, 
      axis lines=middle,
      ylabel style={align=left},
      width=3.5cm,height=1.7cm,
      ymin=0,ymax=5,
      xmin=0.5,xmax=3.5,
      axis line style=thick,
      xtick={1,2,3},
      ytick={0,2,4},
      xticklabels from table={sim1_time.dat}{Label},
      xticklabels={{Typical\\strategy}, {Conservative\\strategy}, {Proposed}},
      xticklabel style={align=center, font=\scriptsize},
      yticklabel style={font=\scriptsize}, 
      every axis x label/.style={at={(current axis.right of origin)},anchor=west, font=\scriptsize},
      every axis y label/.style={at={(current axis.north)},above=0.5mm, font=\scriptsize},
      axis on top,
      nodes near coords,
      nodes near coords style={font=\scriptsize}, 
      nodes near coords align={vertical},
      ]
      \addplot+[
            ybar,
            bar width=0.32cm,  
            fill=blue!30,
            draw=blue,
            nodes near coords,
            point meta=explicit symbolic,
        ] table [meta=label] {
            x     y     label
            1.02   5.0   $\infty$
            1.3   2.93  2.93
        };
      \addplot+[
            ybar,
            fill=blue!30,     
            draw=blue, 
            bar width=0.6,  
            nodes near coords,
            every node near coord/.append style={color=blue},
            ]table [ 
                  x expr=\coordindex+1.65, 
                  y=Height,
              ] {sim1_time.dat}; 



    \end{axis}
  \end{tikzpicture}\label{fig:sim1_time}}%
    \end{minipage}%
    \caption{Simulation results for scenarios described in Section~\ref{sec:latentrisk}.}
    \label{fig:sim}
\end{figure*}

\section{Conclusion}
This study presents insights into an efficient learning methodology and an enhanced safety mechanism in the presence of latent risks. The proposed framework decomposes complex tasks into easier subtasks and optimizes subtasks' parameters. 
Additionally, the framework demonstrates its capability to identify latent risks, ensuring safety without compromising operational efficiency. These are achieved by integrating LLM, numerical optimization, and optimization-based control in a synergistic manner.




\bibliographystyle{unsrt}  
\bibliography{reference}

\begin{thebibliography}{10}

\bibitem{Yu2023}
Wenhao Yu et~al.
\newblock Language to rewards for robotic skill synthesis.
\newblock In Jie Tan, Marc Toussaint, and Kourosh Darvish, editors, {\em Conference on Robot Learning, CoRL 2023, 6-9 November 2023, Atlanta, GA, USA}, volume 229 of {\em Proceedings of Machine Learning Research}, pages 374--404. PMLR, 2023.

\bibitem{zhao2023expel}
Andrew Zhao, Daniel Huang, Quentin Xu, Matthieu Lin, Yong-Jin Liu, and Gao Huang.
\newblock Expel: {LLM} agents are experiential learners.
\newblock {\em arXiv preprint arXiv:2308.10144}, 2023.

\bibitem{huang2022inner}
Wenlong Huang, Fei Xia, Ted Xiao, Harris Chan, Jacky Liang, Pete Florence, Andy Zeng, Jonathan Tompson, Igor Mordatch, Yevgen Chebotar, et~al.
\newblock Inner monologue: Embodied reasoning through planning with language models.
\newblock {\em arXiv preprint arXiv:2207.05608}, 2022.

\bibitem{huang2023voxposer}
Wenlong Huang, Chen Wang, Ruohan Zhang, Yunzhu Li, Jiajun Wu, and Li~Fei-Fei.
\newblock Voxposer: Composable 3d value maps for robotic manipulation with language models.
\newblock {\em arXiv preprint arXiv:2307.05973}, 2023.

\bibitem{yang2023plug}
Ziyi Yang, Shreyas~S Raman, Ankit Shah, and Stefanie Tellex.
\newblock Plug in the safety chip: Enforcing constraints for llm-driven robot agents.
\newblock {\em arXiv preprint arXiv:2309.09919}, 2023.

\bibitem{liu2023reflect}
Zeyi Liu, Arpit Bahety, and Shuran Song.
\newblock Reflect: Summarizing robot experiences for failure explanation and correction.
\newblock {\em arXiv preprint arXiv:2306.15724}, 2023.

\bibitem{zhao2024expelllmagentsexperiential}
Andrew Zhao, Daniel Huang, Quentin Xu, Matthieu Lin, Yong-Jin Liu, and Gao Huang.
\newblock Expel: Llm agents are experiential learners, 2024.

\bibitem{ren2023robots}
Allen~Z Ren, Anushri Dixit, Alexandra Bodrova, Sumeet Singh, Stephen Tu, Noah Brown, Peng Xu, Leila Takayama, Fei Xia, Jake Varley, et~al.
\newblock Robots that ask for help: Uncertainty alignment for large language model planners.
\newblock {\em arXiv preprint arXiv:2307.01928}, 2023.

\bibitem{huang2021toward}
Chao Huang, Chen Lv, Peng Hang, and Yang Xing.
\newblock Toward safe and personalized autonomous driving: Decision-making and motion control with dpf and cdt techniques.
\newblock {\em IEEE/ASME Transactions on Mechatronics}, 26(2):611--620, 2021.

\bibitem{marvi2021safe}
Zahra Marvi and Bahare Kiumarsi.
\newblock Safe reinforcement learning: A control barrier function optimization approach.
\newblock {\em International Journal of Robust and Nonlinear Control}, 31(6):1923--1940, 2021.

\bibitem{ahangar2021survey}
M~Nadeem Ahangar, Qasim~Z Ahmed, Fahd~A Khan, and Maryam Hafeez.
\newblock A survey of autonomous vehicles: Enabling communication technologies and challenges.
\newblock {\em Sensors}, 21(3):706, 2021.

\bibitem{muller2022motion}
Johannes M{\"u}ller, Jan Strohbeck, Martin Herrmann, and Michael Buchholz.
\newblock Motion planning for connected automated vehicles at occluded intersections with infrastructure sensors.
\newblock {\em IEEE Transactions on Intelligent Transportation Systems}, 23(10):17479--17490, 2022.

\bibitem{wang2023physics}
Zhuoyuan Wang, Reece Keller, Xiyu Deng, Kenta Hoshino, Takashi Tanaka, and Yorie Nakahira.
\newblock Physics-informed representation and learning: Control and risk quantification.
\newblock {\em arXiv preprint arXiv:2312.10594}, 2023.

\bibitem{hoshino2024physics}
Hikaru Hoshino and Yorie Nakahira.
\newblock A physics-informed reinforcement learning framework for risk probability estimation.
\newblock {\em American Control Conference}, 2024.

\bibitem{Wright2024propagation}
Oren Wright, J.~M.~F. Moura, and Yorie Nakahira.
\newblock An analytic solution to covariance propagation in neural networks.
\newblock {\em The 27th International Conference on Artificial Intelligence and Statistics}, 2024.

\bibitem{wang2023generalizable}
Zhuoyuan Wang and Yorie Nakahira.
\newblock A generalizable physics-informed learning framework for risk probability estimation.
\newblock In {\em Learning for Dynamics and Control Conference}, pages 358--370. PMLR, 2023.

\bibitem{gangadhar2023occlusion}
Siddharth Gangadhar, Zhuoyuan Wang, Kofi Poku, Naoya Yamada, Kohei Honda, Yorie Nakahira, Hiroyuki Okuda, and Tatsuya Suzuki.
\newblock An occlusion-and interaction-aware safe control strategy for autonomous vehicles.
\newblock In {\em 2023 22nd IFAC World Congress}, 2023.

\bibitem{cao2022trustworthy}
Zhong Cao, Shaobing Xu, Xinyu Jiao, Huei Peng, and Diange Yang.
\newblock Trustworthy safety improvement for autonomous driving using reinforcement learning.
\newblock {\em Transportation research part C: emerging technologies}, 138:103656, 2022.

\bibitem{miyaoka2024cbfllmsafecontrolllm}
Yuya Miyaoka and Masaki Inoue.
\newblock Cbf-llm: Safe control for llm alignment, 2024.

\bibitem{biondi2019human}
Francesco Biondi, Ignacio Alvarez, and Kyeong-Ah Jeong.
\newblock Human--vehicle cooperation in automated driving: A multidisciplinary review and appraisal.
\newblock {\em International Journal of Human--Computer Interaction}, 35(11):932--946, 2019.

\bibitem{lu2023multimodal}
Yujie Lu, Pan Lu, Zhiyu Chen, Wanrong Zhu, Xin~Eric Wang, and William~Yang Wang.
\newblock Multimodal procedural planning via dual text-image prompting.
\newblock {\em arXiv preprint arXiv:2305.01795}, 2023.

\bibitem{shinn2023reflexion}
Noah Shinn, Federico Cassano, Beck Labash, Ashwin Gopinath, Karthik Narasimhan, and Shunyu Yao.
\newblock Reflexion: Language agents with verbal reinforcement learning, 2023.
\newblock {\em URL https://arxiv. org/abs/2303.11366}, 2023.

\bibitem{wu2024toolplanner}
Qinzhuo Wu, Wei Liu, Jian Luan, and Bin Wang.
\newblock Toolplanner: A tool augmented llm for multi granularity instructions with path planning and feedback.
\newblock {\em arXiv preprint arXiv:2409.14826}, 2024.

\bibitem{wang2023describe}
Zihao Wang, Shaofei Cai, Anji Liu, Xiaojian Ma, and Yitao Liang.
\newblock Describe, explain, plan and select: Interactive planning with large language models enables open-world multi-task agents.
\newblock {\em arXiv preprint arXiv:2302.01560}, 2023.

\bibitem{feng2023llama}
Yicheng Feng, Yuxuan Wang, Jiazheng Liu, Sipeng Zheng, and Zongqing Lu.
\newblock Llama rider: Spurring large language models to explore the open world.
\newblock {\em arXiv preprint arXiv:2310.08922}, 2023.

\bibitem{wang2023voyager}
Guanzhi Wang, Yuqi Xie, Yunfan Jiang, Ajay Mandlekar, Chaowei Xiao, Yuke Zhu, Linxi Fan, and Anima Anandkumar.
\newblock Voyager: An open-ended embodied agent with large language models, 2023.
\newblock {\em URL https://arxiv. org/abs/2305.16291}, 2023.

\bibitem{lin2023swiftsage}
Bill~Yuchen Lin, Yicheng Fu, Karina Yang, Faeze Brahman, Shiyu Huang, Chandra Bhagavatula, Prithviraj Ammanabrolu, Yejin Choi, and Xiang Ren.
\newblock Swiftsage: A generative agent with fast and slow thinking for complex interactive tasks.
\newblock {\em Advances in Neural Information Processing Systems}, 36:23813--23825, 2023.

\bibitem{singh2023progprompt}
Ishika Singh, Valts Blukis, Arsalan Mousavian, Ankit Goyal, Danfei Xu, Jonathan Tremblay, Dieter Fox, Jesse Thomason, and Animesh Garg.
\newblock Progprompt: Generating situated robot task plans using large language models.
\newblock In {\em 2023 IEEE International Conference on Robotics and Automation (ICRA)}, pages 11523--11530. IEEE, 2023.

\bibitem{song2023llm}
Chan~Hee Song, Jiaman Wu, Clayton Washington, Brian~M Sadler, Wei-Lun Chao, and Yu~Su.
\newblock Llm-planner: Few-shot grounded planning for embodied agents with large language models.
\newblock In {\em Proceedings of the IEEE/CVF International Conference on Computer Vision}, pages 2998--3009, 2023.

\bibitem{chen2021psi}
Tina Chen, Taotao Jing, Renran Tian, Yaobin Chen, Joshua Domeyer, Heishiro Toyoda, Rini Sherony, and Zhengming Ding.
\newblock Psi: A pedestrian behavior dataset for socially intelligent autonomous car.
\newblock {\em arXiv preprint arXiv:2112.02604}, 2021.

\bibitem{yu2023language}
Wenhao Yu, Nimrod Gileadi, Chuyuan Fu, Sean Kirmani, Kuang-Huei Lee, Montse~Gonzalez Arenas, Hao-Tien~Lewis Chiang, Tom Erez, Leonard Hasenclever, Jan Humplik, Brian Ichter, Ted Xiao, Peng Xu, Andy Zeng, Tingnan Zhang, Nicolas Heess, Dorsa Sadigh, Jie Tan, Yuval Tassa, and Fei Xia.
\newblock Language to rewards for robotic skill synthesis, 2023.

\bibitem{Newsflare2022}
Newsflare.
\newblock Insane dash cam moments | near misses caught on camera, May 2022.
\newblock YouTube video.

\bibitem{newyork}
J~Utah.
\newblock Driving downtown - new york city 4k - usa, March 2018.
\newblock YouTube video.

\bibitem{pittsburgh}
J~Utah.
\newblock Driving pittsburgh 4k hdr - coal \& steel city - usa, August 2023.
\newblock YouTube video.

\bibitem{chongqing}
China~Street View.
\newblock Driving in chongqing - this is a city with the most complicated traffic in china, July 2022.
\newblock YouTube video.

\end{thebibliography}

\clearpage
\appendices
\onecolumn
\makeatletter
\@addtoreset{figure}{subsection}
\makeatother
\renewcommand{\thefigure}{\thesubsection-\arabic{figure}}
\section{Extended Simulation Results: Robot Arm Pushing Performance}\label{sec:robot_simulation}
The extended description and results for the simulation presented in Section~\ref{sec:fastadaptation} are provided below.
\subsection{Robot system dynamics}\label{sec:robot_dynamics}
In this paper, we adopt a simplified kinematic model for the two-degree-of-freedom robot arm, which is derived as follows:
\begin{equation}
\begin{bmatrix}
x_a(t+1) \\
v_a(t+1)
\end{bmatrix}
=
\begin{bmatrix}
\mathbf{I} & \Delta t \cdot \mathbf{I} \\
\mathbf{0} & \mathbf{0}
\end{bmatrix}
\begin{bmatrix}
x_a(t) \\
v_a(t)
\end{bmatrix}
+
\begin{bmatrix}
\mathbf{0} \\
\mathbf{J}(q_1, q_2)
\end{bmatrix}
u(t),
\end{equation}
where $x_a(t) \in \mathbb{R}^2$ and $v_a(t) \in \mathbb{R}^2$ denote the end-effector position and velocity of the robot arm, respectively, $u(t)\in\mathbb{R}^2$ is the control input representing the control velocity applied to the robot joints, $\mathbf{I}\in\mathbb{R}^{2\times2}$ and $\mathbf{0}\in\mathbb{R}^{2\times2}$ denote the identity and zero matrices, respectively, $\Delta t$ is a fixed sampling interval for simulation and control, and $\mathbf{J}(q_1, q_2)\in\mathbb{R}^{2\times2}$ denotes the Jacobian matrix, which can be derived as follows:
\begin{equation}
    \mathbf{J}(q_1, q_2) =
\begin{bmatrix}
    -l_1 \sin(q_1) - l_2 \sin(q_1 + q_2) & -l_2 \sin(q_1 + q_2) \\
    l_1 \cos(q_1) + l_2 \cos(q_1 + q_2) & l_2 \cos(q_1 + q_2)
\end{bmatrix},
\end{equation}
where $q_1$ and $q_2$ represent the joint angles of the two-degree-of-freedom robot arm, and $l_1$ and $l_2$ denote the lengths of the two robot links, respectively.

\subsection{Simulation settings and results}
We consider a simulation in which a 2-degree-of-freedom robot arm is tasked with pushing a box (Fig.~\ref{fig:sim2_env}). The full system state at time $t$ is represented by $x(t) = [x_a(t),\, v_a(t),\, x_b(t),\, v_b(t),\, \mathbf{a}_b(t),\,\mathbf{F}(t)]^\top$, where $x_a(t) \in \mathbb{R}^2$ and $v_a(t) \in \mathbb{R}^2$ denote the end-effector position and velocity of the robot arm, respectively. The state of the box is defined by its position $x_b(t) \in \mathbb{R}^2$, velocity $v_b(t) \in \mathbb{R}^2$, and acceleration $\mathbf{a}_b(t) \in \mathbb{R}^2$. The contact force between the robot arm and the box is represented by $\mathbf{F}(t) \in \mathbb{R}^2$. This formulation results in a 16-dimensional state space that captures both the internal configuration of the robot and the external dynamics of the object it interacts with. 
The system dynamics are defined in continuous time. The box follows second-order Newtonian dynamics: $\dot{x}_b(t) = v_b(t), \dot{v}_b(t) = \mathbf{a}_b(t)$. 
We model this interaction as $\mathbf{F}(t) - \mathbf{F}_s = m_b \cdot \mathbf{a}_b(t)$, where $\mathbf{F}_s$ denotes the frictional force between the box and the floor and $m_b$ denotes the box mass. 
The control input $u(t) =v_a(t)$ represents the desired joint velocities.
These continuous-time dynamics are discretized using a fixed sampling interval $\Delta t$ for simulation and control.

\textbf{Experimental settings.}
The simulation task is to push the box along predefined waypoints $x_d(t)$. Performance is evaluated using the following metric: $\mathcal{R} = \|x_d(t) - x_b(t)\|_2$. 
We run the simulation under 20 randomly sampled initial configurations. Specifically, we define the sampling set for initial box positions as: $\Lambda = \left\{ x = [x_1, x_2] : x_1 \in [-0.8, -0.6],\ x_2 \in [0.7, 0.9] \right\}$, and draw samples $x_b(0) \overset{\text{i.i.d.}}{\sim} \mathcal{U}(\Lambda)$ independently from the uniform distribution over $\Lambda$. Similarly, the box mass is initialized by sampling independently from the interval $\mathcal{M} = [0.1, 0.3]$, as $m_b \overset{\text{i.i.d.}}{\sim} \mathcal{U}(\mathcal{M})$.

\textbf{Results.} 
The results in Fig.~\ref{fig:sim2_cost} show that approach \textit{a)} without task decomposition fails to complete the task. In contrast, LLM-assisted approaches (approaches \textit{b) - f)}) overcome this limitation by decomposing the task into two subtasks:
(1) during subtask window $\omega^{[1]}$, the goal is to move the robot arm to the box using reward $R^{[1]} = -\lVert \theta^{[1]} - x_{a}(t) \rVert_2$, where $\theta^{[1]} = x_b(0)$ is the box’s initial position;
(2) during subtask window $\omega^{[2]}$, the goal is to push the box with a safe contact force using reward $R^{[2]} = -\lVert \theta^{[2]}_d - x_{a}(t) \rVert_2$ and constraint $\Phi^{[2]} = \theta^{[2]}_f -  |\mathbf{F}|$. Here, $\theta^{[2]}_d$ and $\theta^{[2]}_f$ denote the target location and the safe contact force threshold, respectively. The subtask termination conditions $\varphi^{[1]}$ and $\varphi^{[2]}$ are both set to zero to ensure that the robot arm makes contact with the box and pushes it to the desired location.
Fig.~\ref{fig:sim2_param} illustrates the evolution of the constraint parameter $\theta^{[2]}_f$ across six methods. In the single-turn LLM approach (approach b), $\theta^{[2]}_f$ is selected once by the LLM and remains fixed, resulting in limited adaptability. The optimization-only method (approach c) refines $\theta^{[2]}_f$ using zeroth-order gradient descent but lacks high-level task understanding, leading to slower convergence. The multi-turn zero-shot LLM (approach d) allows iterative refinement through feedback, but without memory across turns, often results in unstable updates. The multi-turn in-context LLM (approach e) improves upon this by incorporating prior interactions, generating more coherent updates to $\theta^{[2]}_f$, though it still lacks precise tuning.
In contrast, the proposed method combines the strengths of reasoning and numerical optimization. It refines $\theta^{[2]}_f$ over three in-context LLM-guided reasoning rounds, each followed by five zeroth-order optimization steps. This yields a step-shaped trajectory that reflects both high-level semantic adaptation and low-level numerical refinement. The LLM provides informed initialization via $\mathtt{LLM_{correction}}$, while the embedded optimization tunes the parameter toward task satisfaction. These results highlight the importance of integrating multi-turn reasoning with direct optimization for effective learning of constraint parameters in complex robotic control tasks.
\begin{figure}[htbp]
    \centering
    \makebox[\textwidth][c]{%
        \subfloat[Environment.]{%
              \begin{tikzpicture}
    \begin{axis}[
        axis on top,
        width=0.2\linewidth,
        height=0.2\linewidth,
        scale only axis,
        enlargelimits=false,
        xmin=-2,
        xmax=1,
        ymin=-1,
        ymax=2,
        axis equal=true,
        every axis label/.append style={font=\scriptsize},
        every tick label/.append style={font=\scriptsize}
        ]
      \addplot[thick,blue] graphics[xmin=-2,ymin=-1,xmax=1,ymax=2]{figures/env.pdf};
      \node[fill = white, minimum width = 0.8cm,] at (axis cs: -1.6,0.8) {\scriptsize Box};
      \node[fill = white, minimum width = 1.3cm, minimum height = 1cm] at (axis cs: 0.5,0.8){};
      \node[] at (axis cs: 0.2,1) {\scriptsize Robot};
      \node[] at (axis cs: 0.2,0.75) {\scriptsize arm};

      \node[fill = white, minimum width = 0.5cm, minimum height = 0.6cm] at (axis cs: -0.7,0.1){};
      \node[] at (axis cs: -0.7,0.1) {\scriptsize $\theta^2_f$};
    \end{axis}
  \end{tikzpicture}
            \label{fig:sim2_env}
        }\hspace{2em}%
        \subfloat[Illustration of 6 approaches.]{%
            \includegraphics[width=0.4\linewidth]{figures/legend.pdf}
            \label{fig:cube}
        }%
    }\par\medskip

    \makebox[\textwidth][c]{%
        \subfloat[Parameter $\theta^{[2]}_f$ vs. \# data point.]{%
                \begin{tikzpicture}
        \begin{axis}[
            xlabel={\#Data point},
            ylabel={Parameter $\theta^{[2]}_f$},
            ylabel shift=-6pt,
            grid=both,
            grid style={dashed,gray!30},
            legend pos=north east,
            width=0.4\linewidth,
            height=0.3\linewidth,
            xmin=1, xmax=15,
            legend style={font=\scriptsize,draw=none, fill = none},
            legend style={font=\scriptsize, at={(axis cs:3.2,1.5)},anchor=south west},
            every axis label/.append style={font=\scriptsize},
            every tick label/.append style={font=\scriptsize}
        ]
        \addplot[dashdotted, violet, line width=1.2pt, domain=1:15] {0.55};

        \addplot [newgreen!300, mark=*, mark size=1, forget plot] table [x=iter, y=s3] {sim2_theta.dat}; 
        
        \addplot [magenta, mark=*, mark size=1, forget plot] table [x=iter, y=s4] {sim2_theta.dat}; 
        \addplot [newblue!300, mark=*, mark size=1, forget plot] table [x=iter, y=s5] {sim2_theta.dat}; 
        \addplot [red, mark=*, mark size=1, forget plot] table [x=iter, y=s6] {sim2_theta.dat}; 
        
        \end{axis}
    \end{tikzpicture}
            \label{fig:sim2_param}
        }\hspace{2em}%
        \subfloat[Cost $\mathcal{L}$ vs. \# data point.]{%
                \begin{tikzpicture}
        \begin{axis}[
            xlabel={\# Data point},
            ylabel={Cost $\mathcal{L}$},
            ylabel shift=-6pt,
            grid=both,
            grid style={dashed,gray!30},
            legend pos=north east,
            width=0.4\linewidth,
            height=0.3\linewidth,
            xmin=1, xmax=15,
            legend style={font=\scriptsize,draw=none, fill = none},
            legend style={font=\scriptsize, at={(axis cs:3.2,1.5)},anchor=south west},
            every axis label/.append style={font=\scriptsize},
            every tick label/.append style={font=\scriptsize}
        ]
        \addplot[dashdotted, olive, line width=1.2pt, domain=1:15] {0.24865221377805}; 
        \addplot[dashdotted, violet, line width=1.2pt, domain=1:15] {0.221053862864606}; 

        \addplot [newgreen!300, mark=*, mark size=1, forget plot] table [x=iter, y=s3] {sim2_loss.dat}; 

        \addplot [magenta, mark=*, mark size=1, forget plot] table [x=iter, y=s4] {sim2_loss.dat}; 
        \addplot [newblue!300, mark=*, mark size=1, forget plot] table [x=iter, y=s5] {sim2_loss.dat}; 
        \addplot [red, mark=*, mark size=1, forget plot] table [x=iter, y=s6] {sim2_loss.dat}; 
        
        \end{axis}
    \end{tikzpicture}
            \label{fig:sim2_cost}
        }%
    }

    \caption{Simulation results and three-layer framework visualization for scenarios described in Section~\ref{sec:fastadaptation}. In (b), the red star denotes the proposed method. Colored rings indicate ablation baselines, corresponding to the curves in (c) and (d).}
\end{figure}
\subsection{LLM prompts}\label{sec:robot_llmprompt}
\noindent The prompt structure in this work is inspired by~\cite{yu2023language}. Below is the prompt $\mathcal{P}_{\mathtt{coder}}$ used to instruct the LLM to generate reward and constraint functions for the robot arm scenario (Section~\ref{sec:fastadaptation}): 
\begin{lstlisting}
We have a plan of a robot arm with palm to manipulate objects and we want you to turn that into 
the corresponding program with following functions:

Rules:
1. Always format the code in code blocks. In your response execute_plan should be called exactly once at the end.
2. Do not invent new functions or classes. The only allowed functions you can call are the ones listed below. Do not leave unimplemented code blocks in your response.
3. The only allowed library is numpy. Do not import or use any other library.
4. If you are not sure what value to use, just use your best judge. Do not use None for anything.
5. Do not calculate the position or direction of any object (except for the ones provided above). Just use a number directly based on your best guess.
6. You do not need to make the robot do extra things not mentioned in the plan such as stopping the robot.

```
def safe_distance_constraint(name_obj, safe_distance=0.3)
```
this constraint is to assure the distance between the robot and name_obj must not below "safe_distance"
name_obj (e.g., human, glass, vase, obstacle) could be anything depneding on context of the provided motion plan

```
def safe_contact_force_constraint(name_obj, safe_force=10.0)
```
the constraint to assure that the contact force between the robot and name_obj must below "safe_force"
name_obj (e.g., human, glass, vase, obstacle) could be anything depneding on context of the provided motion plan

```
def safe_speed_constraint(max_speed=0.1)
```
set the constraint on the robot speed. The robot velocity must not exceed the "max_speed"
default "max_speed"=0.1, but it can be modified according to distances between robots and external objects

```
def set_l2_distance_reward(name_obj_A, name_obj_B)
```
where name_obj_A and name_obj_B are selected from ['palm', 'arm', 'apple', 'banana', 'box', 'bowl', 'drawer_handle', 'faucet_handle', 'drawer_center', 'rest_position', 'target_position'].
This term sets a reward for minimizing l2_distance between name_obj_A and name_obj_B so they get closer to each other.
rest_position is the default position for the palm when it's holding in the air.
target_position is the default position for the object to be puhsed or shifted on the x-y 2D plane

```
def execute_plan(duration=4) -> (rewards, forces)
```
This function sends the parameters to the robot and execute the plan for `duration` seconds, default to be 2. The function should return rewards

```
def reset_reward()
```
This function resets the reward to default values.

Example plan:
To perform this task, the manipulator's arm/palm should move away from the obstacle.
safe distance between the manipulator and obstacle must be larger than 0.5m
robot must keep the speed below 0.2m/s for a safety reason.

Example answer code:
"""
set_l2_distance_reward("palm", "target_position")
safe_distance_constraint("obstacle", 0.5)
safe_speed_constraint(0.2)

rewards = execute_plan(4)
"""

Example plan:
To perform this task, the manipulator's arm should move close to bottle.
safe contact force between the manipulator and bottle must be below 4N
safe distance between the manipulator and bottle must be larger than 0.1m
robot must keep the speed below 0.5m/s for a safety reason.

Example answer code:
"""
import numpy as np

set_l2_distance_reward("arm", "bottle")
safe_contact_force_constraint("bottle", 4.0)
safe_distance_constraint("bottle", 0.1)
safe_speed_constraint(0.5)

rewards = execute_plan(3)
"""

If you understand, say Yes.
\end{lstlisting}

\noindent Below is the prompt $\mathcal{P}_{\mathtt{correction}}$ used to instruct the LLM to evaluate task satisfaction and provide corresponding solutions if the task is determined to have failed in the robot arm scenario (Section~\ref{sec:fastadaptation}):

\begin{lstlisting}
You are an assistant tasked with evaluating whether a robot has fulfilled a task based on a provided task objective and feedback. 
Your role is to analyze the feedback, determine whether the task was fulfilled, and if not, identify exactly one underlying cause for task failure and propose one corresponding solution. 
The cause and solution must be selected from the following sets:

- Cause Set:
  - The object doesn't move at all.  
  - The object lags behind the target waypoints.  
  - The object leads ahead of (moves faster than) the target waypoints.

- Solution Set (with Dynamic Force Adjustment):
  - If the object lags behind, recommend: "Increase the pushing force to [new_force]N," where [new_force] is calculated based on the deviation amount.
  - If the object leads ahead, recommend: "Reduce the pushing force to [new_force]N," where [new_force] is calculated based on the deviation amount.
  - If the current force is already at the maximum allowable value (2N), reduce the force instead of increasing it.
      -In this case, recommend: "Reduce the pushing force to [reduced_force]N," where [reduced_force] is an appropriate decrease to stabilize motion.

User Instructions:
Both the task objective and feedback are provided. The task objective describes the desired action and reference waypoints, while the feedback includes the measured pushing force and tracked waypoints. When making your assessment, note that:
- The task is considered fulfilled if the tracked waypoints match the reference waypoints within a small tolerance. Specifically:
    - For waypoint coordinates, differences within 0.005m are acceptable.
    - For force, treat values such as 0.5, 0.5111, and 0.4999 as equivalent.
- Even if the pushing force is slightly lower than expected, if the tracked waypoints are within the acceptable tolerance, the task should be considered fulfilled.
- The task is considered unfulfilled if the tracked waypoints deviate significantly from the reference beyond the allowed tolerance.
- Dynamic Adjustment: When deviations are observed:
    - For lagging: Determine how far the object lags behind the reference and recommend an increase in pushing force accordingly.
    - For leading: Determine how far the object is ahead of the reference and recommend a reduction in pushing force accordingly.
    - Express the new recommended force as an absolute value (for example, "Increase the pushing force to 1.0N" if the current force is 0.5N and the deviation warrants an additional 0.5N).
- If the current force is already at 2N, do not increase it further. Instead, reduce it.

Rules:
1. Assess task fulfillment based on both the tracked waypoints and the applied force, considering the tolerances defined above.
2. If the tracked waypoints match the reference waypoints (within tolerance), report:
    - Task fulfillment: Yes
    - No solutions are needed.
3. If the tracked waypoints do not match the reference waypoints beyond the allowed tolerance, report:
    - Task fulfillment: No
    - Select exactly one cause from the following set:  
      - The object doesn't move at all.  
      - The object lags behind the target waypoints.  
      - The object leads ahead of the target waypoints.
    - Select exactly one corresponding solution from the following set:  
      - If the object lags behind, output: "Increase the pushing force to [new_force]N," where [new_force] is based on the deviation.  
      - If the object leads ahead, output: "Reduce the pushing force to [new_force]N," where [new_force] is based on the deviation.
4. Format your answer as follows:
   ```
   Task fulfillment: Yes or No
   Cause:
   - [Selected Cause]
   Solution:
   - [Selected Solution]
   ```

Examples:

Successful Example:

**Query:**  
Task objective: Push the bottle of water (mass = 200 grams) via the following waypoints {x: [0.1, 0.2, 0.3, 0.4, 0.5], t: [1, 2, 3, 4, 5]}  
Feedback:  
- The pushing force is 0.2N.  
- The tracked waypoints are {x: [0.11, 0.205, 0.298, 0.402, 0.49], t: [1, 2, 3, 4, 5]}.

**Answer:**  
Task fulfillment: Yes
Cause:
- The tracked waypoints are within 0.005m of the reference, indicating acceptable performance.
Solution:
- No solution needed.

Failure Example:

**Query:**  
Task objective: Push the bottle of water (mass = 200 grams) via the following waypoints {x: [0.1, 0.2, 0.3, 0.4, 0.5], t: [1, 2, 3, 4, 5]}  
Feedback:  
- The pushing force is 0.2N.  
- The tracked waypoints are {x: [0.1, 0.15, 0.3, 0.35, 0.5], t: [1, 2, 3, 4, 5]}.

**Answer:**  
Task fulfillment: No
Cause:
- The object lags behind the target waypoints.
Solution:
- Increase the pushing force to at least 1N.

**Query:**  
Task objective: Push the bottle of water (mass = 200 grams) via the following waypoints {x: [0.1, 0.2, 0.3, 0.4, 0.5], t: [1, 2, 3, 4, 5]}
Feedback:
- The pushing force is 2.0N.
- The tracked waypoints are {x: [0.1, 0.3, 0.4, 0.5, 0.6], t: [1, 2, 3, 4, 5]}.

**Answer:**  
Task fulfillment: No
Cause:
- The object leads ahead of the target waypoints.
Solution:
- Reduce the pushing force to 1.5N.

---

If you understand my requirements, say Yes.
\end{lstlisting}

\subsection{Feedback template}
\noindent In Algorithm~\ref{alg:proposed_framework}, we refer to the use of $\mathtt{FeedbackTemplate}$ to map the numerical state and action information $(x_{0:T}, u_{0:T})$ into a textual format. Below is the specific $\mathtt{FeedbackTemplate}$ used in the robot arm pushing simulation.
\begin{lstlisting}
    feedback_text = f"""
    Task objective: {task_description}
    Feedback: 
    - The pushing force is {pushing_force}N
    - The tracked waypoints is {tracked_waypoints}
    Based on the feedback, has the robot already fulfilled the task? 
    If it hasn't, what are the possible reasons that the robot was not able to fulfill the task, and what are your solutions?
    """
\end{lstlisting}

\section{Extended Simulation Results: Robot Arm Object Grasping Performance}\label{sec:grasp_simulation}
We incorporate the robot arm object grasping task into both the online framework described in Algorithm~\ref{alg:online_framework} and the offline framework described in Algorithm~\ref{alg:proposed_framework}.
\subsection{System model and experiment settings}\label{sec:grasp_experiment}

We consider a simulation in which a 2-degree-of-freedom robot arm is tasked with grasping an object to a location.
The full system state at time $t$ is represented by 
$x(t) = [x_a(t), v_a(t), \{x_o^{(i)}(t), \mathbf{F}^{(i)}(t), \mathtt{status}^{(i)}(t)\}_{i=1}^{N_g}]^\top$, 
where $x_a(t) \in \mathbb{R}^2$ and $v_a(t) \in \mathbb{R}^2$ denote the end-effector position and velocity of the robot arm, respectively. Each object $i$ is characterized by its position $x_o^{(i)}(t) \in \mathbb{R}^2$, the contact force $\mathbf{F}^{(i)}(t) \in \mathbb{R}^2$ exerted by the end-effector, and a status flag $\mathtt{status}^{(i)}(t) \in \{\mathtt{free}, \mathtt{grasped}, \mathtt{pushed}\}$, which determines its current interaction mode with the robot.
This formulation results in a $5N_g+4$-dimensional state space that captures both the internal configuration of the robot and the external dynamics of the objects it interacts with. 
The control input $u(t) =v_a(t)$ represents the desired joint velocities. The controller operates on the robot internal state $[x_a(t), u_a(t)]$, follow the formulation described in~\ref{sec:robot_dynamics}.
These continuous-time dynamics are discretized using a fixed sampling interval $\Delta t$ for simulation and control.
For simplification, we assume that the robot arm consists of two functional components: the arm and the palm. The palm is capable of grasping and transporting objects, while the arm can exert pushing forces through contact. However, due to high friction between the objects and the surface, we assume that pushing via the arm does not result in object motion. As a result, objects can only be moved when explicitly grasped by the palm. This modeling choice allows us to simplify the object dynamics and focus on grasp-based interactions for task execution.
\subsection{Online Showcases}\label{sec:grasp_results}
\begin{figure}[ht]
    \centering
        \input{figures/workflow_online}
        \label{fig:online_framework}
    \caption{Online framework.}
\end{figure}

We present three showcase examples of grasping experiments, each conducted under different settings and initializations.
\subsubsection{Setting 1}
\begin{figure}[!ht]
    \centering
    \includegraphics[width=0.7\linewidth]{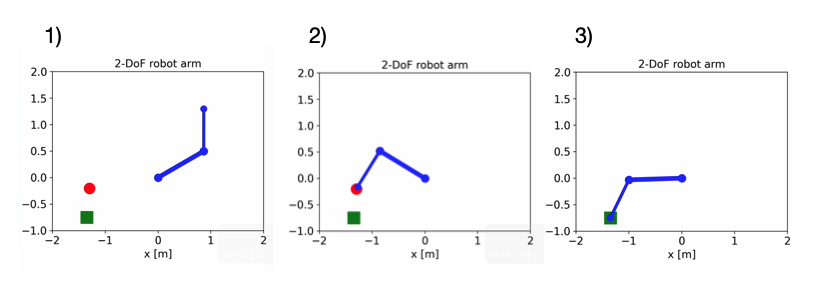}
    \caption{Setting 1: From left to right — 1) scenario initialization; 2) palm grasping the apple; and 3) palm placing the apple into the box.}
    \label{fig:grasp_s1}
\end{figure}

In this scenario, a red apple (represented by a red circle) is placed at position $[-1.3, -0.2]$, and a green box (represented by a green square) is placed at position $[-1.35, -0.75]$. A 2-DoF robot arm (depicted in blue) is also present in the playground. We consider two different user instructions as task objectives.

For Setting 1.1, the user input is \textit{“move the apple to the box.”}
Then, the motion descriptor $\mathtt{LLM_{riskhandling}}$ provides:
\begin{lstlisting}
[start of plan]
To perform this task, the manipulator's palm should move close to apple.
object1=apple should be close to object2=box.
robot should release the object1=apple in the end
[end of plan]
\end{lstlisting}
Following the motion description, the $\mathtt{LLM_{coder}}$ generates the corresponding code block:
\begin{lstlisting}
```python
import numpy as np

reset_reward()  # This is a new task so reset reward; otherwise we don't need it
set_l2_distance_reward("palm", "apple")
set_l2_distance_reward("apple", "box")

rewards = execute_plan(4)
release_object("apple")
```
\end{lstlisting}
Upon execution, the system returns the following feedback via the $\mathtt{FeedbackTemplate}$:
\begin{lstlisting}
Task objective: move the apple to the box
Feedback: 
- The current position of the robot arm's end-effector is: [-1.35000002 -0.75      ]
- The current position of the robot arm's end-effector is: [-1.35000002 -0.75      ]
- The apple's initial position is [-1.3 -0.2] and its current position of  is [-1.35 -0.75]
- The apple's initial position is [-1.3 -0.2] and its current position of  is [-1.35 -0.75]
- The box's initial position is [-1.35 -0.75] and its current position of  is [-1.35 -0.75]
\end{lstlisting}
Finally, the $\mathtt{LLM_{correction}}$ module evaluates the task completion status:
\begin{lstlisting}
Task fulfillment: Yes
Task completed!
\end{lstlisting}

For Setting 1.2, the user input is \textit{“move the apple to the box carefully.”} In this case, we focus specifically on the responses from the motion descriptor $\mathtt{LLM_{riskhandling}}$ and $\mathtt{LLM_{coder}}$. Since the instruction includes the word \textit{“carefully”}, the generated response incorporates safety constraints accordingly, as shown below.
\begin{lstlisting}
[start of plan]
To perform this task, the manipulator's palm should move close to apple.
object1=apple should be close to object2=box.
safe contact force between the manipulator and obstacle=apple must be below 0.5N
safe contact force between the manipulator and obstacle=box must be below 0.5N
robot must keep the speed below 0.1m/s for a safety reason.
robot should release the object1=apple in the end
[end of plan]
\end{lstlisting}

\begin{lstlisting}
```python
import numpy as np

reset_reward()  # This is a new task so reset reward; otherwise we don't need it
set_l2_distance_reward("palm", "apple")
set_l2_distance_reward("apple", "box")
safe_contact_force_constraint("apple", 0.5)
safe_contact_force_constraint("box", 0.5)
safe_speed_constraint(0.1)

rewards = execute_plan(4)
release_object("apple")
```
\end{lstlisting}
For both Setting 1.1 and Setting 1.2, the overall task is decomposed into two subtasks: the first is to move the \textit{'palm'} to the \textit{'apple'} and the second is to move the \textit{'apple'} to the \text{'box'}. The execution of these subtasks is illustrated in Fig.~\ref{fig:grasp_s1}.

\begin{figure}[!ht]
    \centering
    \includegraphics[width=0.8\linewidth]{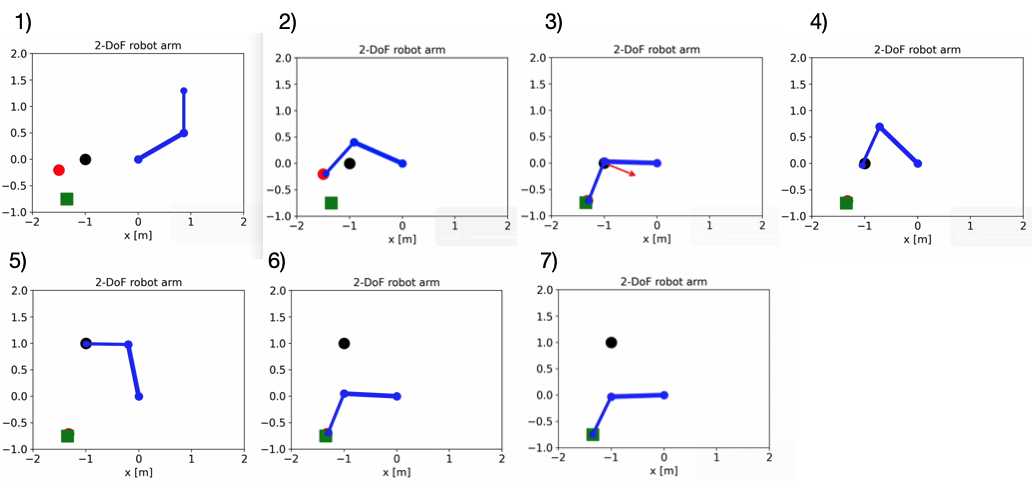}
    \caption{Setting 2: Task execution from beginning to end, illustrated over subfigures 1)–7). Subfigures 1)–3) depict the first-round attempt based on the initial instruction. Subfigures 4)–7) show the second round of recovery triggered by a collision with the cup, where the robot removes the cup and successfully completes the task by moving the apple to the box.}
    \label{fig:grasp_s2}
\end{figure}
\subsubsection{Setting 2}
In this scenario, a red apple (represented by a red circle) is placed at position $[-1.5, -0.2]$, a green box (represented by a green square) is placed at position $[-1.35, -0.75]$, and a chair (represented by a black circle) at position $[-1.0, 0]$. A 2-DoF robot arm (depicted in blue) is also present in the playground. We consider the user input for task objectives as \textit{"move the apple to the box carefully"}.

After receiving the task instruction, the motion descriptor, the motion descriptor $\mathtt{LLM_{riskhandling}}$ helps describe the task by giving:
\begin{lstlisting}
[start of plan]
To perform this task, the manipulator's palm should move close to apple.
object1=apple should be close to object2=box.
robot must keep the speed below 0.2m/s for a safety reason.
robot should release the object1=apple in the end
[end of plan]
\end{lstlisting}
Following the motion description, the $\mathtt{LLM_{coder}}$ generates the corresponding code block for those 2 subtasks: \textit{"move palm to the apple"} and \textit{"move apple to the box"} and also assign a safe speed constraint:
\begin{lstlisting}
```python
import numpy as np

reset_reward() # This is a new task so reset reward; otherwise we don't need it
set_l2_distance_reward("palm", "apple")
set_l2_distance_reward("apple", "box")
safe_speed_constraint(0.2)

rewards = execute_plan(4)
release_object("apple")
```
\end{lstlisting}
Upon execution, the system returns the following feedback via the $\mathtt{FeedbackTemplate}$:
\begin{lstlisting}
Task objective: move the apple to the box carefully
Feedback: 
- The current position of the robot arm's end-effector is: [-1.33147295 -0.70848672]
- The current position of the robot arm's end-effector is: [-1.33147295 -0.70848672]
- The apple's initial position is [-1.5 -0.2] and its current position of  is [-1.33 -0.71]
- The apple's initial position is [-1.5 -0.2] and its current position of  is [-1.33 -0.71]
- The box's initial position is [-1.35 -0.75] and its current position of  is [-1.35 -0.75]
- The pushing force between the robot arm and cup is 5.00000022796337N
- The cup's initial position is [-1.  0.] and its current position of  is [-1.  0.]
Warning: The robot arm collided with the cup (contact force: 5.00000022796337N)
\end{lstlisting}
After collecting the feedback in natural language, the $\mathtt{LLM_{correction}}$ module evaluates the task completion status and provides the corresponding correction command:
\begin{lstlisting}
Task fulfillment: No  
Cause: Collision with the cup blocked the robot's movement.
The robot was tasked to move the apple to the box carefully, but ["Collision with the cup blocked the robot's movement."]. What the robot arm should do to overcome this situation?
Solution: relocate the cup to a different position
\end{lstlisting}
As the task is not fulfilled, $\mathtt{LLM_{correction}}$ module proposed a solution as \textit{"relocate the cup to a different position"}. This solution is then fed back into the beginning of the LLM module chain-of-thought, initiating a second round of planning and reasoning. Below is an example of the updated task instruction with the correction applied. The updated instruction is constructed using a predefined template:
\begin{lstlisting}
Round 2 - : The overall task is: move the apple to the box carefully
However, the robot failed due to an issue: ["Collision with the cup blocked the robot's movement."]
To overcome this, follow this intermediate step: ['relocate the cup to a different position']
After that, continue to complete the original task.
\end{lstlisting}
Then the motion descriptor $\mathtt{LLM_{riskhandling}}$ and coder $\mathtt{LLM_{coder}}$ then proceeds to the second round of analysis based on the updated task instruction:
\begin{lstlisting}
[start of plan]
To perform this task, the manipulator's palm should move close to cup.
object1=cup should be close to object2=target_position.
target_position is set as [-1, 1] to relocate the object safely.  
robot should release the object1=cup in the end

To perform this task, the manipulator's palm should move close to apple.
object1=apple should be close to object2=box.
robot must keep the speed below 0.2m/s for a safety reason.
robot should release the object1=apple in the end
[end of plan]
\end{lstlisting}

\begin{lstlisting}
```python
import numpy as np

reset_reward() # This is a new task so reset reward; otherwise we don't need it
set_l2_distance_reward("palm", "cup")
set_target_position([-1, 1])
set_l2_distance_reward("cup", "target_position")

rewards_1 = execute_plan(4)
release_object("cup")

reset_reward() # This is a new task so reset reward; otherwise we don't need it
set_l2_distance_reward("palm", "apple")
set_l2_distance_reward("apple", "box")
safe_speed_constraint(0.2)

rewards_2 = execute_plan(4)
release_object("apple")

rewards = rewards_1 + rewards_2
```
\end{lstlisting}
Upon execution, the system returns the second round feedback via the FeedbackTemplate, and then call the $\mathtt{LLM_{correction}}$ module to evaluate the task completion status:
\begin{lstlisting}
Task objective: move the apple to the box carefully
Feedback: 
- The current position of the robot arm's end-effector is: [-1.35 -0.75]
- The current position of the robot arm's end-effector is: [-1.35 -0.75]
- The apple's initial position is [-1.5 -0.2] and its current position of  is [-1.35 -0.75]
- The apple's initial position is [-1.5 -0.2] and its current position of  is [-1.35 -0.75]
- The box's initial position is [-1.35 -0.75] and its current position of  is [-1.35 -0.75]
- The box's initial position is [-1.35 -0.75] and its current position of  is [-1.35 -0.75]
- The cup's initial position is [-1.  0.] and its current position of  is [-1.  1.]
\end{lstlisting}

\begin{lstlisting}
Task fulfillment: Yes
Task completed!
\end{lstlisting}
The full execution of all the subtasks is illustrated in Fig.~\ref{fig:grasp_s2}.

\begin{figure}[!ht]
    \centering
    \includegraphics[width=0.8\linewidth]{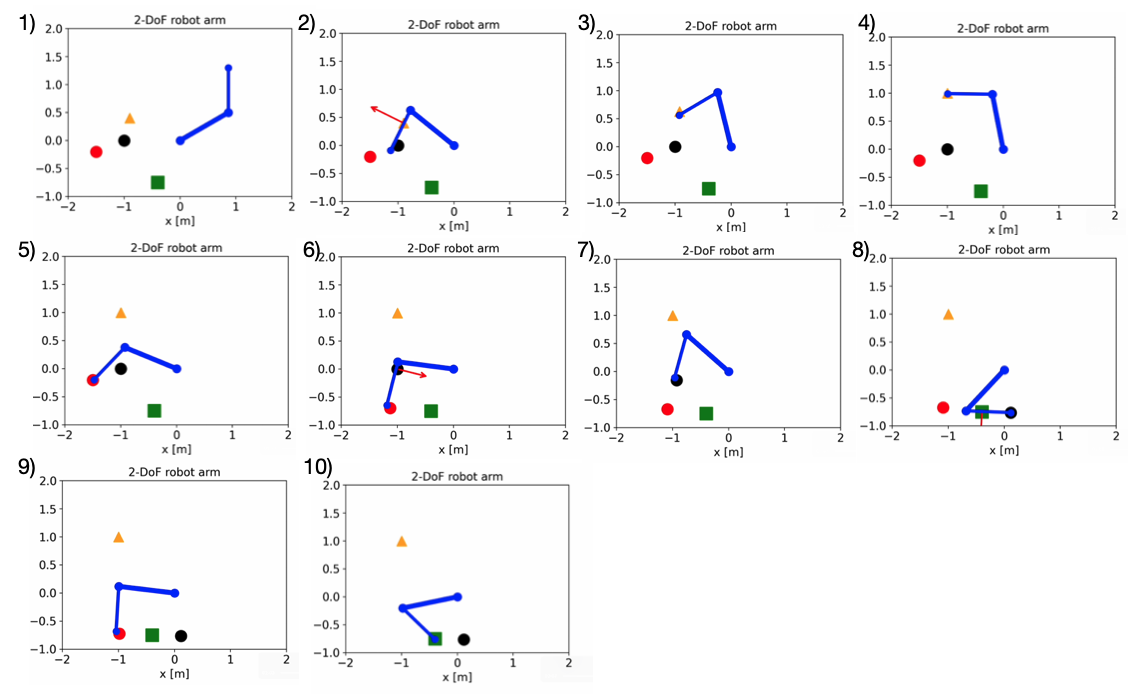}
    \caption{Setting 3: Task execution from beginning to end, illustrated over subfigures 1)–10). Subfigures 1)–2) depict the first-round attempt based on the initial instruction. Subfigures 3)–6) show the second round of recovery triggered by a collision with the cup, where the proposed solution involves removing the cup. Subfigures 7)–10) illustrate the third round of recovery after a collision with the obstacle, where the robot removes the obstacle and successfully completes the task by moving the apple to the box.}
    \label{fig:grasp_s3}
\end{figure}
\subsubsection{Setting 3}
In this scenario, a red apple (represented by a red circle) is placed at position $[-1.5, -0.2]$, a green box (represented by a green square) is located at $[-0.4, -0.75]$, a chair (represented by a black circle) is positioned at $[-1.0, 0]$, and an obstacle (represented by a yellow cone) is placed at $[-0.9, 0.4]$. A 2-DoF robot arm (depicted in blue) is also present in the playground. The user provides the following task instruction: \textit{“move the apple to the box carefully.”}

After receiving the task instruction, the motion descriptor $\mathtt{LLM_{riskhandling}}$ helps describe the task, while $\mathtt{LLM_{coder}}$ generates the executable functions along with the corresponding variables:
\begin{lstlisting}
[start of plan]
To perform this task, the manipulator's palm should move close to apple.
object1=apple should be close to object2=box.
robot must keep the speed below 0.2m/s for a safety reason.
robot should release the object1=apple in the end
[end of plan]
\end{lstlisting}

\begin{lstlisting}
```python
import numpy as np

reset_reward() # This is a new task so reset reward; otherwise we don't need it
set_l2_distance_reward("palm", "apple")
set_l2_distance_reward("apple", "box")
safe_speed_constraint(0.2)

rewards = execute_plan(4)
release_object("apple")
```
\end{lstlisting}
Upon execution, the system returns the following feedback via the $\mathtt{FeedbackTemplate}$:
\begin{lstlisting}
Task objective: move the apple to the box carefully
Feedback: 
- The current position of the robot arm's end-effector is: [-1.13025824 -0.14490971]
- The apple's initial position is [-1.5 -0.2] and its current position of  is [-1.5 -0.2]
- The box's initial position is [-0.4  -0.75] and its current position of  is [-0.4  -0.75]
- The cup's initial position is [-1.  0.] and its current position of  is [-1.  0.]
- The pushing force between the robot arm and cup is 2.1111703898725N
Warning: The robot arm collided with the cup (contact force: 2.1111703898725N)
- The obstacle's initial position is [-0.9  0.4] and its current position of  is [-0.9  0.4]
- The pushing force between the robot arm and obstacle is 4.9999997601877855N
Warning: The robot arm collided with the obstacle (contact force: 4.9999997601877855N)
\end{lstlisting}
After collecting the feedback in natural language, the $\mathtt{LLM_{correction}}$ module evaluates the task completion status and provides the corresponding correction command:
\begin{lstlisting}
Task fulfillment: No  
Cause: Collision with the obstacle blocked the robot's motion.
The robot was tasked to move the apple to the box carefully, but ["Collision with the obstacle blocked the robot's motion."]. What the robot arm should do to overcome this situation?
Solution: move the obstacle away to free path
\end{lstlisting}
As the task is not fulfilled, $\mathtt{LLM_{correction}}$ module proposed a solution as \textit{"move the obstacle away to free path"}. This solution is then fed back into the beginning of the LLM module chain-of-thought, initiating a second round of planning and reasoning. Below is an example of the updated task instruction with the correction applied. The updated instruction is constructed using a predefined template:
\begin{lstlisting}
Round 2 - : The overall task is: move the apple to the box carefully
However, the robot failed due to an issue: ["Collision with the obstacle blocked the robot's motion."]
To overcome this, follow this intermediate step: ['move the obstacle away to free path']
After that, continue to complete the original task.
\end{lstlisting}
Then the motion descriptor $\mathtt{LLM_{riskhandling}}$ and coder $\mathtt{LLM_{coder}}$ then proceeds to the second round of analysis based on the updated task instruction:
\begin{lstlisting}
[start of plan]
To perform this task, the manipulator's palm should move close to obstacle.
object1=obstacle should be close to object2=target_position.
target_position is set as [-1, 1] to relocate the obstacle safely.  

To perform this task, the manipulator's palm should move close to apple.
object1=apple should be close to object2=box.
robot must keep the speed below 0.2m/s for a safety reason.
robot should release the object1=apple in the end
[end of plan]
\end{lstlisting}

\begin{lstlisting}
```python
import numpy as np

reset_reward() # This is a new task so reset reward; otherwise we don't need it
set_target_position([-1, 1])
set_l2_distance_reward("palm", "obstacle")
set_l2_distance_reward("obstacle", "target_position")

rewards_1 = execute_plan(4)

reset_reward() # This is a new task so reset reward; otherwise we don't need it
set_l2_distance_reward("palm", "apple")
set_l2_distance_reward("apple", "box")
safe_speed_constraint(0.2)

rewards_2 = execute_plan(4)
release_object("apple")

rewards = rewards_1 + rewards_2
```
\end{lstlisting}
This code instructs the system to move the obstacle to a different location first, with the target position also suggested by the LLM. Upon execution, the system returns the second round feedback via the FeedbackTemplate, and then call the $\mathtt{LLM_{correction}}$ module to evaluate the task completion status:
\begin{lstlisting}
Task objective: move the apple to the box carefully
Feedback: 
- The current position of the robot arm's end-effector is: [-1.09298125 -0.67117154]
- The apple's initial position is [-1.5 -0.2] and its current position of  is [-1.09 -0.67]
- The box's initial position is [-0.4  -0.75] and its current position of  is [-0.4  -0.75]
- The cup's initial position is [-1.  0.] and its current position of  is [-1.  0.]
- The pushing force between the robot arm and cup is 4.9999999137492726N
Warning: The robot arm collided with the cup (contact force: 4.9999999137492726N)
- The obstacle's initial position is [-0.9  0.4] and its current position of  is [-1.  1.]
\end{lstlisting}
\begin{lstlisting}
Task fulfillment: No  
Cause: Collision with the cup hindered the robot's operation.
The robot was tasked to move the apple to the box carefully, but ["Collision with the cup hindered the robot's operation."]. What the robot arm should do to overcome this situation?
Solution: lift the cup and place it aside to clear the path.
\end{lstlisting}
As the task is still not fulfilled, $\mathtt{LLM_{correction}}$ module proposed a solution as \textit{"lift the cup and place it aside to clear the path"}. This solution is then fed back into the beginning of the LLM module chain-of-thought, initiating a third round of planning and reasoning. Below is an example of the updated task instruction with the correction applied. The updated instruction is constructed using a predefined template:
\begin{lstlisting}
Round 3 - : The overall task is: move the apple to the box carefully
However, the robot failed due to an issue: ["Collision with the cup hindered the robot's operation."]
To overcome this, follow this intermediate step: ['lift the cup and place it aside to clear the path.']
After that, continue to complete the original task.
\end{lstlisting}
Then the motion descriptor $\mathtt{LLM_{riskhandling}}$ and coder $\mathtt{LLM_{coder}}$ then proceeds to the third round of analysis based on the updated task instruction:
\begin{lstlisting}
Motion Descriptor:
[start of plan]
To perform this task, the manipulator's palm should move close to cup.
object1=cup should be close to object2=target_position.
target_position is set as [2, -2] to relocate the cup safely.  

To perform this task, the manipulator's palm should move close to apple.
object1=apple should be close to object2=box.
robot must keep the speed below 0.2m/s for a safety reason.
robot should release the object1=apple in the end
[end of plan]
\end{lstlisting}

\begin{lstlisting}
```python
import numpy as np

reset_reward() # This is a new task so reset reward; otherwise we don't need it
set_target_position([2, -2])
set_l2_distance_reward("palm", "cup")
set_l2_distance_reward("cup", "target_position")

rewards_1 = execute_plan(4)

reset_reward() # This is a new task so reset reward; otherwise we don't need it
set_l2_distance_reward("palm", "apple")
set_l2_distance_reward("apple", "box")
safe_speed_constraint(0.2)

rewards_2 = execute_plan(4)
release_object("apple")

rewards = rewards_1 + rewards_2
```
\end{lstlisting}
Upon execution, the system returns the third round feedback via the FeedbackTemplate, and then call the $\mathtt{LLM_{correction}}$ module to evaluate the task completion status:
\begin{lstlisting}
Task objective: move the apple to the box carefully
Feedback: 
- The current position of the robot arm's end-effector is: [-0.4  -0.75]
- The apple's initial position is [-1.5 -0.2] and its current position of  is [-0.4  -0.75]
- The box's initial position is [-0.4  -0.75] and its current position of  is [-0.4  -0.75]
- The cup's initial position is [-1.  0.] and its current position of  is [ 0.12 -0.76]
- The obstacle's initial position is [-0.9  0.4] and its current position of  is [-1.  1.]
\end{lstlisting}

\begin{lstlisting}
Task fulfillment: Yes
Task completed!
\end{lstlisting}

The full execution of all the subtasks is illustrated in Fig.~\ref{fig:grasp_s3}.

\subsection{Offline Showcase}\label{sec:grasp_results_offline}

Offline results for \textit{Setting 2} and \textit{Setting 3} are presented, with the environment being reset at the start of each turn. Each line of figures corresponds to one turn of interaction.

\begin{figure}[!ht]
    \centering
    \includegraphics[width=0.7\linewidth]{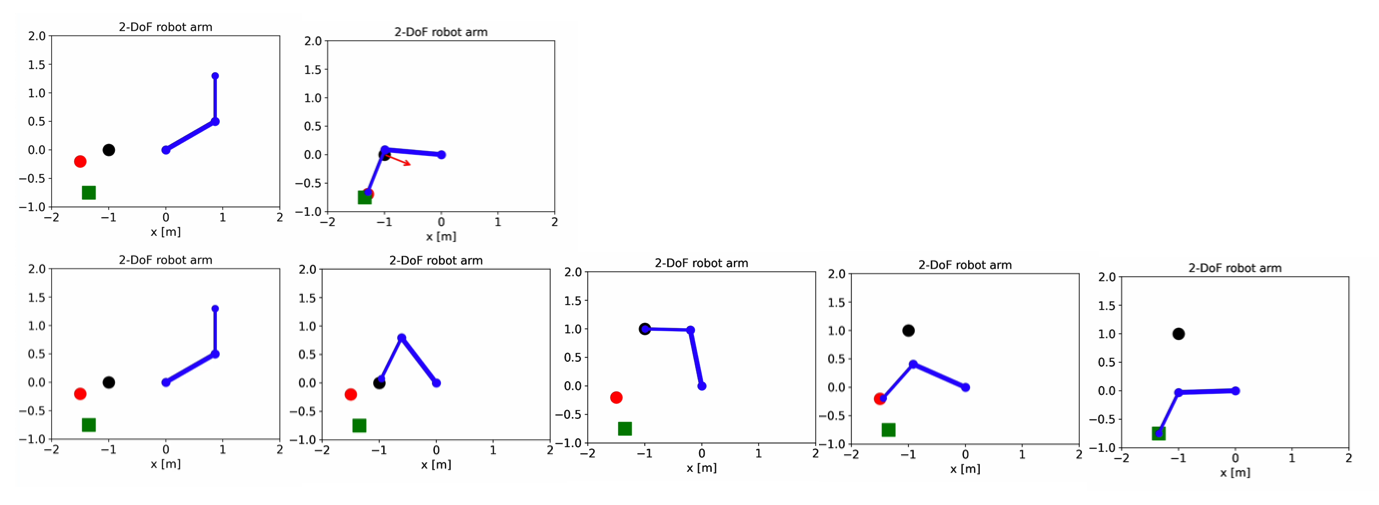}
    \caption{Setting 2 for offline framework.}
    \label{fig:grasp_offline2}
\end{figure}

\begin{figure}[!ht]
    \centering
    \includegraphics[width=0.9\linewidth]{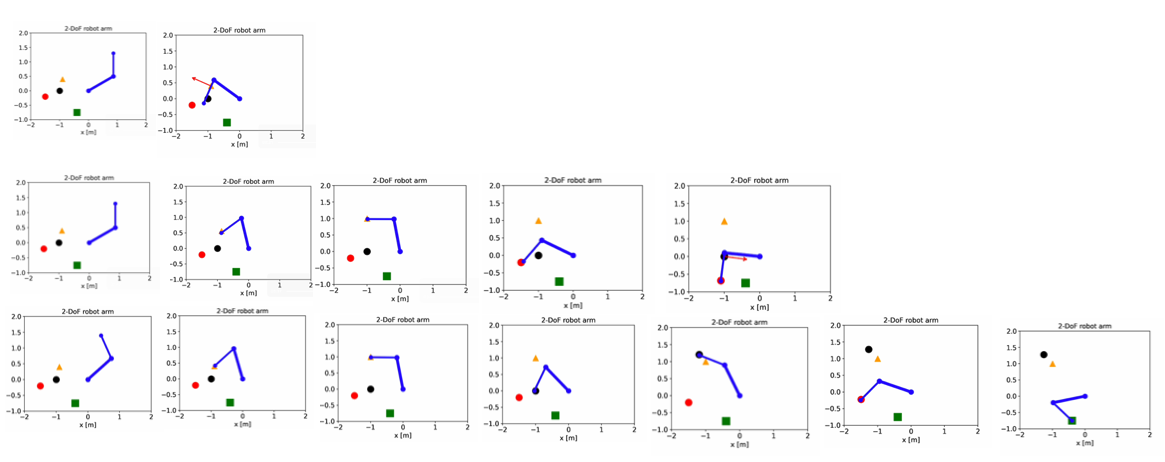}
    \caption{Setting 3 for offline framework.}
    \label{fig:grasp_offline3}
\end{figure}

\subsection{LLM prompts}\label{sec:grasp_llmprompt}
In this simulation, we incorporate human input in the form of task instructions. A descriptor module with prompt $\mathcal{P}_{\mathtt{riskhandling}}$ maps these natural language commands into corresponding language-based reward signals and constraint suggestions $s$. Here is the example for $\mathcal{P}_{\mathtt{riskhandling}}$. 
\begin{lstlisting}
We have a (mobile) manipulator and we want you to help plan how it should move to perform tasks using the following template. You must also obey the following Rules:
You are supposed to create a motion plan that is safe and should not damage or hurt external objects (like humans) or environments

Rules:
1. You should strictly follow the template, do not invent unnessary commands
2. If you see phrases like {NUM: default_value}, replace the entire phrase with a numerical value. The numerical value must be larger than zero.
2.1 The value for the safe distance between the robot and target object can be any resonable number, but the minimum one is 0.3m.
2.2 The value for the safe contact force between the robot and target object can be any resonable number.
2.3 The value for target position {NUM: [x, y]} can be any resonable number in the frame to avoid further collision, therefore should be a far away position.
3. If you see phrases like {CHOICE: choice1, choice2, ...}, it means you should replace the entire phrase with one of the choices listed.
4. If you see [optional], it means you only add that line if necessary for the task, otherwise remove that line.
5. If you see phrases like [obstacle_A] and [obstacle_B], replace the entire phrase with a resonable object mentioned in the user's commands or instructions and must be specific.
6. If the robot are tasked to manipulate an [object], do NOT create safe distance constraint associated with such the [object]
7. You can resonably create as many constraints as possible
8. I will tell you a behavior/skill/task that I want the manipulator to perform and you will provide the full plan, even if you may only need to change a few lines. Always start the description with [start of plan] and end it with [end of plan].
9. You can assume that the robot is capable of doing anything, even for the most challenging task.\
10. If `target_position` is used in the plan, make sure to define its value explicitly using the line:  
    `target_position is set as {NUM: [x, y]}`  
    Replace the value with a 2D coordinate. This line must appear before any reward using `target_position`.

[start of plan]
To perform this task, the manipulator's {CHOICE: palm} should move close to {CHOICE: apple, banana, box, bowl, drawer_handle, faucet_handle, drawer_center}.
object1={CHOICE: apple, banana, box, bowl, drawer_handle, faucet_handle, drawer_center} should be close to object2={CHOICE: apple, banana, box, bowl, target_position}.
[optional] safe distance between the manipulator and {CHOICE: glass, bottle, human, obstacle} must be larger than {NUM: NULL}
[optional] safe contact force between the manipulator {CHOICE: glass, bottle, human, obstacle} must be below {NUM: NULL}
[optional] robot must keep the speed below {NUM: 0.0}  for a safety reason
[optional] define the target_position as {NUM: [x, y]} to avoid potential collisions.
robot should release the object in the end
[end of plan]

Query:
move apple to the box, assure safe distances between manipulator and humans, while allowing light touches with nearby cups
Answer:
[start of plan]
To perform this task, the manipulator's palm should move close to apple.
object1=apple should be close to object2=box.
safe distance between the manipulator and obstacle=human must be larger than 0.3m
safe contact force between the manipulator and obstacle=cup must be below 1.0N
robot must keep the speed below 0.2m/s for a safety reason.
robot should release the object1=apple in the end
[end of plan]

Query:
Remove the [object] or Move the [object] away [something].
Answer:
[start of plan]
To perform this task, the manipulator's palm should move close to [object].
object1=[object] should be close to object2=target_position.
target_position is set as [-1, 1] to relocate the object safely.  
[end of plan]

Query:
Touch the [object]
Answer:
[start of plan]
To perform this task, the manipulator's palm should move close to [object].
safe contact force between the manipulator and obstacle=[object] must be below {NUM: 0.0}N
safe distance between the manipulator and obstacle=[object] must be larger than {NUM: 0.0}N
[end of plan]

Query:
Grasp the [object]
Answer:
[start of plan]
To perform this task, the manipulator's palm should move close to [object].
safe contact force between the manipulator and obstacle=[object] must be below {NUM: 0.0}N
safe distance between the manipulator and obstacle=[object] must be larger than {NUM: 0.0}N
[end of plan]
\end{lstlisting}

\noindent Below is the prompt $\mathcal{P}_{\mathtt{coder}}$ used to instruct the LLM to connect the language suggestions with executable codes.

\begin{lstlisting}
We have a plan of a robot arm with palm to manipulate objects and we want you to turn that into 
the corresponding program with following functions:

Rules:
1. Always format the code in code blocks. Each block should start with reset_reward. You may call execute_plan multiple times if the plan contains multiple distinct subtasks. Each call simulates one phase of the plan.
2. Do not invent new functions or classes. The only allowed functions you can call are the ones listed below. Do not leave unimplemented code blocks in your response.
3. The only allowed library is numpy. Do not import or use any other library.
4. If you are not sure what value to use, just use your best judge. Do not use None for anything.
5. Do not calculate the position or direction of any object (except for the ones provided above). Just use a number directly based on your best guess.
6. You do not need to make the robot do extra things not mentioned in the plan such as stopping the robot.

```
def safe_distance_constraint(name_obj, safe_distance=0.3)
```
this constraint is to assure the distance between the robot and name_obj must not below "safe_distance"
name_obj (e.g., human, glass, vase, obstacle) could be anything depneding on context of the provided motion plan

```
def safe_contact_force_constraint(name_obj, safe_force=10.0)
```
the constraint to assure that the contact force between the robot and name_obj must below "safe_force"
name_obj (e.g., human, glass, vase, obstacle) could be anything depneding on context of the provided motion plan

```
def safe_speed_constraint(max_speed=0.1)
```
set the constraint on the robot speed. The robot velocity must not exceed the "max_speed"
default "max_speed"=0.1, but it can be modified according to distances between robots and external objects

```
def set_target_position(target_position=[-0.5, 0.2])
```
set the target 2D coordinate on the plane used by reward functions (e.g., for placing or moving objects).
default target_position = [-0.5, 0.2], but it can be modified depending on the task context.

```
def set_l2_distance_reward(name_obj_A, name_obj_B)
```
where name_obj_A and name_obj_B are selected from ['palm', 'arm', 'apple', 'banana', 'box', 'bowl', 'drawer_handle', 'faucet_handle', 'drawer_center', 'rest_position', 'target_position'].
This term sets a reward for minimizing l2_distance between name_obj_A and name_obj_B so they get closer to each other.
rest_position is the default position for the palm when it's holding in the air.
target_position is the position for the object to be puhsed or shifted on the x-y 2D plane.
 


```
def query_state(name_obj):
```
this function returns the state = (position, velocity) of name_obj, including a 2D position and its velocity
e.g: (position, velocity) = query_state("apple")

```
def release_object(name_obj)
```
This function is used to release a currently grasped object. When the robot is holding an object (e.g., after moving it to a target location), call this function to let go of it.
For example: release_object("apple") will release the apple from the robot's grasp.

```
def execute_plan(duration=4) -> (rewards, forces)
```
This function sends the parameters to the robot and execute the plan for `duration` seconds, default to be 2. The function should return rewards

```
def reset_reward()
```
This function resets the reward to default values.


Example plan:
To perform this task, the manipulator's palm should move close to object1=apple.
object1 should be close to object2=bowl.
object2 needs to be rotated by 30 degrees along x axis.
object2 needs to be lifted to a height of 1.0.

This is the first plan for a new task.

Example answer code:
"""
import numpy as np

reset_reward() # This is a new task so reset reward; otherwise we don't need it
set_l2_distance_reward("palm", "apple")
set_l2_distance_reward("apple", "bowl")
set_obj_orientation_reward("bowl", np.deg2rad(30))
set_obj_z_position_reward("bowl", 1.0)

rewards = execute_plan(3)
release_object("apple")
release_object("bowl")
"""

Example plan:
To perform this task, the manipulator's arm/palm should move away from the obstacle.
safe distance between the manipulator and obstacle must be larger than 0.5m
robot must keep the speed below 0.2m/s for a safety reason.

Example answer code:
"""
reset_reward() # This is a new task so reset reward; otherwise we don't need it
set_target_position([-0.1, 1])
set_l2_distance_reward("palm", "target_position")
safe_distance_constraint("obstacle", 0.5)
safe_speed_constraint(0.2)

rewards = execute_plan(4)
release_object("obstacle")
"""

Example plan:
To perform this task, the manipulator's arm should move close to box.
safe contact force between the manipulator and obstacle=box must be below 4N
safe distance between the manipulator and obstacle=box must be larger than 0.1m
robot must keep the speed below 0.5m/s for a safety reason.

Example answer code:
"""
import numpy as np

reset_reward()
set_l2_distance_reward("arm", "box")
safe_contact_force_constraint("box", 4.0)
safe_distance_constraint("box", 0.1)
safe_speed_constraint(0.5)

rewards = execute_plan(3)
"""

Example plan:
To perform this task, the robot should first move the cup away to clear a path, and then move the apple to the box.

Example answer code:
"""
import numpy as np

reset_reward() # This is a new task so reset reward; otherwise we don't need it
set_target_position([-0.1, 1])
set_l2_distance_reward("palm", "cup")
set_l2_distance_reward("cup", "target_position")
safe_distance_constraint("apple", 0.3)
rewards_1 = execute_plan(2)

reset_reward() # This is a new task so reset reward; otherwise we don't need it
set_l2_distance_reward("palm", "apple")
set_l2_distance_reward("apple", "box")
safe_distance_constraint("cup", 0.3)
rewards_2 = execute_plan(4)

rewards = rewards_1 + rewards_2
"""
\end{lstlisting}

When a failure is detected during simulation, we initiate a two-step correction procedure guided by chain-of-thought reasoning that incorporates image feedback. In this process, the current state and failure context—along with corresponding visual observations—are passed to a reasoning module with image understanding capabilities. This module first analyzes the root cause of the failure and then proposes corrective actions to realign the system with the intended task.
Below, we present the prompts used for the correction prompt module, denoted as $\mathcal{P}_{\mathtt{correction}}$:

\begin{lstlisting}
You are an assistant designed to reason about the most likely cause for a robot's task failure.
The robot task objective and the robot's feedback about the environment state are provided as input. Your role is to analyze the feedback and determine whether the task was fulfilled. If the task failed, identify the single most likely reason for the failure.

The feedback includes:
- The current position of the robot's end-effector
- Pushing forces between the robot arm and objects due to collision
- The initial and current positions of objects (e.g., apple, box)
- Grasping status inferred from robot-object proximity and force
- Warnings indicate critical events like collisions or blocked paths. Prioritize them when analyzing failure causes.


Rules:
1. Decide whether the robot has fulfilled the task or not, based on the textual feedback.
2. If task failed, please answer the most likely failure cause. Do not repeat the task objective as the cause. Focus on physical constraints and warning information. Answer a single cause only.
3. Try to make your answer as concise as possible. Limits your answer below 15 words for the cause. 
4. If the answer to the Task fulfillment: Yes, you don't need to provide the cause.

Examples:
Query: 
Answer:
Task fulfillment: Yes or No
Cause: the obstacle blocks the robot motion.
\end{lstlisting}

\begin{lstlisting}
You would be an assistant to provide solutions to a cause or situation that impede a robot to finish its task

The solution you provided would be for a 2DoF robot arm tasked to do manipulation task and handling objects
The failure cause and an image of a current environment will be given as input.

The provided image represents a simulation environment of 2-Degree of Freedom (DoF) robot arm with the following graphical representations:
- The 2-DoF robot arm is depicted by two blue lines connected to each other, simulating the arm's segments and joints. 
- The red-filled circle represents an apple.
- The green-filled square represents a box.
- The black-filled circle represents a cup.
- The red arrow indicates the contact between the robot and respective object in the scene. 

Rules:
1. Please answer the most (only one) likely solution or action. Answer a single solution only.
2. You should repharse your solutions as if you are giving instructions to the robot so that your answer can be used to control the robot in a closed loop.
3. You don't need to explain for your solutions, make it as a command and as concise as possible. Limits your answer below 15 words for the solution.
4. Do not provide general solutions, the solution should be relate directly to low-level control action and robot motion.
5. Do not provide solutions associated with collision avoidance.
6. Do not provide solutions associated with retour the robot path or change its path, there is no other ways it can go.
7. Do not provide solutions associated with the robot motion itself. 


Examples:
Query: 
The robot was tasked to move apple to the box, but the obstacle blocks the robot motion. What the robot arm should do to overcome this situation?
Answer:
Solution: move the obstacle away to free path
\end{lstlisting}

We also define two templates: one that converts the system states into natural language feedback, and another that wraps the proposed solution and integrates it with the task objective to guide the next round of operation. Below are examples of each template, respectively:

\begin{lstlisting}
def generate_state_feedback_text(td, controller):
    """
    @Parameters:
    - td: task description
    - controller: controller object to get the states of environment
    """

    feedback_list = ["Feedback: "]

    robot_pos = controller.get_predict_eff_pos()
    # generate feedback data for robot arm
    text = "- The current position of the robot arm's end-effector is: {}".format(robot_pos)
    feedback_list.append(text)
    grasped_object = None
    force_threshold = 0.5
    contact_threshold = 1e-3
    mass_threshold = 50


    objects_list = controller.objects_list
    for obj_name in objects_list:
        obj_attribute = getattr(controller, obj_name)
        init_pos = np.round(obj_attribute["center"], 2)
        state = np.round(obj_attribute["state"], 2)
        current_pos = np.round(state[:2], 2)
        current_vel = np.round(state[2:], 2)
        current_speed = np.round(np.linalg.norm(current_vel), 2)
        # print("obj_name: {}, grasp: {}".format(obj_name, obj_attribute["grasp"]))
        force_magnitude = np.linalg.norm(obj_attribute["force"])
        text = "- The {}'s initial position is {} and its current position of  is {}".format(obj_name, 
                                                                               init_pos, 
                                                                               current_pos)
        feedback_list.append(text)
        if obj_attribute["grasp"] and np.linalg.norm(robot_pos - current_pos) <= contact_threshold:
            text = "The robot has sucessfully grasped the {}".format(obj_name)
            feedback_list.append(text)
            grasped_object = obj_name
        
        if obj_attribute["grasp"] and obj_attribute["mass"] > mass_threshold:
            feedback_list.append(
                f"Warning: The robot has grasped the {obj_name}, but it may be too heavy to move."
            )

        if force_magnitude != 0:
            text = "- The pushing force between the robot arm and {object} is {force}N".format(
                object=obj_name, 
                force=force_magnitude
            )
            feedback_list.append(text)
        
        if obj_name != grasped_object and force_magnitude >= force_threshold:
            feedback_list.append(f"Warning: The robot arm collided with the {obj_name} (contact force: {force_magnitude}N)")

        
        # text = "- The current {} speed is {}".format(obj_name, current_speed)
        # feedback_list.append(text)

    if hasattr(controller, "target_position"):
        target_state = controller.target_position.get("state", None)
        if target_state is not None:
            target_pos = np.round(target_state[:2], 2)
            text = "- The target position is set at {}, if the collision still happened, try to set another position".format(target_pos)
            feedback_list.append(text)

    td_text = "Task objective: {}".format(td)
    constraint_text = """
    Environment constraints:
    - safe distance between the manipulator and obstacle
    - safe contact force between the manipulator and obstacle
    """
    fb_state_text = '\n'.join(feedback_list)
    question_text = "Based on the feedback, has the robot already fulfilled the task? If it hasn't, what are the two most factors or causes preventing the robot task to complete?"    
    query_text = td_text + "\n" + fb_state_text + "\n" + question_text
    # query_text = td_text + "\n" + constraint_text + "\n" + fb_state_text + "\n" + question_text

    return query_text
\end{lstlisting}

\begin{lstlisting}
if solution_ouput:
        user_like_input = (
            f"The overall task is: {user_cmd}\n"
            f"However, the robot failed due to an issue: {cause_output}\n"
            f"To overcome this, follow this intermediate step: {solution_ouput}\n"
            f"After that, continue to complete the original task.\n"
        )
\end{lstlisting}

\section{Extended Simulation Results: AV Navigation}\label{sec:av_simulation}
As discussed in Section~\ref{sec:latentrisk}, we consider an AV operating in 10 distinct scenarios that involve latent risks. We assess its performance under uncertainty by varying the probability that a latent risk becomes an apparent (observable) risk.  Example video data from the driver's perspective is sourced from the PSI dataset~\cite{chen2021psi}, and youtube videos~\cite{Newsflare2022, newyork, pittsburgh, chongqing}.
\subsection{LLM prompts}\label{sec:av_llmprompt}
\noindent Below is the prompt $\mathcal{P}_{\mathtt{vision}}$ used to instruct the LLM to analyze the scene:
\begin{lstlisting}
From the driver's perspective, describe what happens in the video. Focus on vulnerable pedestrians or other latent risk-citic objects that could affect safe driving.
\end{lstlisting}
\noindent Below is the prompt $\mathcal{P}_{\mathtt{latentobject}}$  was used to instruct the LLM to analyze the latent risks and their likelihood to become apparent risk:
\begin{lstlisting}
In this task, a user will provide you a description of images captured from driver view of an grounded autonomous vehicle.
You are responsible to provide insights into things or latent risks that the autonomous car should be catious about when driving down the road.

Rules:
1. Please provide your answers by bullet points for each latent risk or object that the car should be catious about.
2. For each bullet point, there should be a pair of {keyword} and {description}, separted by a colon, for example {keyword}: {description}
3. The {keyword} should name the object that the driver should be careful about and {action} should tell what driver could do to avoid or mitigate the risk.
4. The keyword should be limited to the list of following objects of interest: [Pedestrians, Other Vehicles, Children, Teenagers, Elderly, People With Disability, Bicycles, Motorcycles, Traffic Lights, Stop Signs, Yield Signs, Crosswalks, Road Signs, Lane Markings, Roadside Obstacles, Construction Zones, Traffic Cones, Speed Bumps, Parked Cars, Oncoming Traffic, Intersections, Railroad Crossings, Pedestrian Crossings, Emergency Vehicles, Driveways, Roundabouts, School Zones, Disabled Vehicles, Curb and Sidewalks, Cyclists, Street Lights]
5. If you can identify the specific objects such as Teenagers, try NOT use the general keywords such as Pedestrians in the list.
6. The description should start with the estimated probability that object would introduce apparent risk in the following very short time. The value of probability should be placed inside the square brackets, for example [probability].
7. The probability should be chosen from following categories: [low, middle, high].
8. Please assign specific objects such as Teenagers with higher probability than the general objects such as Pedestrains.
9. Your repsonse should be limited into the most three important things that the car should be cautious about.
10. The points should be listed in the order of decreasing probability.
11. The latent risk should be infer from the observed objects, for example, if the observed objects are School Zones, the latent risk object should be Childrens or Teenagers. An other example is the if the observed objects are Intersections or Stop Signs, the latent risk objects should be Pedestrian Crossings.
12. The moving objects should be assign higher probability than statics objects, for example, Children should be assigned higher probability than School Zones.
Examples:

Query: 
Description: 
Based on the description, what are latent risks an autonomous car should be cautious about when driving down this road? 

Answer:
1. {keyword}:
2. {keyword}:
3. {keyword}:

If you understand my requirements, say yes.
\end{lstlisting}

\noindent$\mathcal{P}_{\mathtt{riskhandling}}$ used to instruct the LLM to analyze the scene:
\begin{lstlisting}
In this task, a user will provide you an object name an grounded autonomous vehicle may encounter while driving.
You should reason and suggest what the low-level controller of the autonomous car should do to assure the safety of the provided object.

Rules:
1. Please provide your answers by bullet points for each action as concise and descriptive as possible.
2. Your response should be limited into the most two important actions that the car should be cautious about.
3. The points should be listed in the order of importance.
4. You must yield to pedestrians when they are: a. crossing at any intersection without a traffic light (with or without a crosswalk); b. crossing the roadway in marked crosswalks, whether or not at an intersection; c. walking on a sidewalk crossing a driveway or alley; and d. when the driver is turning a corner and pedestrians are crossing with the light. You must always yield to any blind pedestrian carrying a white cane or being led by a guide dog.
5. Take care of the disadvantaged pedestrians: [Children, Elderly, Teenagers, People With Disabilities]
6. Keep the right-of-way, do NOT slow down if the object doesn't introduce risk.

Examples:

Query: 
Object: childrens
And what the car should do in the next sequence of control actions to assure safety for the children and also maintain the right-of-way? 

Answer:
1. 
2. 

If you understand my requirements, say yes.
\end{lstlisting}

\noindent Below is the prompt $\mathcal{P}_{\mathtt{coder}}$  was used to instruct the LLM to provide the suggestions for the AV:
\begin{lstlisting}
We have a plan and contraint motion of a autonomous car and we want you to turn that into 
the corresponding program to assure safe driving with following functions:

Rules:
1. Always format the code in code blocks. In your response execute_plan should be called exactly once at the end.
2. Do not invent new functions or classes. The only allowed functions you can call are the ones listed below. Do not leave unimplemented code blocks in your response.
3. Do not invent your implementation for the functions by yourself.
4. Do not answer any text or exclamation except for the generated Python code block
4. The only allowed library is numpy. Do not import or use any other library.
5. If you are not sure what value to use, just use your best judge. Do not use None for anything.
6. You do not need to make the robot do extra things not mentioned in the plan such as stopping the robot.

```
def safe_distance_constraint(name_obj, safe_distance=0.3)
```
this constraint is to assure the distance between the car and name_obj must not below "safe_distance"
name_obj (e.g., children, bus, pedestrian) could be anything depneding on context of the provided motion plan

```
def safe_speed_constraint(max_speed=1)
```
set the constraint on the car speed. The car velocity must not exceed the "max_speed"
default "max_speed"=1, but it can be modified according to distances between robots and external objects

```
def execute_plan(duration=4) -> (rewards)
```
This function sends the parameters to the robot and execute the plan for `duration` seconds, default to be 2. The function should return rewards


Example plan:
[start of plan]
safe distance between the car and children must be larger than 0.3m
car must reduce the speed to below 16m/s for a safety reason
[end of plan]

Example answer code:
"""
import numpy as np

safe_distance_constraint("children", 0.8)
safe_speed_constraint(3.5)

rewards = execute_plan(3)
"""

If you understand, say Yes.
\end{lstlisting}
\subsection{Chain-of-Thought reasoning: scenarios and outcomes}\label{sec:av_scenarios}
\begin{figure}[H]
    \centering
    \includegraphics[width=0.3\linewidth]{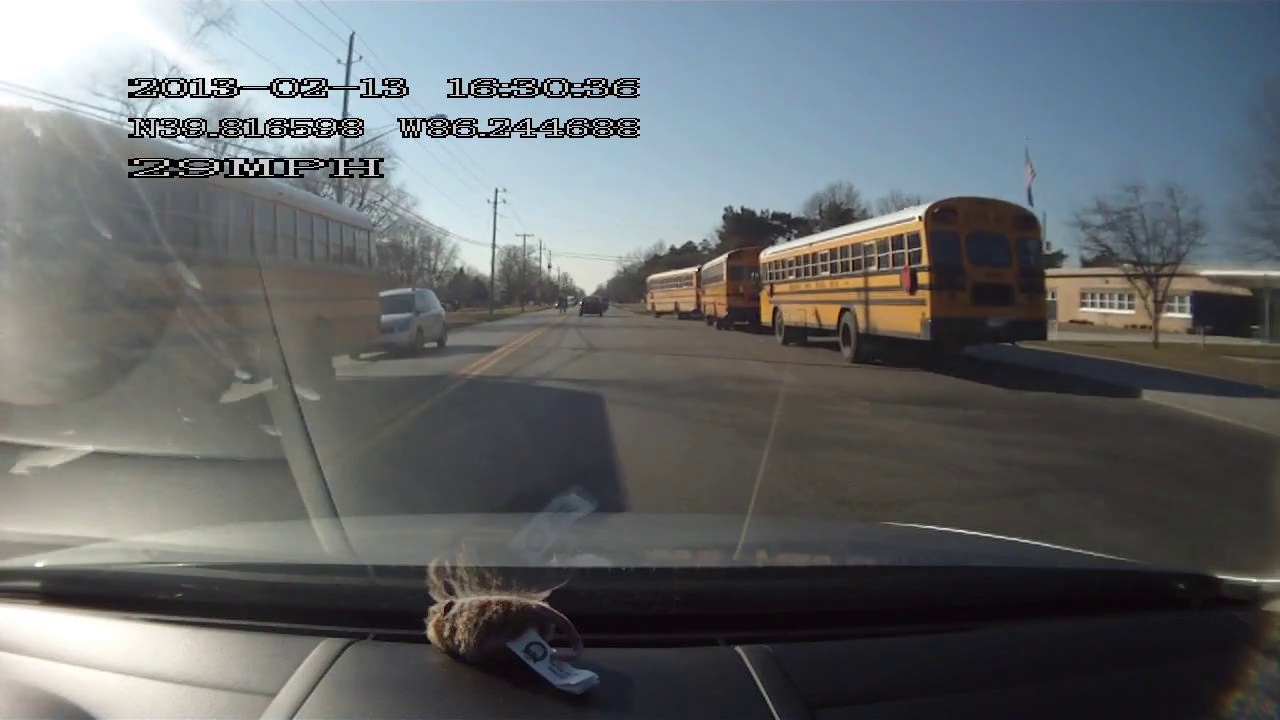}
    \caption{Facing a school bus.}
    \label{fig:f1_schoolbus}  
\end{figure}
\noindent$\mathtt{LLM_{vision}}:$
\begin{lstlisting}
From the driver's perspective, various potential hazards can be identified in the video. 

1. **School Buses**: Several yellow school buses are present on the side of the road, indicating that children may be nearby. This heightens the risk, as children are often unpredictable and may suddenly cross the street.

2. **Children Walking**: If there are children getting on or off the buses or walking near them, the driver must remain vigilant. Since visibility can be obstructed by the buses, there is a risk of not seeing a child who may step out unexpectedly.

3. **Other Vehicles**: A vehicle is seen parked on the side, which could pose a risk when it maneuvers back onto the road. The driver must be alert to the possibility of other cars moving in and out of traffic, especially in a school zone.
\end{lstlisting}
$\mathtt{LLM_{latentobjects}}:$
\begin{lstlisting}
1. {Children}: [high] The presence of school buses indicates that children may be nearby and could unexpectedly enter the roadway.

2. {Crosswalks}: [middle] The area likely has crosswalks where students may be crossing, requiring careful monitoring to ensure pedestrian safety.

3. {Other Vehicles}: [low] The behavior of other vehicles, such as sudden stops or lane changes without signaling, can pose risks and should be closely observed.
\end{lstlisting}
$\mathtt{LLM_{riskhandling}}:$
\begin{lstlisting}
Object: Children
1. Slow down and prepare to stop if children are detected near or on the roadway, especially in school zones or residential areas.
2. Yield to children crossing the road, ensuring they have safely crossed before proceeding.
\end{lstlisting}

$\mathtt{LLM_{coder}}:$
\begin{lstlisting}
```python
import numpy as np

safe_distance_constraint("children", 0.5)
safe_speed_constraint(3)

rewards = execute_plan(4)
```
\end{lstlisting}

\begin{figure}[H]
    \centering
    \includegraphics[width=0.3\linewidth]{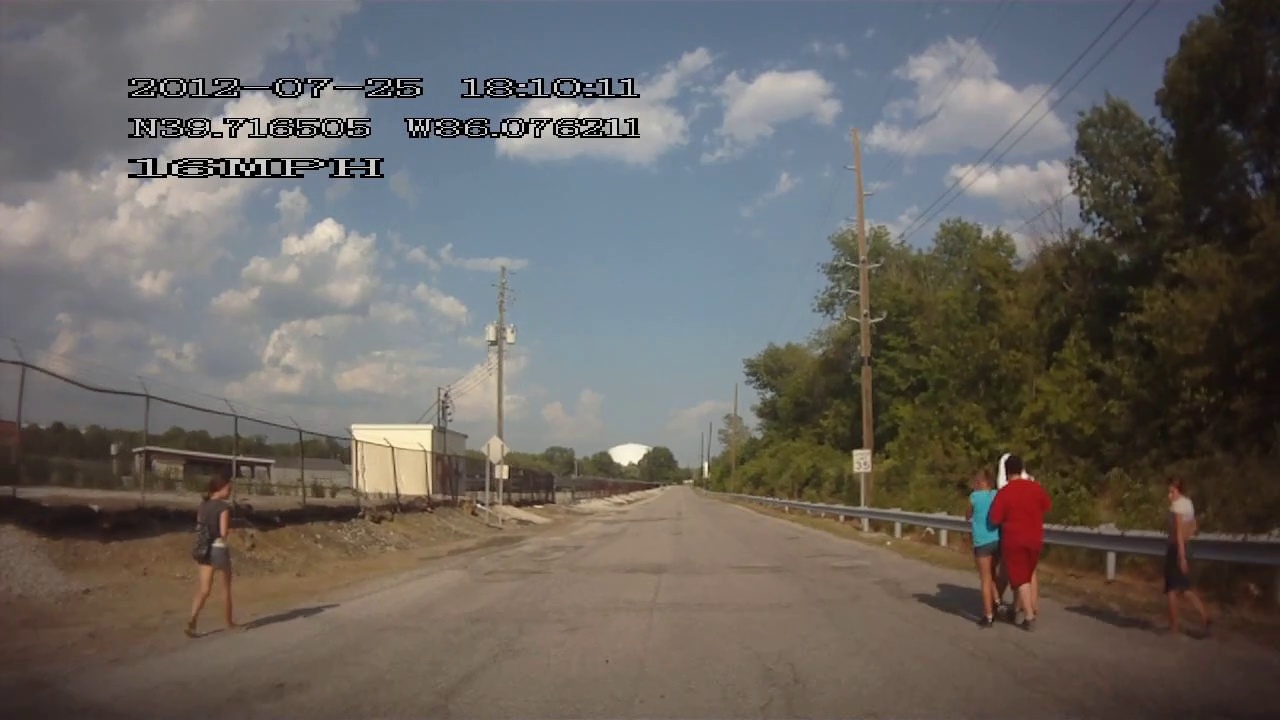}
    \caption{Facing teenagers on roadside.}
    \label{fig:f2_children}  
\end{figure}
\noindent$\mathtt{LLM_{vision}}:$
\begin{lstlisting}
From the driver's perspective, as the scene unfolds, several vulnerable pedestrians can be observed along the roadside. 

On the left, a teenage girl walks closely to the road, her position making her potentially at risk if the driver is not attentive. On the right, there are a few more individuals, including a person in red who may require additional consideration due to their slower movement or possible mobility issues. 

The driver must remain vigilant, monitoring these pedestrians movements carefully. Any sudden shifts in the pedestrians direction could lead to a hazardous situation. The open road and surrounding environment seem quiet, but the presence of these individuals necessitates a cautious approach at all times.

As the vehicle approaches, the driver should reduce speed and be prepared to stop if any pedestrian appears to cross the road. The lack of sidewalks in some sections poses an additional risk, as pedestrians may step closer to the road unexpectedly.
\end{lstlisting}
$\mathtt{LLM_{latentobjects}}:$
\begin{lstlisting}
1. {Teenagers}: [high] A teenage girl is walking closely to the road, which increases the risk of her stepping into the path of the vehicle. The car should reduce speed and maintain a safe distance.

2. {People With Disability}: [middle] A person in red may have mobility issues and could move unpredictably. The car should be prepared to stop or slow down as needed to ensure their safety.

3. {Pedestrians}: [low] Other individuals along the roadside could also pose a risk if they move closer to the road. The car should remain vigilant and adjust its speed to accommodate any sudden pedestrian movements.
\end{lstlisting}
$\mathtt{LLM_{riskhandling}}:$
\begin{lstlisting}
Object: Teenagers
1. Slow down and be prepared to stop if teenagers are near the roadway or appear to be crossing, as they may act unpredictably.
2. Yield to teenagers crossing at intersections, crosswalks, or any other designated pedestrian areas, ensuring their safety.
\end{lstlisting}
$\mathtt{LLM_{coder}}:$
\begin{lstlisting}
```python
import numpy as np

safe_distance_constraint("teenagers", 0.5)
safe_speed_constraint(3.0)

rewards = execute_plan(4)
```
\end{lstlisting}

\begin{figure}[H]
    \centering
    \includegraphics[width=0.3\linewidth]{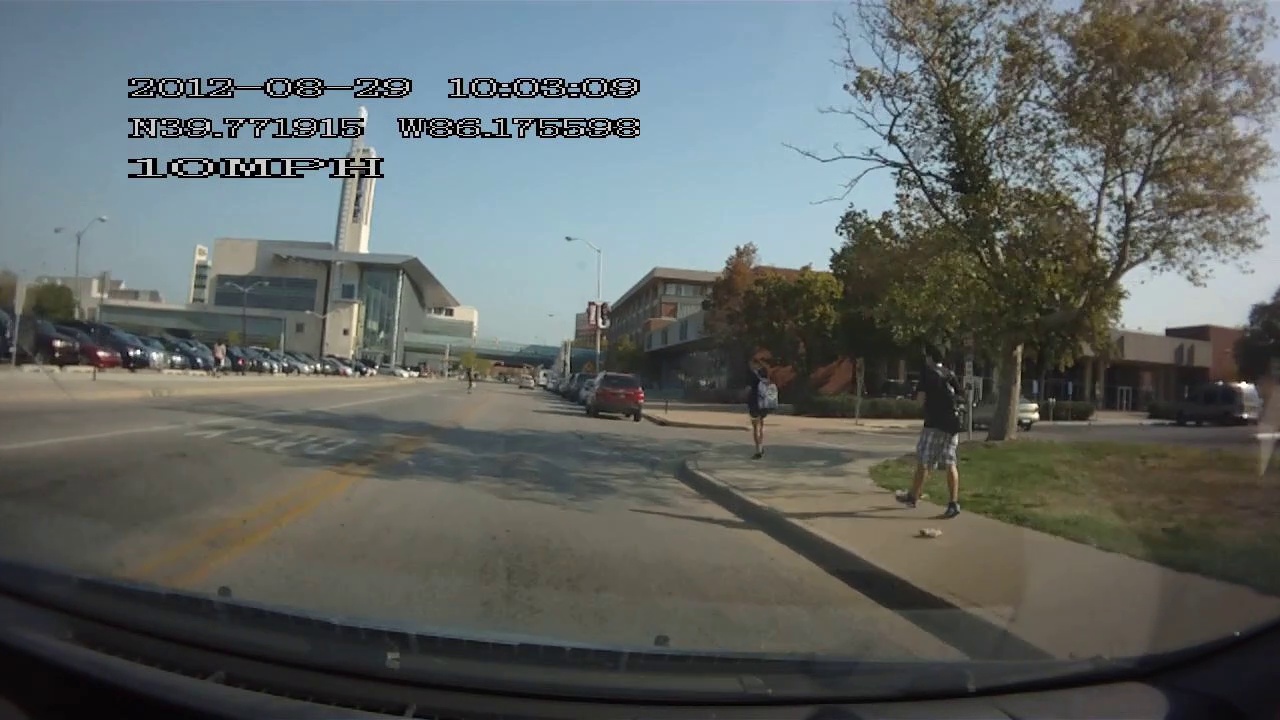}
    \caption{Facing adults on roadside.}
    \label{fig:f3_adults}  
\end{figure}
\noindent$\mathtt{LLM_{vision}}:$
\begin{lstlisting}
From the drivers perspective, the video captures a scene in which a vehicle is traveling on a road with visible pedestrians nearby. As the driver approaches, attention is drawn to several factors that could influence safe driving, particularly regarding vulnerable pedestrians.

1. **Presence of Pedestrians:** The video shows individuals, including a teenager walking along the sidewalk. This could pose a risk, especially if they unexpectedly cross the street or are distracted.

2. **Speed Consideration:** The vehicle appears to be moving at a low speed (10 mph), which is advisable in areas where pedestrians are present. This speed allows for better reaction time if a pedestrian were to step into the road.

3. **Unpredictable Movements:** The pedestrians could potentially enter the roadway without warning, particularly if they engage in activities like using a phone or talking to each other.
\end{lstlisting}
$\mathtt{LLM_{latentobjects}}:$
\begin{lstlisting}
1. **Pedestrians**: [middle] General presence of pedestrians who might enter the roadway without warning, especially if they are engaged in activities like using a phone or talking.

2. **Parked Cars**: [low] Parked cars lining the street could obscure the view of pedestrians emerging from between them.
\end{lstlisting}
$\mathtt{LLM_{riskhandling}}:$
\begin{lstlisting}
No latent viables detected. Keep Driving!
\end{lstlisting}

\begin{figure}[H]
    \centering
    \includegraphics[width=0.3\linewidth]{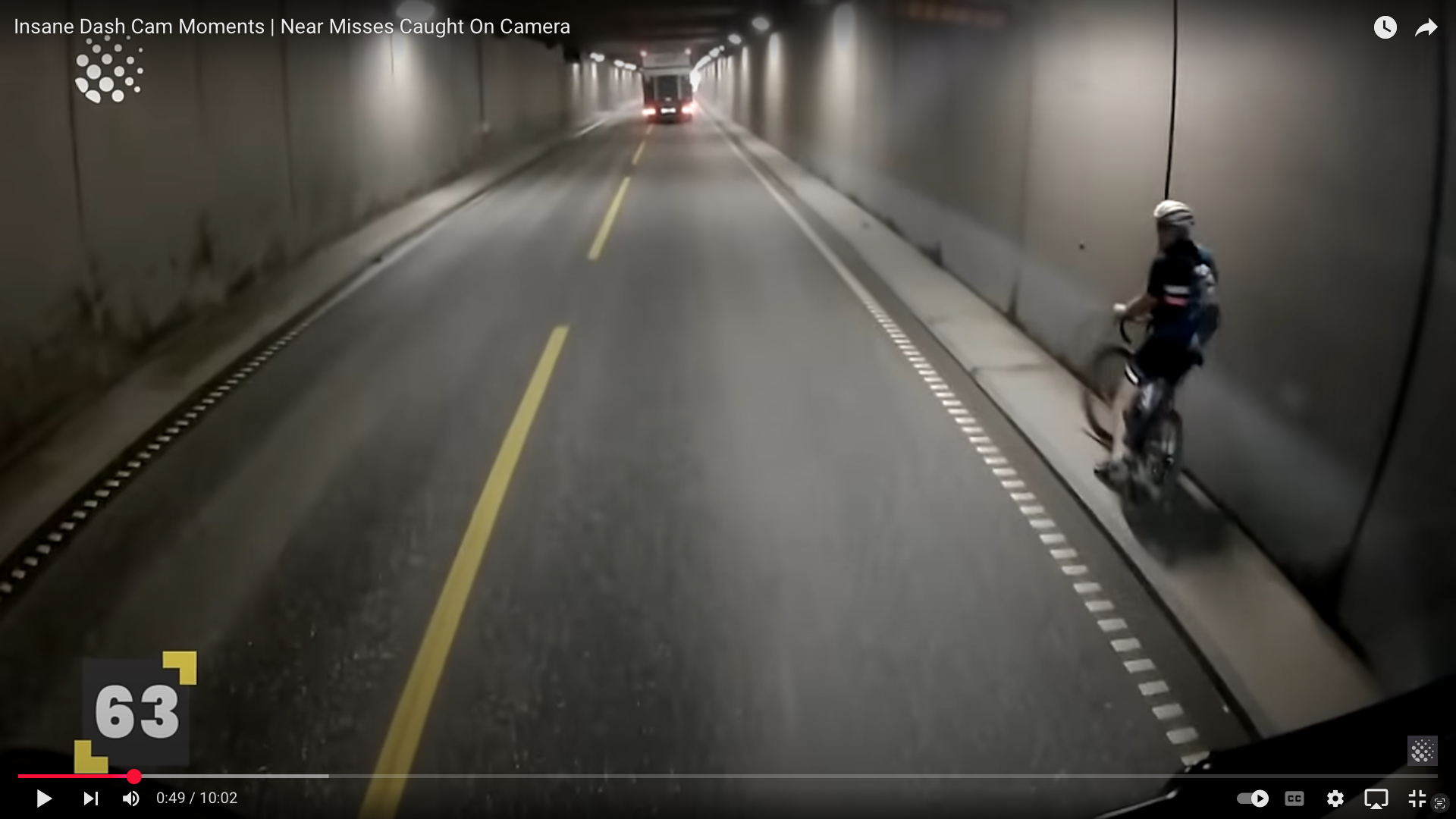}
    \caption{Facing a cyclist in the tunnel.}
    \label{fig:f4_biker}  
\end{figure}
\noindent$\mathtt{LLM_{vision}}:$
\begin{lstlisting}
From the drivers perspective, several potential hazards could arise in the tunnel scenario. 

1. **Cyclist Presence**: The cyclist on the side of the road poses a risk. If they were to lose control or make an unexpected movement, such as swerving into the lane, it could lead to a collision. Drivers need to remain vigilant and give adequate space when passing.

2. **Tunnel Environment**: The confined space of the tunnel may limit visibility and can affect the drivers ability to see potential hazards until they are very close. Drivers should be cautious, especially as speed may increase in such areas.

3. **Surface Conditions**: If the roadway is wet or slick, stopping distances increase. This means the driver needs to maintain a greater distance from the cyclist and any potential obstacles.
\end{lstlisting}
$\mathtt{LLM_{latentobjects}}:$
\begin{lstlisting}
1. {Cyclists}: [high] The presence of a cyclist on the side of the road poses a significant risk, especially if they make unexpected movements. The car should maintain a safe distance and be prepared to react quickly.

2. {Tunnel Environment}: [middle] The confined space of the tunnel can limit visibility and make it difficult to spot hazards until they are very close. The car should reduce speed and increase alertness to ensure safety.

3. {Surface Conditions}: [middle] If the road is wet or slick, it increases the stopping distance required. The car should adjust its speed accordingly and maintain a safe distance from the cyclist and any other potential obstacles.
\end{lstlisting}
$\mathtt{LLM_{riskhandling}}:$
\begin{lstlisting}
Object: Cyclists
1. Maintain a safe distance: Ensure there is adequate space between the car and the cyclist, adjusting speed if necessary to maintain a safe buffer zone.
2. Monitor cyclist movements: Continuously track the cyclist's position and speed, anticipating any sudden changes in direction or speed to react promptly.
\end{lstlisting}
$\mathtt{LLM_{coder}}:$
\begin{lstlisting}
```python
import numpy as np

safe_distance_constraint("cyclist", 0.5)
safe_speed_constraint(3)

rewards = execute_plan()
```
\end{lstlisting}

\begin{figure}[H]
    \centering
    \includegraphics[width=0.3\linewidth]{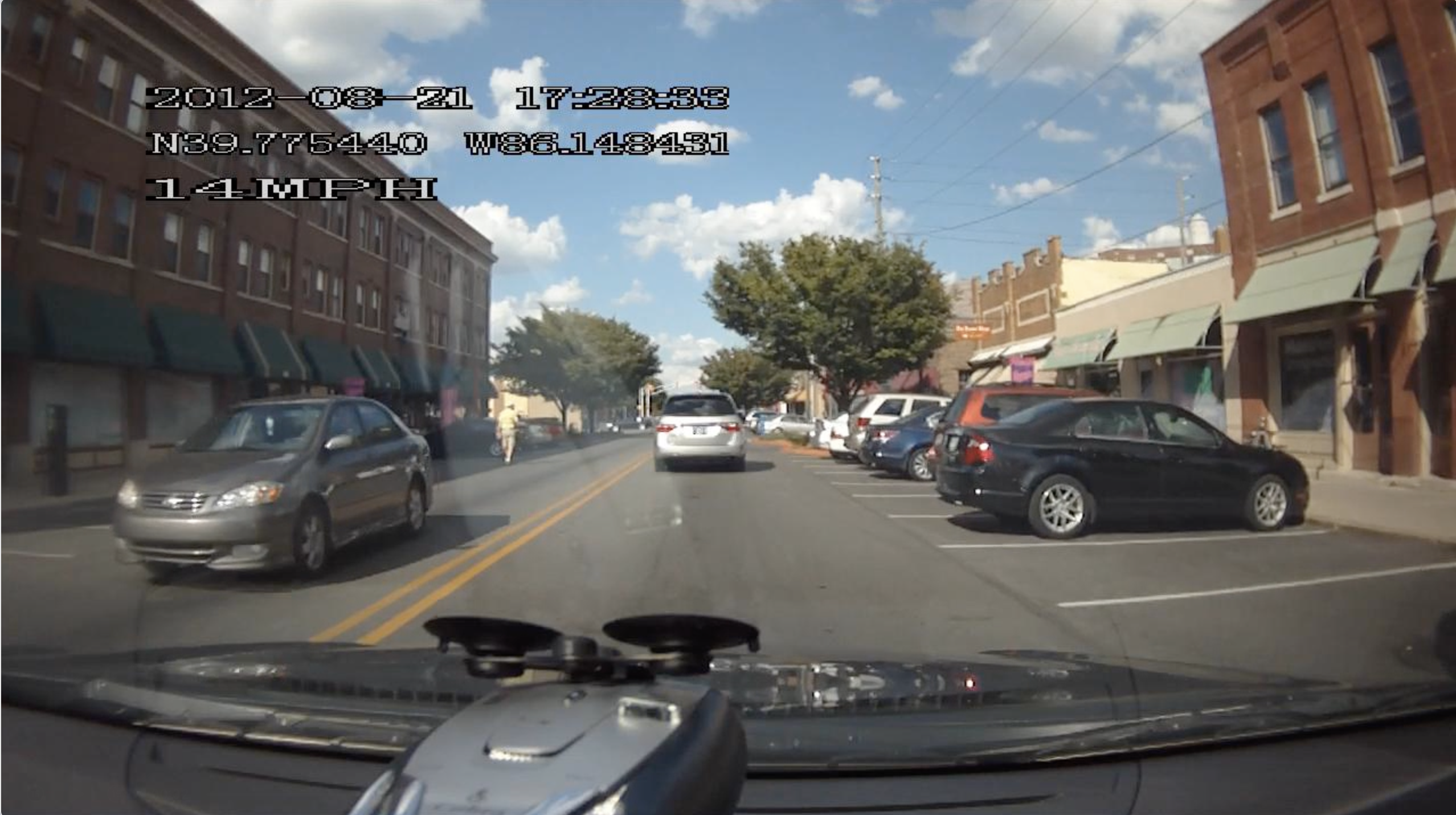}
    \caption{Facing a parking lot.}
    \label{fig:f5_parkinglot} 
\end{figure}
\noindent$\mathtt{LLM_{vision}}:$
\begin{lstlisting}
From the driver's perspective in the video, several factors contribute to the potential for increased risk while driving. 

1. **Vulnerable Pedestrians**: There may be pedestrians walking close to the road, particularly along the sidewalks. Drivers should be attentive to individuals who may step off the curb or cross the street unexpectedly, especially in busy areas.

2. **Parked Vehicles**: The presence of parked cars along the street poses a risk, as drivers exiting these vehicles may not be visible until they open a door or step into the road. This can create sudden hazards for passing vehicles.

3. **Traffic Dynamics**: The vehicle directly in front is slowing down or stopped, potentially indicating that it may be yielding to pedestrians or navigating around an obstruction. The driver needs to maintain awareness to avoid collisions.
\end{lstlisting}
$\mathtt{LLM_{latentobjects}}:$
\begin{lstlisting}
1. {Parked Cars}: [middle] Be aware of parked vehicles along the street as drivers may exit and create sudden hazards.

2. {Pedestrians}: [middle] Be cautious of pedestrians walking close to the road who may unexpectedly step off the curb or cross the street.

3. {Bicycles}: [low] Stay attentive to the potential presence of cyclists who may be riding along the road or in bike lanes.
\end{lstlisting}
$\mathtt{LLM_{riskhandling}}:$
\begin{lstlisting}
No latent viables detected. Keep Driving!
\end{lstlisting}

\begin{figure}[H]
    \centering
    \includegraphics[width=0.3\linewidth]{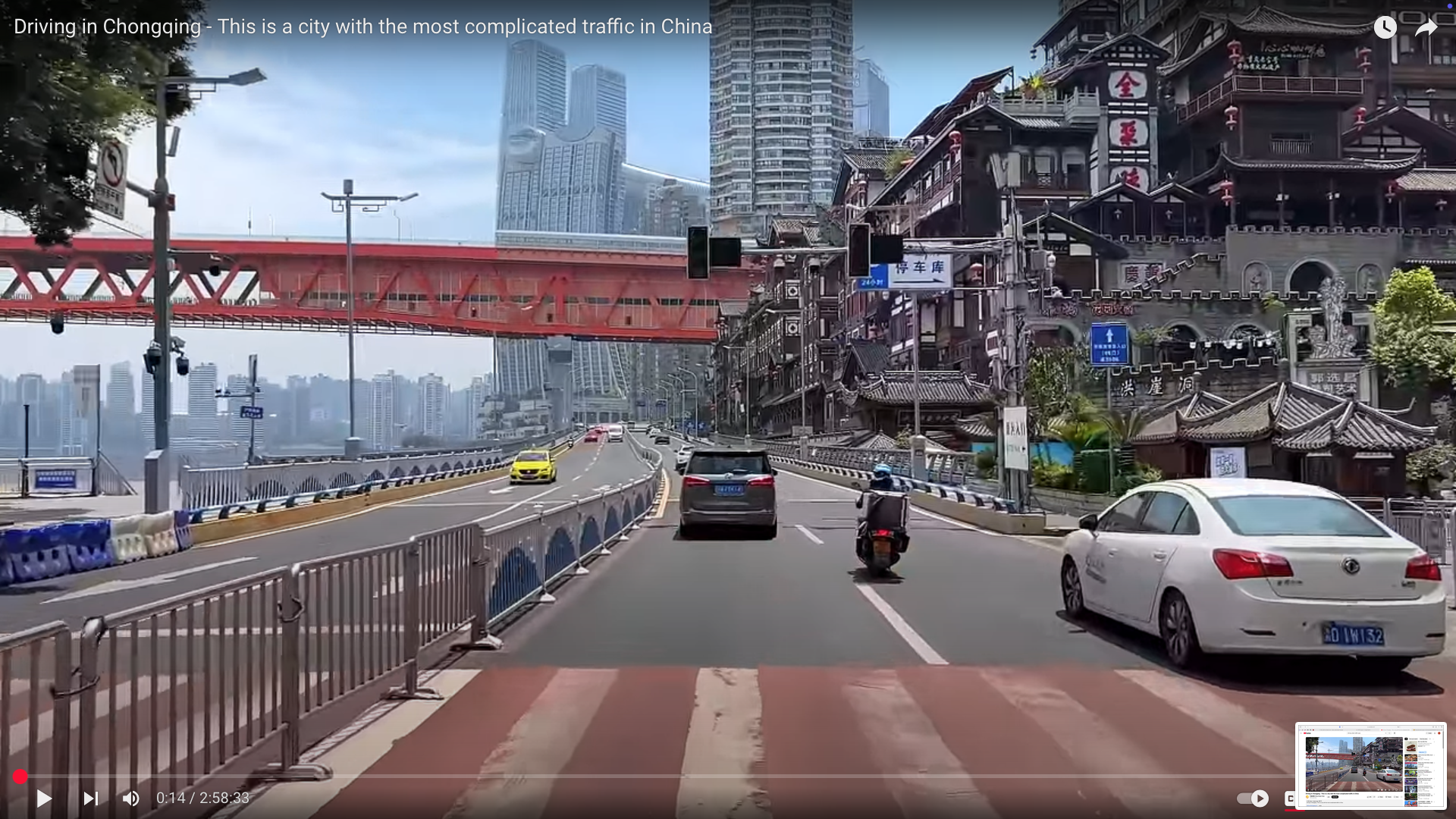}
    \caption{Facing a delivery person on the road.}
    \label{fig:f6_delivery}  
\end{figure}
\noindent$\mathtt{LLM_{vision}}:$
\begin{lstlisting}
In the video, from the driver's perspective, the scene unfolds in a bustling urban environment marked by a mix of modern skyscrapers and traditional architecture. As the vehicle moves through this complex landscape, several factors related to vulnerable pedestrians and potential risks become evident:

1. **Pedestrian Crossings**: The driver must remain vigilant at pedestrian crossings, where individuals might suddenly appear, especially in a busy area. This is crucial as pedestrians may not always adhere to traffic signals.

2. **Cyclists and Motorcyclists**: There are various cyclists and motorcyclists on the road. Their unpredictable movements can pose a risk to the driver, particularly at intersections or when they weave in and out of traffic.

3. **Narrow Roads and Barriers**: The presence of road barriers indicates an attempt to organize traffic flow, but they may also limit visibility around bends. Navigating these areas requires caution to avoid collisions with any pedestrians or vehicles obscured by the barriers.
\end{lstlisting}
$\mathtt{LLM_{latentobjects}}:$
\begin{lstlisting}
1. {Cyclists}: [high] Cyclists are present on the road and may have unpredictable movements, especially at intersections or when weaving through traffic. The car should maintain a safe distance and be prepared to react quickly to sudden changes in their path.

2. {Pedestrian Crossings}: [middle] Pedestrians might suddenly appear at crossings, particularly in busy urban areas where they may not adhere to traffic signals. The car should slow down and be ready to stop if necessary.

3. {Motorcycles}: [middle] Motorcyclists are on the road and can exhibit unpredictable behavior similar to cyclists. The car should be cautious and maintain awareness of their presence, particularly when they are weaving in and out of traffic.
\end{lstlisting}
$\mathtt{LLM_{riskhandling}}:$
\begin{lstlisting}
Object: Cyclists
1. Maintain a safe distance: Ensure a minimum of 3 feet (or as per local regulations) between the car and the cyclist when passing or driving alongside.
2. Adjust speed: Reduce speed when approaching or overtaking cyclists to allow for safe maneuvering and reaction time.
\end{lstlisting}
$\mathtt{LLM_{coder}}:$
\begin{lstlisting}
```python
import numpy as np

safe_distance_constraint("cyclists", 0.9)
safe_speed_constraint(3.0)

rewards = execute_plan(4)
```
\end{lstlisting}

\begin{figure}[H]
    \centering
    \includegraphics[width=0.3\linewidth]{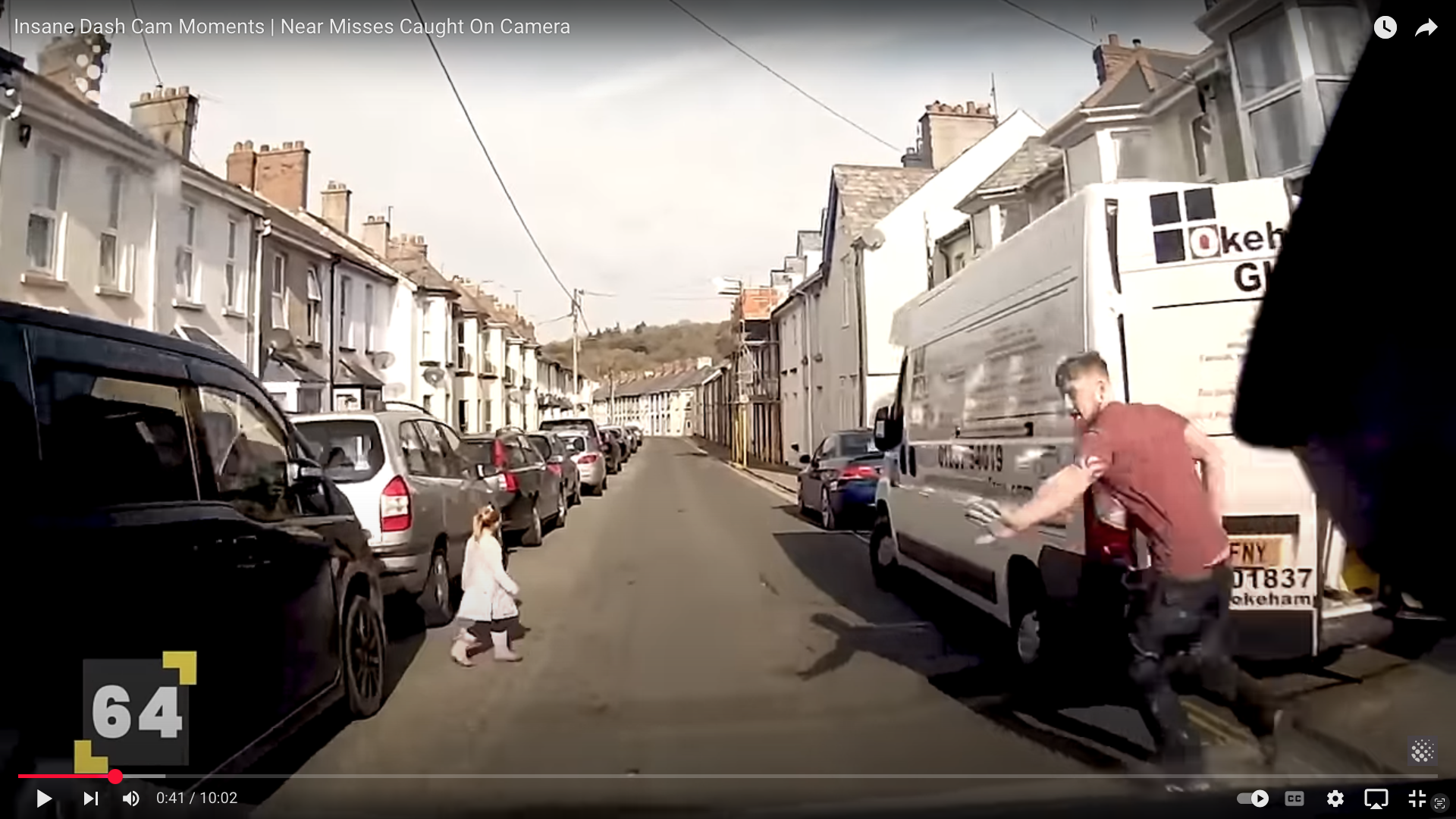}
    \caption{Facing a child and an adult in a narrow alley.}
    \label{fig:f7_children}  
\end{figure}
\noindent$\mathtt{LLM_{vision}}:$
\begin{lstlisting}
In the video from the driver's perspective, several potential risks to safe driving emerge:

1. **Vulnerable Pedestrian**: A small child is walking along the side of the road, which poses a significant risk. The driver must remain vigilant to prevent any accidents, especially since young children can be unpredictable and may suddenly dart into the street.

2. **Obstructed View**: Parked cars line both sides of the narrow street, limiting visibility. The driver may struggle to see any pedestrians or obstacles that could appear suddenly.

3. **Running Adult**: A man is running toward the child, which may indicate that the child is either in danger or needs immediate attention. This sudden movement could catch the driver off guard, requiring quick reflexes to react safely.
\end{lstlisting}
$\mathtt{LLM_{latentobjects}}:$
\begin{lstlisting}
1. {Children}: [high] A small child walking along the side of the road poses a significant risk due to their unpredictability. The car should maintain a safe distance and be prepared to stop suddenly if the child moves toward the street.

2. {Parked Cars}: [middle] Parked cars on both sides of the narrow street limit visibility, making it difficult to see pedestrians or obstacles that may appear suddenly. The car should proceed cautiously and be prepared for unexpected movements.

3. {Pedestrians}: [low] A running adult heading toward the child may indicate a potential emergency or need for immediate attention. The car should be ready to react quickly to any sudden movements or stops that may be required.
\end{lstlisting}
$\mathtt{LLM_{riskhandling}}:$
\begin{lstlisting}
Object: Children
1. Slow down and be prepared to stop if children are detected near or on the roadway, especially in residential areas or near schools.
2. Yield to children crossing the road, ensuring they have safely reached the other side before proceeding.
\end{lstlisting}
$\mathtt{LLM_{coder}}:$
\begin{lstlisting}
```python
import numpy as np

safe_distance_constraint("children", 0.5)
safe_speed_constraint(3.0)

rewards = execute_plan(4)
```
\end{lstlisting}

\begin{figure}[H]
    \centering
    \includegraphics[width=0.3\linewidth]{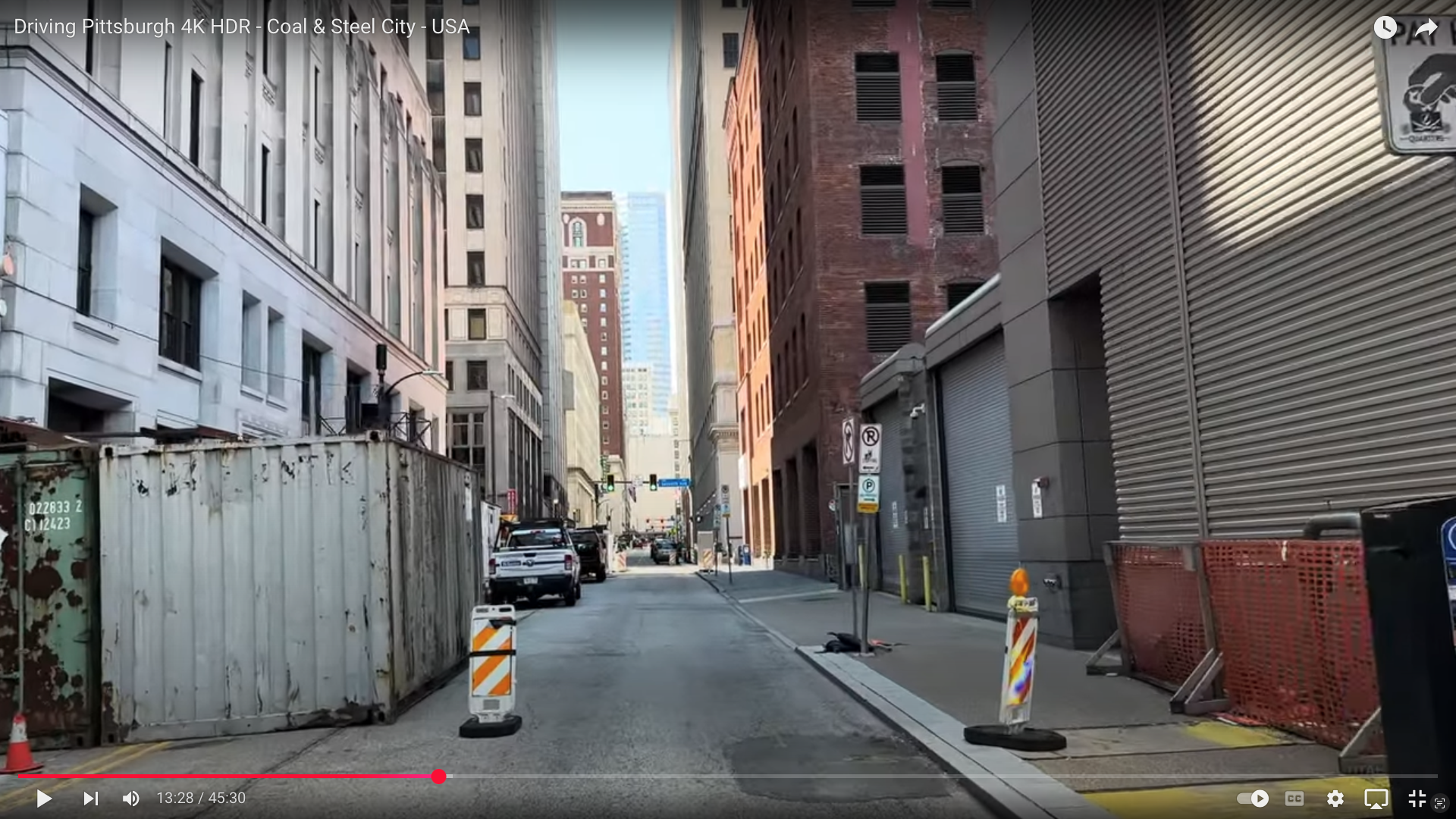}
    \caption{Facing a narrow alley with some road signs.}
    \label{fig:f8_sign}  
\end{figure}
\noindent$\mathtt{LLM_{vision}}:$
\begin{lstlisting}
From the driver's perspective in the video, a few key observations regarding vulnerable pedestrians and other latent risk factors can be highlighted:

1. **Narrow Streets**: The setting appears to be in a narrow urban area, which can increase the likelihood of pedestrians being in close proximity to vehicles. Drivers must be particularly vigilant for individuals who may suddenly step into the roadway.

2. **Construction Barriers and Equipment**: The presence of barriers and containers along the sides of the street can obstruct visibility. This can create blind spots where pedestrians or cyclists might be hidden from view, increasing the risk of accidents.

3. **Sidewalk Conditions**: The quality of the sidewalks is crucial. If they are in poor condition or partially blocked, pedestrians may be compelled to walk closer to the roadway, which increases their vulnerability.
\end{lstlisting}
$\mathtt{LLM_{latentobjects}}:$
\begin{lstlisting}
1. {Pedestrians}: [middle] The narrow streets and potential for pedestrians to step into the roadway require the car to be vigilant and prepared to stop suddenly.

2. {Construction Zones}: [middle] Construction barriers and equipment can create blind spots, making it difficult to see pedestrians or cyclists who might be hidden.

3. {Parked Cars}: [low] Parked cars can obstruct views of oncoming pedestrians or cyclists, increasing the risk of an unexpected encounter.
\end{lstlisting}
$\mathtt{LLM_{riskhandling}}:$
\begin{lstlisting}
No latent viables detected. Keep Driving!
\end{lstlisting}

\begin{figure}[H]
    \centering
    \includegraphics[width=0.3\linewidth]{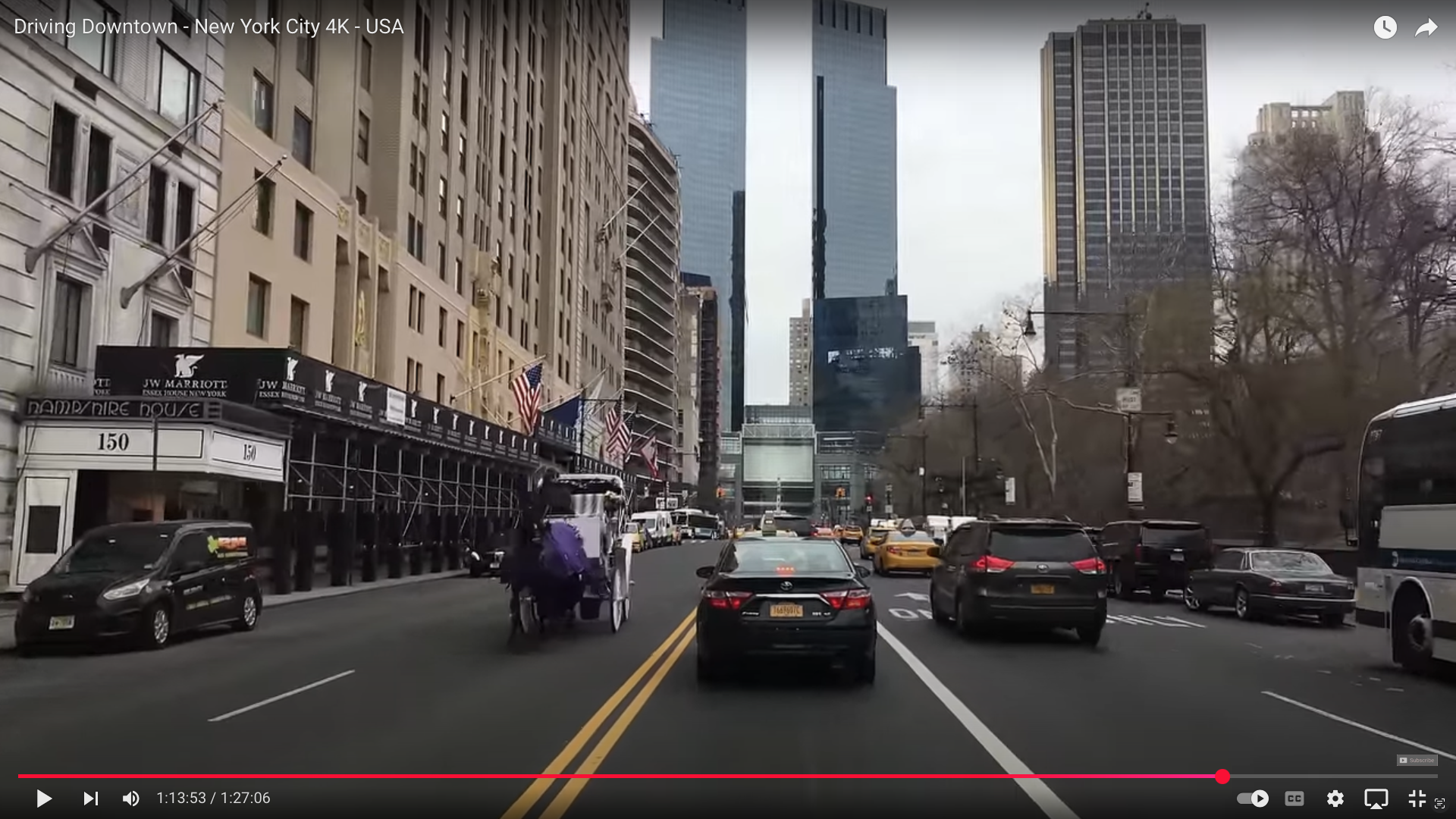}
    \caption{Facing a tourist carriage on the opposite side of the road.}
    \label{fig:f9_animal}  
\end{figure}
\noindent$\mathtt{LLM_{vision}}:$
\begin{lstlisting}
From the driver's perspective in the video, several potential hazards and vulnerable pedestrians can be observed:

1. **Horse-Drawn Carriage**: A visible horse-drawn carriage poses a unique risk. The driver's attention should be heightened around such vehicles, as they can be slower and less predictable than motorized traffic.

2. **Pedestrians**: With the busy streets of New York City, there are likely many pedestrians crossing at various points, especially at intersections. Drivers must remain vigilant for jaywalkers and those stepping off the curb, particularly when distracted by nearby attractions or vehicles.

3. **Bicyclists and Cyclists**: Shared roadways might also accommodate cyclists, who can suddenly appear in the driver's path. Maintaining a safe distance and being aware of bike lanes is crucial.
\end{lstlisting}
$\mathtt{LLM_{latentobjects}}:$
\begin{lstlisting}
1. **Horse-Drawn Carriage**: [high] The presence of a horse-drawn carriage poses a risk due to its slower speed and unpredictability compared to motorized vehicles. The car should be prepared to adjust its speed and maintain a safe distance.

2. **Pedestrians**: [high] There are many pedestrians, especially near intersections, who may jaywalk or step off the curb unexpectedly. The car should be ready to stop or slow down to avoid collisions.

3. **Bicycles**: [middle] Cyclists may suddenly appear in the path of the vehicle. The car should maintain a safe distance and be aware of bike lanes to ensure the safety of cyclists.
\end{lstlisting}
$\mathtt{LLM_{riskhandling}}:$
\begin{lstlisting}
Object: Horse-drawn Carriages
1. Maintain a safe following distance to allow for any sudden stops or changes in speed by the horse-drawn carriage.
2. Reduce speed and prepare to yield if the carriage is turning or maneuvering, ensuring the safety of both the carriage and its occupants.
\end{lstlisting}
$\mathtt{LLM_{coder}}:$
\begin{lstlisting}
```python
import numpy as np

safe_distance_constraint("horse_drawn_carriage", 0.5)
safe_speed_constraint(3)

rewards = execute_plan(4)
```
\end{lstlisting}

\begin{figure}[H]
    \centering
    \includegraphics[width=0.3\linewidth]{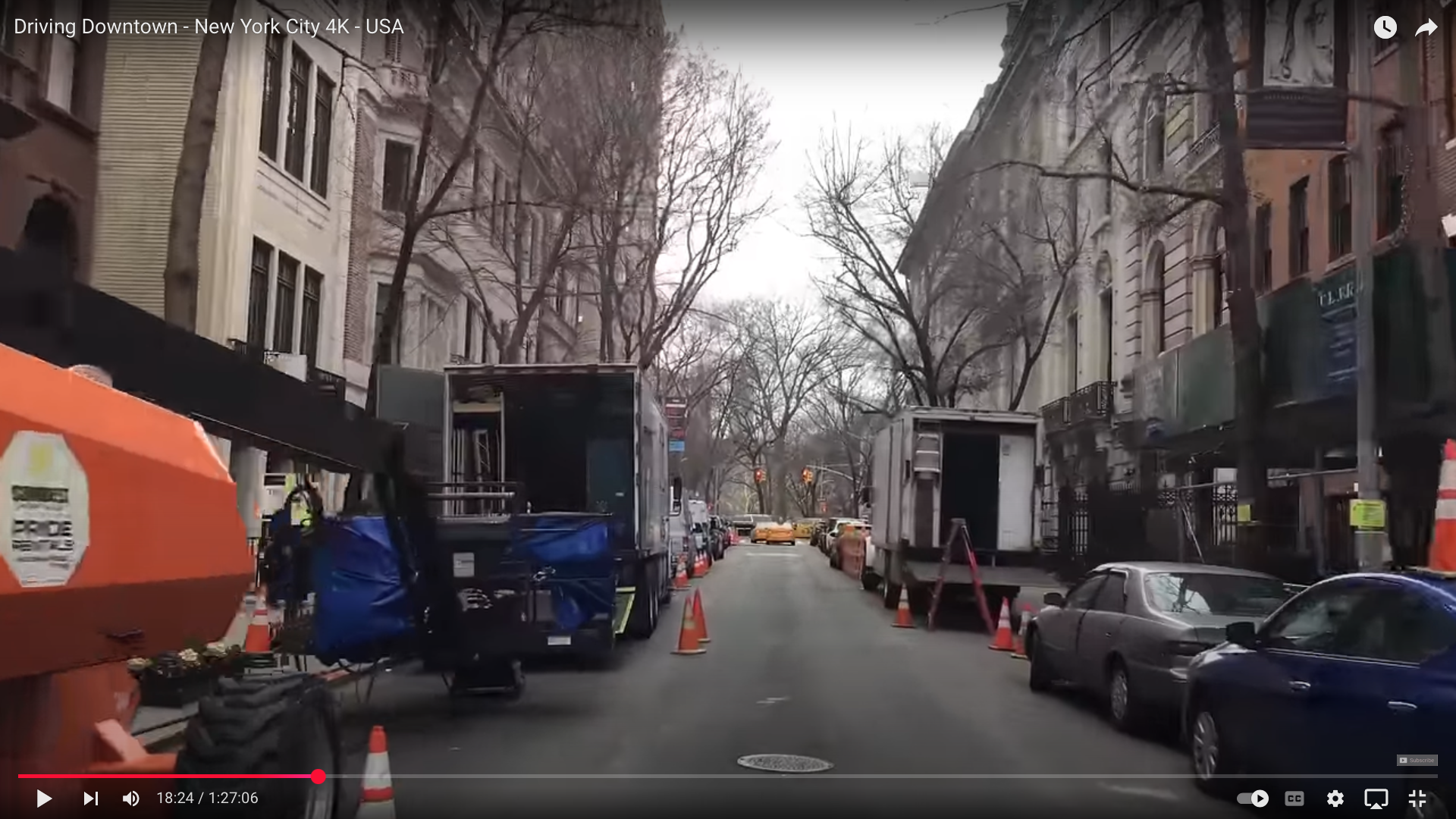}
    \caption{Facing a school bus}
    \label{fig:f10_truck}  
\end{figure}
\noindent$\mathtt{LLM_{vision}}:$
\begin{lstlisting}
From the driver's perspective, navigating the streets depicted in the video involves several considerations for safety, particularly related to vulnerable pedestrians and latent risk factors. 

As the vehicle approaches the narrow street lined with parked trucks and construction equipment, the driver must remain alert to pedestrians who may be crossing or walking close to the curb. The presence of construction cones suggests that some areas are cordoned off, which can create confusion about where it is safe to walk or drive.

The parked vehicles on both sides of the street limit visibility. This means that pedestrians, especially those pushing strollers or carrying packages, might suddenly appear, necessitating a cautious approach. Additionally, the potential for cyclists weaving through traffic adds another layer of risk.

The driver should also be wary of possible distractions from the equipment and workers present, who may move unexpectedly or fail to notice oncoming vehicles. The street's narrowness indicates limited maneuverability, making it critical for the driver to maintain a slow speed and be prepared to stop if necessary.
\end{lstlisting}
$\mathtt{LLM_{latentobjects}}:$
\begin{lstlisting}
1. {Pedestrians}: [high] The presence of parked trucks and construction equipment suggests that pedestrians may be crossing or walking close to the curb, potentially appearing suddenly from behind obstacles.

2. {Construction Zones}: [middle] The construction cones and equipment indicate areas that are cordoned off, which can create confusion about safe paths for both walking and driving, necessitating careful navigation.

3. {Cyclists}: [middle] The potential for cyclists weaving through traffic in the narrow street adds a layer of risk, requiring the vehicle to be vigilant and prepared for sudden movements.
\end{lstlisting}
$\mathtt{LLM_{riskhandling}}:$
\begin{lstlisting}
Object: Pedestrians
1. Yield: If pedestrians are crossing at any intersection without a traffic light, in marked crosswalks, on a sidewalk crossing a driveway or alley, or when the car is turning a corner and pedestrians are crossing with the light, the car must yield to them.
2. Monitor and Adjust Speed: Continuously monitor pedestrian movement and adjust speed accordingly, especially for disadvantaged pedestrians such as children, elderly, teenagers, and people with disabilities.
\end{lstlisting}
$\mathtt{LLM_{coder}}:$
\begin{lstlisting}
```python
import numpy as np

safe_distance_constraint("pedestrians", 0.5)
safe_speed_constraint(3.0)

rewards = execute_plan(4)
```

\end{lstlisting}

\subsection{Simulation results}\label{sec:av_results}
Among the 10 scenarios presented above, 7 of them involve different safe constraint thresholds that trigger risk-aware behaviors, while the remaining 3 proceed without any detected risks. 
We define $z = 1$ indicate that a latent object becomes an apparent risk, and $z = 0$ to indicate that no such object is observed, where 
\begin{align}
\mathbb{P}(Z = z) = p^z (1 - p)^{1 - z}, \quad \text{for } z \in \{0, 1\} \\
p \overset{\text{i.i.d.}}{\sim} P, \quad \text{where } P = \{0.1, 0.2, \dots, 0.9\}
\end{align}
The distance from the starting point to the destination is fixed at $7\,\mathrm{m}$. During the traveling, occluded obstacles may become visible as apparent objects.
If a latent object does not become an apparent risk, we set the minimum distance to the default value $l_{\text{min}} = 7$. As shown in Fig.~\ref{fig:min_distance_all}, the proposed method maintains a similar minimum distance to risky objects as the conservative baselines, indicating that it achieves comparable safety margins.
Time-to-travel is used to evaluate the performance of each method, measuring the total duration from the starting point to the destination.
As shown in Fig.~\ref{fig:time2travel_all}, the proposed method achieves significantly shorter travel times compared to conservative baselines, demonstrating its ability to balance safety and efficiency under varying levels of risk.

\begin{figure}[ht]
    \centering
    \begin{subfigure}[b]{0.45\textwidth}
        \centering
\begin{tikzpicture}
    \begin{axis}[
        xlabel={Probability (\%)},
        ylabel={Distance (m)},
        ylabel shift=-6pt,
        grid=both,
        grid style={dashed,gray!30},
        legend pos=south west, 
        width=6cm,
        height=6cm,
        xmin=0, xmax=100,
        ymin=2, ymax=7,
        every axis label/.append style={font=\scriptsize},
        every tick label/.append style={font=\scriptsize},
        legend style={font=\scriptsize, at={(axis cs:3.2,2.2)}, anchor=south west}
    ]

    \addplot [blue, dashed, mark=*, mark size=1] table [x=prob, y=typical] {min_distance.dat};
    \addlegendentry{Typical}

    \addplot [red, mark=*, mark size=2] table [x=prob, y=proposed] {min_distance.dat};
    \addlegendentry{Proposed}

    \addplot [black, dashed, mark=*, mark size=1] table [x=prob, y=con] {min_distance.dat};
    \addlegendentry{Conservative}

    \end{axis}
\end{tikzpicture}
        \caption{Average minimum distance to risky-critical objects across varying probabilities of being apparent risks.}
        \label{fig:min_distance_all}
    \end{subfigure}%
    \hfill
    \begin{subfigure}[b]{0.45\textwidth}
        \centering
        \begin{tikzpicture}
    \begin{axis}[
        xlabel={Probability (\%)},
        ylabel={Time-to-travel (s)},
        ylabel shift=-6pt,
        grid=both,
        grid style={dashed,gray!30},
        width=6cm,
        height=6cm,
        xmin=0, xmax=100,
        ymin=1, ymax=4,
        legend style={font=\scriptsize, draw, fill=none, at={(axis cs:3.2,2.4)}, anchor=south west},
        every axis label/.append style={font=\scriptsize},
        every tick label/.append style={font=\scriptsize}
    ]
    
    \addplot [blue, dashed, mark=*, mark size=1] table [x=prob, y=typical] {time2travel.dat};
    \addlegendentry{Typical}
    
    \addplot [black, dashed, mark=*, mark size=1] table [x=prob, y=con] {time2travel.dat};
    \addlegendentry{Conservative}
    
    \addplot [red, mark=*, mark size=2] table [x=prob, y=proposed] {time2travel.dat};
    \addlegendentry{Proposed}

    \end{axis}
\end{tikzpicture}
        \caption{Average time-to-travel to destination across varying probabilities of latent objects being apparent risks.}
        \label{fig:time2travel_all}
    \end{subfigure}
    \caption{Safety and Performance over different probability of latent risk becoming apparent risk.}
    \label{fig:av_all}
\end{figure}

\end{document}